\documentclass[a4paper,11pt]{article}
\pdfoutput=1
\usepackage[utf8]{inputenc}
\usepackage{jheppub}
\usepackage{graphicx}
\usepackage{caption}
\usepackage{subcaption}
\usepackage{hyperref}
\usepackage[T1]{fontenc}
\usepackage{lmodern}
\usepackage{cancel}
\usepackage{color}

\newcommand{\be}{\begin{equation}}
\newcommand{\ee}{\end{equation}}
\newcommand{\bea}{\begin{eqnarray}}
\newcommand{\eea}{\end{eqnarray}}

\newcommand{\nn}{\nonumber\\}
\newcommand{\EF}[0]{$\text{EF}^{+}${}}

\def\cO{{\cal O}}

\def\la{\langle}
\def\ra{\rangle}

\def\tp{\tau_{\Pi}}

\def\CA{\mathcal{A}}

\def\CC{\mathcal{C}}

\def\CN{\mathcal{N}}

\def\CO{\mathcal{O}}

\def\CW{\mathcal{W}}

\def\lgb{\lambda_{\scriptscriptstyle GB}}
\def\ggb{\gamma_{\scriptscriptstyle GB}}

\def\htime{t_{\text{hyd}}}
\def\htau{\tau_{\text{hyd}}}

\def\delp{{\nabla_{\perp}}}

\def\tp{\tau_{\Pi}}

\newcommand{\tl}[0]{L}
\newcommand{\tlgb}[0]{\tilde{L}}
\newcommand{\tpp}[0]{\tau_{+}}
\newcommand{\axf}[0]{G}

\preprint{UTTG-06-17, HIP-2017-16/TH}

\title{Holographic constraints on Bjorken hydrodynamics at finite coupling}

\author[a,b,e]{Brandon S. DiNunno,}
\author[c]{Sa\v{s}o Grozdanov,}
\author[d]{Juan F. Pedraza}
\author[e]{and Steve Young}

\affiliation[a]{Theory Group, Department of Physics, The University of Texas at Austin\\
2515 Speedway, Stop C1608, Austin, TX 78712, USA}
\affiliation[b]{Helsinki Institute of Physics, University of Helsinki\\
P.O. Box 64, Helsinki FIN-00014, Finland}
\affiliation[c]{Instituut-Lorentz for Theoretical Physics, Leiden University \\
Niels Bohrweg 2,  Leiden 2333 CA, The Netherlands }
\affiliation[d]{Institute for Theoretical Physics, University of Amsterdam\\
Science Park 904, 1090 GL Amsterdam, The Netherlands}
\affiliation[e]{Theory Group, Maxwell Analytics LLC\\
600 Sabine Street, Austin, TX 78701, USA}

\emailAdd{bsd86@physics.utexas.edu}
\emailAdd{grozdanov@lorentz.leidenuniv.nl}
\emailAdd{jpedraza@uva.nl}
\emailAdd{scyoung@zippy.ph.utexas.edu}

\abstract{In large-$N_c$ conformal field theories with classical holographic duals, inverse coupling constant corrections are obtained by considering higher-derivative terms in the corresponding gravity theory. In this work, we use type IIB supergravity and bottom-up Gauss-Bonnet gravity to study the dynamics of boost-invariant Bjorken hydrodynamics at finite coupling. We analyze the time-dependent decay properties of non-local observables (scalar two-point functions and Wilson loops) probing the different models of Bjorken flow and show that they can be expressed generically in terms of a few field theory parameters. In addition, our computations provide an analytically quantifiable probe of the coupling-dependent validity of hydrodynamics at early times in a simple model of heavy-ion collisions, which is an observable closely analogous to the hydrodynamization time of a quark-gluon plasma. We find that to third order in the hydrodynamic expansion, the convergence of hydrodynamics is improved and that generically, as expected from field theory considerations and recent holographic results, the applicability of hydrodynamics is delayed as the field theory coupling decreases.}

\begin{document}
\maketitle
\flushbottom

\section{Introduction}
\label{sec:intro}

Hydrodynamics is an effective theory \cite{Dubovsky:2011sj, Endlich:2012vt,
  Grozdanov:2013dba,Nicolis:2013lma,Kovtun:2014hpa,Harder:2015nxa,Grozdanov:2015nea,Crossley:2015evo,Glorioso:2017fpd,Haehl:2015foa,Haehl:2015uoc,Torrieri:2016dko,Glorioso:2016gsa,Gao:2017bqf,Jensen:2017kzi}
of collective long-range excitations in liquids, gases and plasmas. Its applicability across energy
scales has made it a popular and fruitful field of research for over a century. A particularly powerful aspect of hydrodynamics is the fact that it provides a good effective description over a vast range of coupling constant strengths of the underlying microscopic constituents. This is true so long as the mean-free-time between microscopic collisions $t_{\text{mft}}$ is smaller than the typical time scale (of observations) over which hydrodynamics is applicable, $t_{\text{mft}} \ll t_{\text{hyd}}$. At weak coupling, the underlying microscopic dynamics can be described in terms of kinetic theory \cite{ferziger-kaper-book,silin-book,kvasnikov-book,chapman-book,saint-raymond-book,KGB,groot-book,Arnold:2002zm,Arnold:2003zc}, which relies on the concept of quasiparticles. On the other hand, at very strong coupling, the applicability of hydrodynamics to the infrared (IR) dynamics of various systems without quasiparticles has been firmly established much more recently through the advent of gauge-gravity duality (holography) \cite{Policastro:2001yc,Policastro:2002se,Kovtun:2004de,Son:2007vk}. In infinitely strongly coupled CFTs with a simple holographic dual, the mean-free-time is set by the Hawking temperature of the dual black hole, $t_{\text{mft}} \sim \hbar / k_B T$.\footnote{We will henceforth set $\hbar = c = k_B = 1$.} In a CFT in which temperature is the only energy scale, this implies that hydrodynamics universally applies to the IR regime of strongly coupled systems for $\omega / T \ll 1$, where the frequency scales as $\omega \sim 1 / \htime$ (and similarly for momenta, $q / T \ll 1$).

A natural question that then emerges is as follows: how does the range of applicability of hydrodynamics depend on the coupling strength of the underlying microscopic quantum field theory? Qualitatively, using simple perturbative kinetic theory arguments (see e.g. a recent work by Romatschke \cite{Romatschke:2015gic} or Ref. \cite{Florkowski:2017olj}), one expects the reliability of hydrodynamics to decrease (at some fixed $\omega / T$ and $q/T$) with decreasing coupling constant $\lambda$. The reason is that, typically, the mean-free-time increases with decreasing $\lambda$. From the strongly coupled, non-perturbative side, the same picture recently emerged in holographic studies of (inverse) coupling constant corrections to infinitely strongly coupled systems in \cite{Grozdanov:2016vgg,Grozdanov:2016zjj,Andrade:2016rln,Grozdanov:2016fkt},\footnote{Aspects of the coupling constant dependent quasinormal spectrum in $\CN=4$ theory were first analyzed in refs. \cite{Stricker:2013lma,Waeber:2015oka}.} which we will further investigate in this work.

In holography, in the limit of infinite number of colors $N_c$ of the dual gauge theory, inverse 't Hooft coupling constant corrections correspond to higher derivative gravity $\alpha'$ corrections to the classical bulk supergravity. In maximally supersymmetric $\CN = 4$ Yang-Mills (SYM) theory, dual to the IR limit of ten-dimensional type IIB string theory, the leading-order corrections to the gravitational sector (including the five-form flux and the dilaton), are given by the action \cite{Green:2003an,deHaro:2003zd,Green:2005qr,Paulos:2008tn,Myers:2008yi}
\begin{align}
S_{\scriptscriptstyle IIB} = \frac{1}{2\kappa_{10}^2} \int d^{10} x \sqrt{-g} \left( R - \frac{1}{2} \left(\partial \phi\right)^2 - \frac{1}{4\cdot 5!} F_5^2 + \gamma e^{-\frac{3}{2} \phi} \CW + \ldots \right)\,,
\label{eq:10DAct}
\end{align}
compactified on $S^5$, where $\gamma = \alpha'^3 \zeta(3) / 8$, $\kappa_{10} \sim 1/N_c$ and the
term $\CW$ is proportional to fourth-power (eight derivatives of the metric) contractions of the Weyl tensor
\begin{align}
\label{eq:Wterm}
\CW = C^{\alpha\beta\gamma\delta}C_{\mu\beta\gamma\nu} C_{\alpha}^{~\rho\sigma\mu} C^{\nu}_{~\rho\sigma\delta} + \frac{1}{2} C^{\alpha\delta\beta\gamma} C_{\mu\nu\beta\gamma} C_{\alpha}^{~\rho\sigma\mu} C^\nu_{~\rho\sigma\delta}\,.
\end{align}
The 't Hooft coupling of the dual $\CN=4$ CFT is related to $\gamma$ by the following expression: $\gamma  = \lambda^{-3/2}\zeta (3) L^6/8$, where $L$ is the anti-de Sitter (AdS) length scale. For this reason, perturbative corrections in $\gamma \sim \alpha'^3$ are dual to perturbative corrections in $1 / \lambda^{3/2}$.

Another family of theories, which have been proven to be a useful laboratory for the studies of coupling constant dependence in holography, are curvature-squared theories \cite{Grozdanov:2014kva,Grozdanov:2015asa,Grozdanov:2016vgg,Grozdanov:2016zjj,Andrade:2016rln,Grozdanov:2016fkt} with the action given by
\begin{align}
S_{\scriptscriptstyle R^2} = \frac{1}{2 \kappa_5^2 } \int d^5 x \sqrt{-g} \left[ R - 2 \Lambda + L^2 \left( \alpha_1 R^2 + \alpha_2 R_{\mu\nu} R^{\mu\nu} + \alpha_3 R_{\mu\nu\rho\sigma} R^{\mu\nu\rho\sigma}  \right) \right] \, .
\label{eq:R2Th}
\end{align}
Although the dual(s) of \eqref{eq:R2Th} are generically unknown,\footnote{In some cases, such terms
  can be interpreted as $1/N_c$ corrections rather than coupling constant corrections
  \cite{Kats:2007mq,Buchel:2008vz}. See also \cite{Grozdanov:2016fkt} for a recent discussion of
  these issues.} one can treat curvature-squared theories as invaluable bottom-up constructions for
investigations of coupling constant corrections on dual observables of hypothetical
CFTs.\footnote{It is well known that curvature-squared terms appear in various effective IR limits
  of e.g. bosonic and heterotic string theory (see e.g. \cite{Gross:1986mw}).} From this point of view, it is natural to interpret the $\alpha_n$ coefficients as proportional to $\alpha'$. Since the action \eqref{eq:R2Th}
results in higher-derivative equations of motion, the $\alpha_n$ need to be treated perturbatively, i.e. on the same footing as the $\gamma \sim \alpha'^3$ corrections in $\CN =4 $ SYM. The latter restriction can be lifted if one instead considers a curvature-squared action with the $\alpha_n$ coefficients chosen such that $\alpha_1 = -4 \alpha_2 = \alpha_3$. The resulting theory, known as the Gauss-Bonnet theory
\begin{align}
\label{eq:GBaction}
S_{\scriptscriptstyle GB} = \frac{1}{2\kappa_5^2} \int d^5 x \sqrt{-g} \left[ R  + \frac{12}{L^2} + \frac{\lgb L^2}{2} \left( R^2 - 4 R_{\mu\nu} R^{\mu\nu} + R_{\mu\nu\rho\sigma} R^{\mu\nu\rho\sigma} \right) \right] \, ,
\end{align}
results in second-derivative equations of motions, therefore enabling one to treat the
Gauss-Bonnet coupling, $\lgb \in (-\infty, 1/4 ]$, at least formally,
non-perturbatively.\footnote{Note that through the use of gravitational field redefinitions, the
  action \eqref{eq:R2Th} and any holographic results that follow from it can be reconstructed from
  corresponding calculations in $\CN = 4$ theory at infinite coupling ($\alpha_n = 0$) and
  perturbative Gauss-Bonnet results. See e.g. \cite{Brigante:2007nu,Grozdanov:2014kva}.} Even though
this theory is known to suffer from various UV causality problems and instabilities
\cite{Brigante:2007nu,Brigante:2008gz,Hofman:2008ar,Buchel:2009tt,Hofman:2009ug,Buchel:2009sk,Camanho:2009vw,Camanho:2014apa,Reall:2014pwa,Willison:2014era,
  Papallo:2015rna,Willison:2015vaa,Cheung:2016wjt,Andrade:2016yzc,Afkhami-Jeddi:2016ntf,Konoplya:2017ymp,Konoplya:2017lhs,Konoplya:2017zwo},
one may still treat Eq. \eqref{eq:GBaction} as an effective theory which can, for sufficiently low energy and momentum, provide a well-behaved window into non-perturbative coupling constant corrections to the low-energy part of the spectrum. This point of view was advocated and investigated in
\cite{Grozdanov:2014kva,Grozdanov:2015asa,Grozdanov:2016vgg,Grozdanov:2016fkt} where it was found
that a variety of weakly coupled properties of field theories, including the emergence of
quasiparticles, were successfully recovered not only from the type IIB supergravity action \eqref{eq:10DAct} but also from the Gauss-Bonnet theory \eqref{eq:GBaction}.\footnote{We refer the readers to Ref.  \cite{Grozdanov:2016fkt} for a more detailed review of known causality problems and instabilities of the Gauss-Bonnet theory.} An important fact to note is that these weakly coupled predictions follow from the theory with a {\em negative} $\lgb$ coupling (increasing $|\lgb|$).

We can now return to the question of how coupling dependence influences the validity of hydrodynamics as a description of IR dynamics by using the above two classes of top-down and bottom-up higher derivative theories. The first concrete holographic demonstration of the failure of hydrodynamics at reduced (intermediate) coupling was presented in \cite{Grozdanov:2016vgg}. The same qualitative behaviour was observed in both $\CN=4$ and (non-perturbative) Gauss-Bonnet theory. Namely, as one increases the size of higher derivative gravitational couplings (decreases the coupling in a dual CFT), there is an inflow of new (quasinormal) modes along the negative imaginary $\omega$ axis from $- i \infty$. Note that at infinite 't Hooft coupling $\lambda$, these modes are not present in the quasinormal spectrum. However, as $\lambda$ decreases, the leading new mode on the imaginary $\omega$ axis monotonically approaches the regime of small $\omega / T$. In the shear channel,\footnote{See \cite{Kovtun:2005ev} for conventions regarding different channels and the connection between quasinormal modes and hydrodynamics.} which contains the diffusive hydrodynamic mode, the new mode collides with the hydrodynamic mode after which point both modes acquire real parts in their dispersion relations. Before the modes collide, to leading order in $q$, the diffusive and the new mode have dispersion relations \cite{Grozdanov:2016vgg,Grozdanov:2016fkt}
\begin{align}
\omega_1 &= - i \frac{\eta}{\varepsilon + P } q^2 + \cdots, \\
\omega_2 &= \omega_{\mathfrak g} + i \frac{\eta}{\varepsilon + P } q^2 + \cdots,
\end{align}
where the imaginary gap $\omega_{\mathfrak g}$, the shear viscosity $\eta$ and energy density
$\varepsilon$, and pressure $P$ depend on the details of the theory \cite{Grozdanov:2016vgg,Grozdanov:2016fkt}. Note also that both the IIB coupling $\gamma$ and the Gauss-Bonnet coupling $ - \lgb$ have to be taken sufficiently large in order for this effect to be well described by the small-$q$ expansion (see Ref.  \cite{Grozdanov:2016fkt}). In the sound channel,
\begin{align}
\omega_{1,2} &= \pm c_s q - i \Gamma q^2 + \cdots, \\
\omega_3 &= \omega_{\mathfrak g} + 2 i \Gamma q^2 + \cdots,
\end{align}
where $c_s = 1 / \sqrt{3}$ is the conformal speed of sound and $\Gamma = 2 \eta / 3
\left(\varepsilon + P \right)$. In both channels, it is clear that the IR is no longer described by
hydrodynamics. To quantify this, it is natural to define a critical coupling dependent momentum $q_c
(\lambda)$ at which $\text{Im}\left|\omega_1 (q_c) \right| = \text{Im}\left|\omega_2 (q_c) \right|$
in the shear channel, and $\text{Im}\left|\omega_{1,2} (q_c) \right| = \text{Im}\left|\omega_3 (q_c)
\right|$ in the sound channel. With this definition, hydrodynamic modes dominate the IR spectrum for
frequencies $\omega (q)$, so long as $q < q_c (\lambda)$. To leading order in the hydrodynamic
approximation, in $\CN = 4$ theory, $q_c$ scales as $q_c \sim 0.04 \,T / \gamma \sim 0.28 \,
\lambda^{3/2} T$, while in the Gauss-Bonnet theory, $q_c \sim - 3.14 \, T /\lgb$. Even though these
scalings are approximate, they nevertheless reveal what one expects from kinetic theory: the
applicability of hydrodynamics is limited at weaker coupling by a coupling dependent scaling whereas
at strong coupling, hydrodynamics is only limited to the region of small $q / T$, independent of
$\lambda \gg 1$.\footnote{In kinetic theory (within relaxation time approximation), the hydrodynamic pole does not collide with new poles, but rather crosses a branch cut, which on the complex $\omega$ plane runs parallel to the real $\omega$ axis \cite{Romatschke:2015gic}.}

Understanding of hydrodynamics has been important for not only the description of
everyday fluids and gases, but also a nuclear state of matter known as the quark-gluon plasma that is formed after collisions of heavy ions at RHIC and the LHC. Hydrodynamics becomes a good description of the plasma after a remarkably short hydrodynamization time $\htime \sim 1-2 \, \text{fm}/\text{c}$ measured from the moment of the collision \cite{Teaney:2000cw,Heinz:2013wva,Luzum:2008cw,Luzum:2009sb,Schenke:2010rr,Song:2010mg}. In holography, heavy ion collisions have been successfully modelled by collisions of gravitational shock waves \cite{Chesler:2010bi,Grumiller:2008va,Casalderrey-Solana:2013aba,Casalderrey-Solana:2013sxa,Chesler:2015wra,jorge-book,Chesler:2015lsa,Heller:2016gbp}, including the correct order of magnitude result for the hydrodynamization time (at infinite coupling). Coupling constant corrections to holographic heavy ion collisions were studied in perturbative curvature-squared theories (Gauss-Bonnet) in \cite{Grozdanov:2016zjj}, which found that for narrow and wide gravitational shocks, respectively, the hydrodynamization time is
\begin{equation}\label{SGWSTime}
\begin{aligned}
t_{\text{hyd}} T_{\text{hyd}} &= 0.41 - 0.52 \lgb + \CO(\lgb^2)\,, \\
t_{\text{hyd}} T_{\text{hyd}} &= 0.43 - 6.3 \lgb + \CO(\lgb^2)\,,
\end{aligned}
\end{equation}
where $T_{\rm hyd}$ is the temperature of the plasma at the time of hydrodynamization. For $\lgb =
-0.2$, which corresponds to an $80\%$ increase in the ratio of shear viscosity to entropy density,
we thus find a $25\%$ and $290\%$ increase in the hydrodynamization time
\cite{Grozdanov:2016zjj}. Thus, $\htime$ was found to increase for negative values of $\lgb$, which
is consistent with expectations of the behavior of hydrodynamization at decreased field theory coupling. Consistent with these findings, the investigation of \cite{Andrade:2016rln,Atashi:2016fai} further revealed that for negative $\lgb$, the isotropization time of a plasma also increased, again reproducing the expected trend of transitioning from infinite to intermediate coupling.

In this paper, we continue the investigation of coupling constant dependent physics by studying the
simplest hydrodynamic model of heavy ions---the boost-invariant Bjorken flow
\cite{Bjorken:1982qr}---in higher derivative bulk theories of gravity. The Bjorken flow has widely
been used to study the evolution of a plasma (in the mid-rapidity regime) after the collision. While
the velocity profile of the solution is completely fixed by symmetries, relativistic Navier-Stokes
equations need to be used to find the energy density, which is expressed as a series in inverse
powers of the proper time $\tau$. The details of the solution will be described in Section \ref{sec:hydro_bjorken}.

In $\CN=4$ SYM at infinite coupling, the energy density of the Bjorken flow to third order in the hydrodynamic expansion (ideal hydrodynamics and three orders of gradient corrections) takes the following form \cite{Janik:2005zt,Nakamura:2006ih,Janik:2006ft,Heller:2007qt,Heller:2008fg,Heller:2011ju,Booth:2009ct}:
\begin{align}\label{EBjorkenN4}
\langle T_{\tau\tau}\rangle=\varepsilon(\tau) = \frac{6N_c^2 }{\pi^2 }\frac{w^4}{\tau^{4/3}} \left[1 - \frac{1}{3 w \tau^{2/3}} +\frac{1 + 2 \ln 2}{72w^2\tau^{4/3}}- \frac{3 - 2\pi^2 - 24 \ln 2 + 24 \ln^2 2}{3888 w^3 \tau^2} \right] ,
\end{align}
where $w$ is a dimensionful constant.\footnote{Other conventions that appear in the literature use $\Lambda=\frac{2w}{\pi}$ or $\epsilon=\frac{3w^4}{4}$.} Physically, the energy density of the Bjorken flow must be a positive and monotonically decreasing function of the proper time $\tau$, capturing the late-time expansion and cooling of the fluid. For a conformal, boost-invariant system, the energy density (\ref{EBjorkenN4}) uniquely determines all the components of the stress-energy tensor. Energy conditions then imply that the solution becomes unphysical at sufficiently early times, when (\ref{EBjorkenN4}) is negative. For instance, by considering the first two terms in \eqref{EBjorkenN4}, it is clear that the solution becomes problematic at times $\tau < \htau^{\text{1st}}$, where
\be\label{htau1tp}
\htau^{\text{1st}}w^{3/2}=0.19\,.
\ee
Physically, the reason is that for $\tau < \htau$, the first viscous correction becomes large and the hydrodynamic expansion breaks down, making the Bjorken flow unphysical.\footnote{Higher-order hydrodynamic corrections are expected to improve this bound. However, since hydrodynamics is an \emph{asymptotic} expansion, there should be an absolute lower bound for the regime of validity of hydrodynamics (at all orders). Ref. \cite{Jankowski:2014lna} estimated this bound to be $\tau_{\text{hyd}}T_{\text{hyd}}\sim0.6$ by analyzing a large number of far from equilibrium initial states.} Ref.  \cite{Pedraza:2014moa} further analyzed the evolution of non-local observables in a boost-invariant Bjorken plasma, finding stronger constraints on the value of initial $\tau$ for the Bjorken solution. For instance, equal-time two-point functions and space-like Wilson loops are expected to relax at late times as
\be
\frac{\langle\cO(x)\cO(x')\rangle\;\;\;\;\;}{\langle\cO(x)\cO(x')\rangle|_{\text{vac}}}\sim e^{-\Delta f(\tau w^{3/2})}\,,\qquad\quad \frac{ \langle W(\mathcal{C})\rangle\;\;\;\;\;}{\langle W(\mathcal{C})\rangle|_{\text{vac}}}\sim e^{-\sqrt{\lambda}g(\tau w^{3/2})}\,,
\ee
for some $f$ and $g$ such that $f(\tau w^{3/2})\to0$ and $g(\tau w^{3/2})\to0$ as $\tau\to\infty$. In the hydrodynamic regime, both $f$ and $g$ must be positive and monotonically decreasing functions of $\tau$, implying that, as the plasma cools down, these non-local observables relax smoothly from above to the corresponding vacuum values. Such exponential decays have indeed been observed from the full numerical evolution in shock wave collisions \cite{Bellantuono:2016tkh,Ecker:2016thn}. The interesting point here is that, if we were to truncate the hydrodynamic expansion to include only the first few viscous corrections, then $f$ and $g$ may become negative or non-monotonic at some $\tau_{\text{crit}}>\htau$, imposing further constraints on the regime of validity of hydrodynamics. In \cite{Pedraza:2014moa}, it was found that a much stronger constraint (approximately $15$ times stronger than \eqref{htau1tp}) for first-order hydrodynamics comes from the longitudinal two-point function:
\be
\tau^{\text{1st}}_{\text{crit}}w^{3/2}=2.83\,,
\ee
while for Wilson loops, the constraint was weaker:
\be
\tau^{\text{1st}}_{\text{crit}}w^{3/2}=0.65\,.
\ee
In addition, Ref. \cite{Pedraza:2014moa} also studied the evolution of entanglement (or von Neumann) entropy in a Bjorken flow, but found that the bound obtained in that case was equal to $\htau^{\text{1st}}$ given by Eq. (\ref{htau1tp}), i.e. weaker than the two constraints above. The reason for this equality is that in the late-time and slow-varying limit considered for the computation, the entanglement entropy satisfies the so-called first law of entanglement,
\be
S_A(\tau)=\varepsilon(\tau)\frac{V_A}{T_A}\,,
\ee
where $V_A$ is the volume of the subsystem and $T_A$ is a constant that depends on its shape. Such a law holds for arbitrary time-dependent excited states provided the evolution of the system is adiabatic with respect to a reference state \cite{Lokhande:2017jik}.

In this paper, we ask how higher-order hydrodynamic and coupling constant corrections affect the
critical time $\tau_{\text{crit}}$ after which the Bjorken flow yields physically sensible
observables. In particular, we extend the analysis of \cite{Pedraza:2014moa} focusing on equal-time
two-point functions and expectation values of Wilson loops. From the point of view of our discussion
regarding viscous corrections and their role in keeping $\varepsilon(\tau)$ positive, it seems clear
that at decreased coupling, when the viscosity $\eta$ becomes larger, the applicability of the Bjorken solution should become relevant at larger $\tau$. Our calculations provide further details regarding the applicability of hydrodynamics. As a result, we will be computing an observable that is related to a coupling-dependent hydrodynamization time \cite{Grozdanov:2016zjj}, but is analytically-tractable and therefore significantly simpler to analyze, albeit for realistic applications limited to the applicability of the Bjorken flow model. In this way, we obtain new holographic coupling-dependent estimates for the validity of hydrodynamics, analogous to the statement of Eq. \eqref{SGWSTime}, which allow us to compare top-down and bottom-up higher derivative corrections.

We will consider both the effects of higher-order (up to third order \cite{Grozdanov:2015kqa})
hydrodynamics and coupling constant corrections. Up to third order in the gradient expansion, we
find no surprises as the Bjorken flow observables become well defined in higher-order hydrodynamics at earlier times. In other words, no effects of asymptotic expansion divergences \cite{Heller:2013fn} are found to third order. As for coupling dependence, what we find is that the most stringent constraints arise from the calculations of a longitudinal equal-time two-point function, i.e. with spatial insertions along the boost-invariant flow direction. For the two higher-derivative theories, to first order in the coupling and to second order in the hydrodynamic expansion,
\begin{align}
\tau_\text{crit}^{2\text{nd}} \,w^{3/2} &=1.987 +275.079\, \gamma + \CO(\gamma^2) =1.987 + 41.333 \,\lambda^{-3/2}+\CO(\lambda^{-3})\,, \\
\tau_\text{crit}^{2\text{nd}} \,w^{3/2} &=1.987 - 14.876 \,\lgb+\CO(\lgb^2) \,,
\end{align}
where $\tau_\text{crit}$ is the initial critical proper-time. At $\gamma = 6.67\times 10^{-3}$ ($\lambda = 7.98$, having set $L=1$) and at $\lgb = -0.2$ (each increasing $\eta/s$ by $80\%$), we find that $\tau_\text{crit}^{2\text{nd}} \,w^{3/2} $ increases by $92.3\%$ and by $150\%$ in $\CN = 4$ and a linearized dual of Gauss-Bonnet theory, respectively (see Tables \ref{table:CT1} and \ref{table:CT2} for other numerical estimates). In a fully non-perturbative Gauss-Bonnet calculation, the increase is instead found to be $145\%$, which shows a rather quick convergence of the perturbative Gauss-Bonnet series for this observable to the full result at $\lgb = -0.2$ (see also \cite{Grozdanov:2016zjj}). Thus, our results lie inside the interval of increased hydrodynamization time found in narrow and wide shocks obtained from non-linear shock wave simulations \cite{Grozdanov:2016zjj}.

The paper is structured as follows: In Section \ref{sec:hydro_bjorken}, we discuss higher-order
hydrodynamics and details of the hydrodynamic Bjorken flow solution, including all necessary
holographic transport coefficients that enter into the solution. In Section \ref{section:GBG}, we
discuss the construction of holographic dual geometries to Bjorken flow. We focus in particular on
the case of the Gauss-Bonnet theory which, to our understanding, has not been considered in previous
literature.\footnote{The background for type IIB supergravity  $\alpha'$ corrections has been worked out in \cite{Benincasa:2007tp}.} In Section \ref{sec:break}, we analyze the relaxation
properties of two-point functions and Wilson loops, extracting the relevant critical times at which the hydrodynamic approximation breaks down. Finally, Section \ref{Sec:Discussion} is devoted the discussion of our results.


\section{Hydrodynamics and Bjorken flow}
\label{sec:hydro_bjorken}

We begin by expressing the equations that describe the boost-invariant evolution of charge-neutral, conformal relativistic fluids, which will be studied in this work. In the absence of any external sources, the equations of motion (relativistic Navier-Stokes equations) follow from the conservation of stress-energy
\begin{align}\label{TCons}
\nabla_a T^{ab} = 0 \, .
\end{align}
The constitutive relations for the stress-energy tensor of a neutral, conformal (Weyl-covariant) relativistic fluid can be written as (see e.g \cite{Kovtun:2012rj})
\begin{align}
T^{ab} = \varepsilon u^a u^b + P \Delta^{ab} + \Pi^{ab} \, ,
\end{align}
where we have chosen to work in the Landau frame. The transverse projector $\Delta^{ab}$ is defined
as $\Delta^{ab} \equiv g^{ab} + u^a u^b$, with $u^a$ being the velocity field of the fluid flow. In
four spacetime dimensions, the pressure $P$ and energy density $\varepsilon$ are related by the
conformal relation $P = \varepsilon/3$. The transverse, symmetric and traceless tensor $\Pi^{ab}$
can be expanded in a gradient expansion (in gradients of $u^a$ and a scalar temperature field). To third order in derivatives \cite{Baier:2007ix,Bhattacharyya:2008jc,Grozdanov:2015kqa},
\begin{align}\label{PiExp}
\Pi^{ab} =&\, -\eta \sigma^{ab} + \eta \tp \left[ {}^{\langle}D\sigma^{ab\rangle} + \frac{1}{3} \sigma^{ab} \left(\nabla\cdot u\right) \right] + \kappa \left[ R^{\langle ab \rangle} - 2 u_c R^{c \langle ab \rangle d} u_d  \right]  \nn
&+\lambda_1 \sigma^{\langle a}_{~~c} \sigma^{b\rangle c}  +\lambda_2 \sigma^{\langle a}_{~~c} \Omega^{b\rangle c}  +\lambda_3 \Omega^{\langle a}_{~~c} \Omega^{b\rangle c} + \sum_{i=1}^{20} \lambda^{(3)}_i \CO^{ab}_i \, ,
\end{align}
where we have used the longitudinal derivative $D \equiv u^a \nabla_a$ and a short-hand notation
\begin{align}
A^{\langle ab \rangle} \equiv \frac{1}{2} \Delta^{ac} \Delta^{bd} \left(A_{cd} + A_{dc}\right) - \frac{1}{3} \Delta^{ab} \Delta^{cd} A_{cd} \equiv {}^{\langle} A^{ab\rangle},
\end{align}
which ensures that any tensor $A^{\langle ab \rangle}$ is by construction transverse, $u_a A^{\langle ab \rangle} = 0$, symmetric and traceless, $g_{ab} A^{\langle ab \rangle} = 0$. The tensor $\sigma^{ab}$ is a one-derivative shear tensor
\begin{align}
\sigma^{ab} = 2 {}^\langle \nabla^{a}u^{b\rangle} \, .
\end{align}
The vorticity $\Omega^{\mu\nu}$ is defined as the anti-symmetric, transverse and traceless tensor
\begin{align}
\Omega^{ab} = \frac{1}{2} \Delta^{ac} \Delta^{bd} \left(\nabla_c u_d - \nabla_d u_c \right) \, .
\end{align}
The transport coefficients appearing in \eqref{PiExp} are the shear viscosity $\eta$, 5 second order coefficients $\eta \tp$, $\kappa$, $\lambda_1$, $\lambda_2$, $\lambda_3$, and 20 (subject to potential entropy constraints) conformal third order transport coefficients $\lambda_i^{(3)}$, which multiply 20 linearly independent, third order Weyl-covariant tensors $\CO_i^{ab}$ that can be found in \cite{Grozdanov:2015kqa}.

The boost-invariant Bjorken flow \cite{Bjorken:1982qr} is a solution to the hydrodynamic equations
(Eq. \eqref{TCons}), and has been widely used as a simple model of relativistic heavy ion collisions (see \cite{jorge-book}). Choosing the direction of the beam to be the $z$ axis, the Bjorken flow is boost-invariant along $z$, as well as rotationally and translationally invariant in the plane perpendicular to $z$ (denoted by $\vec{x}_\perp$). By introducing the proper time $\tau = \sqrt{t^2 - z^2}$ and the rapidity parameter $y = \text{arctanh} (z/t)$, the velocity field, which is completely fixed by symmetries, and the flat metric can be written as
\begin{align}
&u^a = \left(u^\tau, u^y, \vec{u}^\perp \right) = \left(1,0,0,0 \right), \\
&\eta_{ab} dx^a dx^b= - d\tau^2 + \tau^2 dy^2 + d \vec{x}_\perp^2.\label{flat-metric-T-Y}
\end{align}
Note that the solution is also invariant under discrete reflections $y \to - y$. What remains is for us to find the solution for the additional scalar degree of freedom that is required to fully characterize the flow. In this case, it is convenient to work with a proper time-dependent energy density $\varepsilon(\tau)$ and write Eq. \eqref{TCons} as in \cite{Baier:2007ix}:
\begin{align}\label{Bjorken1}
D \varepsilon + \left(\varepsilon + P \right) \nabla_a u^a + \Pi^{ab} \nabla_a u_b = 0\, .
\end{align}
By using the conformal relation $P = \varepsilon / 3$ and the fact that the only non-zero component of $\nabla_a u_b$ is $\nabla_y u_y = \delp_y u_y  =\tau$, Eq. \eqref{Bjorken1} then gives
\begin{align}\label{EpsBjorkenEq}
\partial_\tau \varepsilon + \frac{4}{3} \frac{\varepsilon}{\tau}  +  \tau \Pi^{yy} = 0 \, ,
\end{align}
with $\Pi^{yy}$ from Eq. \eqref{PiExp} expanded as
\begin{align}\label{Pixixi}
\Pi^{yy} =& - \frac{4 \eta}{3} \frac{1}{\tau^3}  - \left[ \frac{8 \eta \tp}{9} - \frac{8 \lambda_1}{9}\right] \frac{1}{\tau^4}  - \left[ \frac{\lambda^{(3)}_1}{6 } + \frac{4 \lambda^{(3)}_2}{3 } + \frac{4 \lambda^{(3)}_3}{3 } + \frac{5 \lambda^{(3)}_4}{6} + \frac{5 \lambda^{(3)}_5}{6 } + \frac{4 \lambda^{(3)}_6}{3 } \right. \nn
&\left.  -  \frac{\lambda^{(3)}_7}{2 }  + \frac{3\lambda^{(3)}_8}{2 } + \frac{\lambda^{(3)}_9}{2 } - \frac{2\lambda^{(3)}_{10}}{3 } - \frac{11\lambda^{(3)}_{11}}{6 } - \frac{\lambda^{(3)}_{12}}{3 } + \frac{\lambda^{(3)}_{13}}{6} - \lambda^{(3)}_{15} \right] \frac{1}{\tau^5} + \CO\left(\tau^{-6}\right).
\end{align}
Each transport coefficient appearing in \eqref{Pixixi} can only be a function of the single scalar degree of freedom---the energy density---with dependence on $\varepsilon$ determined uniquely by its conformal dimension under local Weyl transformations \cite{Baier:2007ix,Grozdanov:2015kqa}:
\begin{align}\label{BjorkenScaling}
\eta = C \bar{\eta} \left( \frac{\varepsilon}{C} \right)^{3/4} , && \eta \tp = C \bar\eta \bar{\tp} \left( \frac{\varepsilon}{C} \right)^{1/2}, &&  \lambda_1 = C \bar\lambda_{1} \left( \frac{\varepsilon}{C} \right)^{1/2},&& \lambda^{(3)}_n = C \bar\lambda^{(3)}_{n}  \left( \frac{\varepsilon}{C} \right)^{1/4},
\end{align}
where $C$, $\bar\eta$, $\bar{\tp}$ and $\bar\lambda^{(3)}_{n}$ are constants. Finally, the Bjorken
solution to Eq. \eqref{TCons} for the energy density, expanded in powers of $\tau$, becomes
\begin{align}\label{Bjorken3rdFT}
\frac{\varepsilon(\tau)}{C}  =&~ \frac{1}{\tau^{2 - \nu}} - 2 \bar\eta \frac{1}{\tau^2} + \left[ \frac{3 \bar\eta^2 }{2}- \frac{2 \bar\eta \bar{\tp}}{3} + \frac{2 \bar\lambda_{1}}{3} \right] \frac{1}{\tau^{2+\nu}} \nn
&- \left[ \frac{\bar\eta^3}{2} - \frac{7 \bar\eta ^2 \bar{\tp} }{9} + \frac{7 \bar\eta \bar\lambda_1  }{9}  +\frac{\bar\lambda^{(3)}_1}{12 } + \frac{2 \bar\lambda^{(3)}_2}{3 } + \frac{2 \bar\lambda^{(3)}_3}{3 } + \frac{5 \bar\lambda^{(3)}_4}{12} + \frac{5 \bar\lambda^{(3)}_5}{12 } + \frac{2 \bar\lambda^{(3)}_6}{3 } \right.    \nn
&\left.  -  \frac{\bar\lambda^{(3)}_7}{4 }  + \frac{3\bar\lambda^{(3)}_8}{4 } + \frac{\bar\lambda^{(3)}_9}{4 } - \frac{\bar\lambda^{(3)}_{10}}{3 } - \frac{11\bar\lambda^{(3)}_{11}}{12 } - \frac{\bar\lambda^{(3)}_{12}}{6 } + \frac{\bar\lambda^{(3)}_{13}}{12} - \frac{ \bar\lambda^{(3)}_{15}}{2}  \right] \frac{1}{\tau^{2+2\nu}} + \CO\left(\tau^{-2-3\nu}\right),
\end{align}
with $\nu = 2/3$. Terms at order $\CO\left(\tau^{-2-3\nu}\right)$ are controlled by the hydrodynamic expansion to fourth order, which is presently unknown.

In theories of interest to this work, namely in the $\CN=4$ supersymmetric Yang-Mills theory and in hypothetical duals of curvature-squared gravity, all first- and second-order transport coefficients are known. In $\CN=4$ theory (cf. Eq. \eqref{eq:10DAct}), the relevant expressions, including the leading-order 't Hooft coupling corrections are \cite{Buchel:2004di, Benincasa:2005qc, Buchel:2008sh, Buchel:2008ac,Buchel:2008bz,Buchel:2008kd,Saremi:2011nh,Grozdanov:2014kva}
\begin{align}
\eta &= \frac{\pi}{8} N^2_c T^3\, \left( 1 + \frac{135 \zeta (3)}{8}\, \lambda^{-3/2} + \ldots \,  \right)\,, \label{tcsym1} \\
\tau_{\Pi}  &= \frac{ \left( 2 - \ln{2}\right)}{2\pi T}   + \frac{375 \zeta(3)}{32 \pi T} \, \lambda^{-3/2} +\ldots \,, \\
\lambda_1 &=  \frac{N_c^2 T^2}{16}  \left( 1 + \frac{175\zeta (3)}{4} \, \lambda^{-3/2} + \ldots \,  \right)\,.   \label{tcsym4}
\end{align}
In the most general curvature-squared theory (cf. Eq. \eqref{eq:R2Th}), with $\alpha_i$ treated perturbatively to first order \cite{Grozdanov:2014kva},
\begin{align}
\eta &= \frac{r_+^3}{2 \kappa^2_5} \left( 1 - 8 \left(5 \alpha_1 + \alpha_2 \right) \right) + \CO (\alpha_i^2)\,, \label{etaKP2} \\
\eta \tp &= \frac{r_+^2 \left( 2 - \ln 2\right) }{4 \kappa_5^2} \left( 1 - \frac{26}{3} \left(5 \alpha_1 + \alpha_2 \right) \right) - \frac{r_+^2 \left(23 + 5 \ln 2\right)}{ 12 \kappa_5^2} \alpha_3+ \CO(\alpha_i^2)    \, , \label{etatpKP} \\
\lambda_1 &= \frac{r_+^2}{4 \kappa_5^2} \left( 1 - \frac{26}{3} \left(5 \alpha_1 + \alpha_2 \right) \right) - \frac{r_+^2}{12 \kappa_5^2} \alpha_3 + \CO(\alpha_i^2) \, ,\label{l1KP}
\end{align}
where $r_+$ is the position of the event horizon in the bulk, which depends on all three $\alpha_i$ (see Ref.  \cite{Grozdanov:2014kva}). Finally, in a dual of the Gauss-Bonnet theory (cf. Eq. \eqref{eq:GBaction}) all first- and second-order transport coefficients are known non-perturbatively in the coupling $\lgb$ \cite{Grozdanov:2015asa,Grozdanov:2014kva,Grozdanov:2016fkt},
\begin{align}
\eta &=  \frac{\sqrt{2}\pi^3 }{\kappa_5^2} \frac{T^3 \ggb^2}{\left(1+\ggb\right)^{3/2} }   \,, \label{gb-visc} \\
\tp &= \frac{1}{2\pi T} \left( \frac{1}{4} \left(1+\ggb\right) \left( 5+\ggb - \frac{2}{\ggb}\right) - \frac{1}{2} \ln \left[\frac{2 \left(1+\ggb\right)}{\ggb }\right]  \right) , \label{l0}\\
\lambda_1 &= \frac{\eta}{2\pi T } \left( \frac{ \left(1+\ggb\right)\left(3 - 4\ggb +2\ggb^3\right)}{ 2\ggb^2 } \right) , \label{l1}
\end{align}
where we have defined the coupling $\ggb$ as
\begin{align}
\label{Gamma}
\ggb \equiv \sqrt{1-4\lgb}\,.
\end{align}
The relevant linear combination of the third-order transport coefficients appearing in \eqref{Bjorken3rdFT} is to date only known in $\CN = 4$ theory at infinite coupling. The expression was found in \cite{Grozdanov:2015kqa} by using the holographic Bjorken flow result of \cite{Janik:2005zt,Nakamura:2006ih,Janik:2006ft,Heller:2007qt,Heller:2008fg,Heller:2011ju,Booth:2009ct} for $\varepsilon(\tau)$ stated in Eq. \eqref{EBjorkenN4}, giving
\begin{align}
\frac{\lambda^{(3)}_1}{6 } + \frac{4 \lambda^{(3)}_2}{3 } + \frac{4 \lambda^{(3)}_3}{3 } + \frac{5 \lambda^{(3)}_4}{6} + \frac{5 \lambda^{(3)}_5}{6 } + \frac{4 \lambda^{(3)}_6}{3 }   -  \frac{\lambda^{(3)}_7}{2 } & \nn
+ \frac{3\lambda^{(3)}_8}{2 } + \frac{\lambda^{(3)}_9}{2 } - \frac{2\lambda^{(3)}_{10}}{3 } - \frac{11\lambda^{(3)}_{11}}{6 } - \frac{\lambda^{(3)}_{12}}{3 } + \frac{\lambda^{(3)}_{13}}{6} - \lambda^{(3)}_{15}  &=\frac{N_c^2 T}{648 \pi} \left( 15 - 2 \pi^2 - 45 \ln 2 + 24 \ln^2 2 \right) \nn
& ~~~+ \cdots \, ,
\end{align}
where the ellipsis indicates unknown coupling constant corrections.

In this work, we will not look beyond third-order hydrodynamics. What is important to note is that the gradient expansion is believed to be an asymptotic expansion, similar to perturbative expansions. As a result, the Bjorken expansion in proper time formally has a zero radius of convergence \cite{Heller:2013fn}. In practice, this means that at some order, the expansion in inverse powers of $\tau$ breaks down and techniques of resurgence are required for analyzing long-distance transport (see e.g. \cite{Heller:2013fn,Cherman:2014ofa,Cherman:2014xia,Basar:2015ava,Heller:2015dha,Aniceto:2015mto,Buchel:2016cbj}).


\section{Gravitational background in Gauss-Bonnet gravity}
\label{section:GBG}

In this section, we begin our analysis of holographic duals to Bjorken flow. Throughout this paper, we will be interested in three separate cases:
\begin{itemize}
  \item \emph{Einstein gravity.} Bjorken flow in $\CN = 4$ SYM at infinite coupling, expanded to third order in the hydrodynamic series.
  \item \emph{$\alpha'$-corrections.} Bjorken flow in $\CN = 4$ SYM with first-order 't Hooft coupling corrections, $\alpha'^3 \sim 1 / \lambda^{3/2}$, expanded to second order in the hydrodynamic series.
  \item \emph{$\lgb$-corrections.} Bjorken flow in a hypothetical dual of Gauss-Bonnet theory with $\lgb$ coupling corrections, expanded to second order in the hydrodynamic series.
\end{itemize}
In the first case, the holographic dual geometry is well known (see
Refs. \cite{Janik:2005zt,Nakamura:2006ih,Janik:2006ft,Heller:2007qt,Heller:2008fg,Heller:2011ju,Booth:2009ct}). What
one finds is that in the near-boundary region, which is the only region relevant for computing the
non-local observables studied in this paper (two-point correlators of operators with large
dimensions and Wilson loops), the geometries are specified by symmetry and (relevant order)
hydrodynamic transport coefficients.\footnote{The choice of these cases is dictated by our
  present knowledge of transport coefficients (see Section \ref{sec:hydro_bjorken}).} As we will
see, the same conclusions can also be drawn in higher-derivative theories. As a check,  we
derive here the full geometric Bjorken background in non-perturbative Gauss-Bonnet theory. All details of
the perturbative calculations in Type IIB supergravity with $\alpha'$ corrections will be omitted,
but we refer the reader to \cite{Benincasa:2007tp} for the explicit derivation.

\subsection{Static background}

Equations of motion for Gauss-Bonnet gravity in five dimensions can be derived from the action \eqref{eq:GBaction} and take the following form:
\begin{equation}
R_{\mu\nu}-\frac{1}{2}g_{\mu\nu}\left(R+\frac{12}{\tl^2}+\frac{\lgb \tl^2}{2}\mathcal{L}_{\scriptscriptstyle{GB}}\right)+\lgb \tl^2\mathcal{H}_{\mu\nu} = 0 \, , \label{eq:GB-EOM}
\end{equation}
where
\begin{eqnarray*}
\mathcal{L}_{\scriptscriptstyle{GB}}&=& R_{\mu\nu\alpha\beta}R^{\mu\nu\alpha\beta}-4R_{\mu\nu}R^{\mu\nu}+R^2 \,,\\
\mathcal{H}_{\mu\nu} &=& R_{\mu\alpha\beta\rho}R_{\nu}^{~~\alpha\beta\rho}
-2R_{\mu\alpha\nu\beta}R^{\alpha\beta}
-2R_{\mu\alpha}R_{\nu}^{~~\alpha} +R_{\mu\nu}R\, .
\end{eqnarray*}
This set of differential equations admits a well-known (static) asymptotically AdS black brane solution:
\begin{equation}
ds^2 = -\frac{r^2f(r)}{\tlgb^2}d\tau^2+\frac{\tlgb^2}{r^2f(r)}dr^2+\frac{r^2}{\tlgb^2}d\vec{x}^2 \, ,\label{eq:static-bh}
\end{equation}
with the emblackening factor
\begin{equation}
f(r) = \frac{1}{2\lgb}\frac{\tlgb^2}{L^2}\left[1-\sqrt{1-4\lgb\left(1-\frac{r_h^4}{r^4}\right)}~\right].\label{eq:static-bh-Sol}
\end{equation}
In the near-boundary limit, the asymptotically AdS region exhibits the following scaling:
\begin{equation}
\left.ds^2\right|_{r\rightarrow\infty}=\frac{\tlgb^2}{r^2}dr^2+\frac{r^2}{\tlgb^2}\left(-d\tau^2+d\vec{x}^2\right)=\frac{\tlgb^2}{r^2}dr^2+\frac{r^2}{\tlgb^2}\eta_{ab}dx^{a}dx^b \,,\label{eq:GB-AdS-vacuum}
\end{equation}
where $\eta_{ab}$ is the flat metric and the AdS curvature scale, $\tlgb$, is related to the length scale set by the cosmological constant, $L$, via
\begin{equation}
\tlgb^2  = \frac{L^2}{2}\left(1+\sqrt{1-4\lgb}\right) = \frac{L^2}{2}\left(1+\ggb\right) .
\label{eq:GBL-to-AdSL}
\end{equation}
The Hawking temperature, entropy density and energy density of the dual theory are then given by\footnote{We note that our black brane background can be put into the form given by Eq. (2.2) of \cite{Grozdanov:2016fkt} by a simple rescaling of r: $r\rightarrow \tlgb r/L$ with $r_h\rightarrow \tlgb r_+/L$. }
\begin{eqnarray}
T &=& \frac{r_h}{\pi L^2} \label{eq:static-bh-temp} \,, \\
s &=& \frac{4\sqrt{2}\pi}{(1+\ggb)^{3/2}\kappa_5^2}\left(\frac{r_h}{L}\right)^3 ,\label{eq:static-bh-entropy}\\
\varepsilon &=& 3P = \frac{3}{4}T s \,.\label{eq:static-bh-energy-density}
\end{eqnarray}
In what follows, we will set $L=1$ unless otherwise stated.

To make the metric manifestly boost-invariant along the spatial coordinate $z$, we transform \eqref{eq:static-bh} by introducing a proper time coordinate $\tau = \sqrt{t^2 - z^2}$. Next, we perform an additional coordinate transformation to write the metric in terms of ingoing Eddington-Finkelstein (\EF) coordinates with
\begin{equation}
\tau\rightarrow \tau_{+}-\tlgb^2\int^r\frac{d\tilde{r}}{\tilde{r}^2f(\tilde{r})}\,,
\end{equation}
which gives the metric
\begin{equation}
ds^2 = -\frac{r^2}{\tlgb^2}f(r)d\tpp^2 +2d\tpp dr+\frac{r^2}{\tlgb^2}d\vec{x}^2 \,. \label{eq:static-bh-EF}
\end{equation}
It should be noted that the \EF{} time, $\tpp$, mixes the proper time, $\tau$, and $r$ in the bulk. At the boundary, however,
\begin{equation}
\lim_{r\rightarrow\infty}\tpp=\tau \,.\label{eq:EFt-2-tau}
\end{equation}

A static black brane with a constant temperature cannot be dual to an expanding Bjorken fluid, which
has a temperature that decreases with the proper time, $T_{\text{fluid}}\sim\tau^{-1/3}$. As in the fluid-gravity correspondence \cite{Bhattacharyya:2008jc}, where the black brane is boosted along spatial directions, here, one may make an informed guess and allow for the horizon to become time-dependent by substituting
\begin{equation}
r_h \rightarrow w\tpp^{-1/3} \,,
\end{equation}
where $w$ is a constant and $\tpp$ is the fluid's proper time at the boundary. The Hawking temperature is then
\begin{equation}
T = \frac{w}{\pi L^2}\tpp^{-1/3} \,, \label{eq:hawking-temp_boosted}
\end{equation}
and the static black brane metric \eqref{eq:static-bh-EF} takes the form
\begin{equation}
\label{eq:static-bh-boosted}
ds^2 = -\frac{r^2}{\tlgb^2}\left(\frac{1}{1-\ggb}\right)\left[1-\ggb\sqrt{1-\left(1-\frac{1}{\ggb^2}\right)\frac{w^4}{v^4}}~ \right]d\tpp^2+2d\tpp dr+\frac{r^2}{\tlgb^2}d\vec{x}^2 \,,
\end{equation}
with $v$ defined as
\begin{equation}
v \equiv r \tpp^{1/3} \,. \label{eq:late-time-v}
\end{equation}
Of course, as in the fluid-gravity correspondence, Eq. \eqref{eq:static-bh-boosted} is not a solution to the Gauss-Bonnet equations of motion. As will be shown below, however, the background solution asymptotes to \eqref{eq:static-bh-boosted} at late times, i.e. Eq. \eqref{eq:static-bh-boosted} is (approximately) dual to Bjorken flow in the regime dominated by ideal hydrodynamics.

\subsection{Bjorken flow geometry}

The full (late-time) geometry is systematically constructed following the procedure outlined in Ref. \cite{Kinoshita:2008dq} (see also \cite{Heller:2008mb}). In \EF coordinates, the most general metric respecting the symmetries of Bjorken flow is
\begin{equation}
\label{eq:EF_met_ansatz}
ds^2 = -\frac{r^2}{\tlgb^2}a d\tpp^2+2d\tpp dr+\frac{1}{\tlgb^2}\left(\tlgb^2+r\tpp\right)^2e^{2(b-c)}dy^2+\frac{r^2}{\tlgb^2}e^{c}dx_{\perp}^2 \,,
\end{equation}
where $a$, $b$, $c$ are functions of $r$ and $\tpp$ and our boundary geometry is expressed in proper time--rapidity coordinates (see the discussion above Eq. (\ref{flat-metric-T-Y})).

At late times, the equations of motion \eqref{eq:GB-EOM} can be solved order-by-order in powers of $\tpp^{-2/3}$, provided the $\tpp\rightarrow\infty$ expansion is carried out holding $v\equiv r\tpp^{1/3}$ fixed. To perform the late time expansion, we will change coordinates from $\{\tpp,r\}\rightarrow\{v,u\}$, where
\begin{align}
\label{eq:scaling-vars}
v\equiv r\tpp^{1/3}, && u\equiv\tpp^{-2/3} \,,
\end{align}
and assume the metric functions $a$, $b$ and $c$ can be expanded as
\begin{equation}
\label{eq:metric-function-expansions}
a(u,v) = a_0(v)+a_1(v)u+a_2(v)u^2+\ldots \,.
\end{equation}
We then solve the equations order-by-order in powers of $u$ and impose Dirichlet boundary conditions (at the boundary) at every order:
\begin{eqnarray}
\lim_{v\rightarrow\infty}a_0 &=& 1 \,,\nonumber\\
\lim_{v\rightarrow\infty}\{a_{i>0},b_i,c_i\} &=& 0 \,.
\label{eq:AdS-boundary-conditions}
\end{eqnarray}

At a given order, $i$, the equations of motion form a system of second-order differential equations for $a_i$, $b_i$ and $c_i$ along with two constraint equations. We therefore have six integration constants at each order. One integration constant is related to a residual diffeomorphism invariance of our metric under the coordinate transformation \cite{Kinoshita:2008dq}
\begin{equation}
r\rightarrow r+f(\tpp) \,,
\end{equation}
and can be freely specified without affecting the physics of our boundary field theory---a feature
that will be exploited to simplify the solutions. Three of the five remaining integration constants
can fixed by requiring the bulk geometry to be free of singularities (apart from at $v=0$) and
imposing the asymptotic AdS boundary conditions above. In practice, to the order considered, we find
that the integration constant which ensures bulk regularity can be set by requiring $\partial_v c_i$
to be regular at a particular value\footnote{With the next section in mind, we require
  $\displaystyle\lim_{v\rightarrow w^+} \partial_v c_i < \infty$.} of $v$. The remaining integration
constants are specified by the two constraint equations. For $i>0$, one of the constraint equations
can specify a constant at order $i$, while the other specifies a constant at order $i-1$.

\subsection{Solutions}
\label{section:bg-sols}

We now present the full zeroth- and first-order solutions in the late-time (hydrodynamic gradient) expansion. At second order, we were unable to find closed-form solutions analytically that would extend throughout the entire bulk. However, sufficiently complete solutions for the purposes of this work can be found non-perturbatively in $\lgb$ near the boundary, or perturbatively in the full bulk.

\subsubsection*{Zeroth Order}
At zeroth order in the hydrodynamic expansion (ideal fluid order), the equations of motion are
solved by\footnote{We note that this is not the most general solution to the equations of motion at
  this order---there is an additional nonphysical integration constant corresponding to a gauge
  degree of freedom. A simple coordinate transformation \cite{Kinoshita:2008dq} brings the solution
  into the form presented here. Similar remarks apply for the first-order solution.}
\begin{eqnarray}
a_0 &=& \left(\frac{1}{1-\ggb}\right)\left[1-\ggb\sqrt{1-\left(1-\frac{1}{\ggb^2}\right)\frac{w^4}{v^4}}~ \right] \,,\nonumber\\
b_0 &=& 0 \,, \nonumber\\
c_0 &=& 0\,.\label{eq:GB-ord0-sol}
\end{eqnarray}
One can see immediately that the zeroth-order solution is the boosted black brane metric given by Eq. (\ref{eq:static-bh-boosted}). Near the boundary we find
\begin{eqnarray}
a_0&=& 1-\left(\frac{1+\ggb}{2\ggb}\right)\left(\frac{w}{v}\right)^4+\mathcal{O}(v^{-5}) \,,\nn
b_0 &=& 0 \,,\nn
c_0 &=& 0 \,.
\end{eqnarray}

\subsubsection*{First Order}
At first (dissipative) order, our equations of motion are solved by
\begin{eqnarray}
a_1&=&\frac{\ggb(1+\ggb)}{3}\left[\left(\frac{1}{1-\ggb}\right)\frac{1}{v}+\frac{v}{\axf}\left(\frac{w^3}{v^3}-\frac{1}{1-\ggb}\right)\right] \,, \nonumber\\
b_1 &=& 0 \,, \nonumber\\
c_1 &=& \frac{\ggb(1+\ggb)}{3}\int^{v}\frac{d\tilde{v}}{\tilde{v}^2}\left(\frac{1}{\tilde{v}^2-\ggb\axf}\right)\left\{
\left[1-(1-\ggb)\left(\frac{w}{\tilde{v}}\right)^3
\right]\axf-\tilde{v}^2\right\} \,, \label{eq:ord1-sol}
\end{eqnarray}
where
\begin{equation}
\label{eq:aux-function}
\axf(v) \equiv v^2\sqrt{1-\left(1-\frac{1}{\ggb^2}\right)\frac{w^4}{v^4}} \,.
\end{equation}
For simplicity, here we have presented $c_1$ in an integral representation. An explicit evaluation of the integral would result in an Appell hypergeometric function (see Ref. \cite{Grozdanov:2016fkt}).\footnote{We note that upon integration, the integration constant is fixed by requiring $\displaystyle\lim_{v\rightarrow\infty}c_1 = 0$.} Near the boundary,
\begin{eqnarray}
a_1 &=&\frac{\ggb(1+\ggb)}{3w}\left(\frac{w}{v}\right)^4+\mathcal{O}(v^{-5}) \,,\nn
b_1 &=& 0 \,, \nn
c_1&=&\frac{\ggb(1+\ggb)}{12 w}\left(\frac{w}{v}\right)^4+\mathcal{O}(v^{-5}) \,.
\end{eqnarray}

\subsubsection*{Second Order}
As in Gauss-Bonnet fluid-gravity calculations \cite{Grozdanov:2016fkt}, at second order in the hydrodynamic expansion, one is required to solve non-homogeneous differential equations with sources depending on complicated expressions involving Appell hypergeometric functions. For this reason, we were only able to find non-perturbative solutions (in $\lgb$) near the boundary and solve the full equations perturbatively.

Near the boundary we find
\begin{eqnarray}
a_2 &=& \mathcal{A}_2\left(\frac{w}{v}\right)^4+\mathcal{O}(v^{-5}) \,, \nonumber\\
b_2&=& \mathcal{O}(v^{-5}) \,, \nonumber\\
c_2&=& \mathcal{C}_2\left(\frac{w}{v}\right)^4+\mathcal{O}(v^{-5})\,, \label{eq:ord2-sol-I}
\end{eqnarray}
where $\mathcal{A}_2$ and $\mathcal{C}_2$ are, as yet,
unspecified constants. To determine them, we would need to know the full bulk solutions and the constants would then follow from horizon regularity. Instead, as will be shown below, we will use known properties of the dual field theory (the transport coefficients and energy conservation) to show that they must take the following values:
\begin{eqnarray*}
\CA_2 &=& -\frac{1}{72w^2}\left(\frac{1+\ggb}{\ggb}\right)\left(6+\ggb^2(1+\ggb)(9\ggb-11)+2\ggb^2\ln\left[2+\frac{2}{\ggb}\right]\right)\,, \\
\CC_2 &=& \frac{\CA_2 }{ 2} \,.
\end{eqnarray*}
Full perturbative first-order (in $\lgb$) solutions are presented in Appendix \ref{appendix-second-order}. Here, we only state their near-boundary forms:
\begin{eqnarray}
a_2&=&-\frac{w^2}{18v^4}\bigg[1+2\ln 2-6\lgb\left(1+\ln 2\right)\bigg]+\mathcal{O}(v^{-5})\,,\nonumber\\
b_2 &=& \mathcal{O}(v^{-5})\,,\nonumber\\
c_2&=& -\frac{w^2}{36v^4}\bigg[1+2\ln 2-6\lgb\left(1+\ln 2\right)\bigg]+\mathcal{O}(v^{-5})\,.\label{eq:perturbative-hydro-ord2s}
\end{eqnarray}

\subsection{Stress-energy tensor and transport coefficients}

We can now compute the boundary stress-energy tensor by following the well-known holographic procedure (see e.g. \cite{Astefanesei:2008wz,Brihaye:2008kh,Grozdanov:2016fkt}), which we review here. First, we introduce a regularized boundary located at $r=r_0=\text{const}$. The induced metric on the regularized boundary is given by $\gamma_{\mu\nu} \equiv g_{\mu\nu}-n_{\mu}n_{\nu}$, where $n^{\mu}\equiv \delta^{\mu}_{~r}/\sqrt{g^{rr}}$ is the outward-pointing unit vector normal to the $r=r_0$ hypersurface. The boundary stress-energy tensor is then
\begin{equation}
T_{\mu\nu} = \frac{1}{\kappa_5^2}\frac{r_0^2}{\tlgb^2}\left[K_{\mu\nu}-K\gamma_{\mu\nu}+3\lgb L^2\left(J_{\mu\nu}-\frac{1}{3}J\gamma_{\mu\nu}\right)+\delta_1\gamma_{\mu\nu}+\delta_2 G^{(\gamma)}_{\mu\nu}\right]_{r_0\rightarrow\infty},\label{eq:GB-boundary-stress}
\end{equation}
where $G^{(\gamma)}_{\mu\nu}$ is the induced Einstein tensor on the regularized boundary, $K_{\mu\nu}$ is the extrinsic curvature\footnote{Here, $\nabla_{\mu}$ is the covariant derivative compatible with the full 5-d metric, $g_{\mu\nu}$.}
\begin{eqnarray}
K_{\mu\nu} &=& -\frac{1}{2}\left(\nabla_{\mu}n_{\nu}+\nabla_{\nu}n_{\mu}\right),
\end{eqnarray}
$K=g^{\mu\nu}K_{\mu\nu}$, $J_{\mu\nu}$ is defined by
\begin{eqnarray}
J_{\mu\nu} &\equiv& \frac{1}{3}\left(
2 K K_{\mu\rho}K^{\rho}_{~\nu}+K_{\rho\sigma}K^{\rho\sigma}K_{\mu\nu}-2K_{\mu\sigma}K^{\sigma\rho}K_{\rho\nu}-K^2 K_{\mu\nu}
\right),
\end{eqnarray}
and $J= g^{\mu\nu}J_{\mu\nu}$. The constants $\delta_1$ and $\delta_2$, fixed by holographic renormalization, are given by
\begin{align}
\delta_1 = -\sqrt{2}\left(\frac{2+\ggb}{\sqrt{1+\ggb}}\right)\,,&& \delta_2 = \frac{(2-\ggb)}{2\sqrt{2}}\sqrt{1+\ggb}\,.
\end{align}
For the background derived in Section \ref{section:bg-sols}, the non-zero components of the four dimensional boundary stress-energy tensor, $T_{ab}$, are found to be
\begin{eqnarray}
T_{\tpp\tpp} &=& \frac{3\sqrt{2}w^4}{(1+\ggb)^{3/2}\kappa_5^2}\left(\tau^{-4/3}-\frac{2\ggb^2}{3 w}\tau^{-2}-\mathcal{A}_2\left(\frac{2\ggb}{1+\ggb}\right)\tau^{-8/3}\right),\nonumber\\
T_{yy} &=&  \frac{3\sqrt{2}w^4}{(1+\ggb)^{3/2}\kappa_5^2}\left(\frac{1}{3}\tau^{2/3}-\frac{2\ggb^2}{3 w}-
\frac{2\ggb}{3}\left(\frac{\mathcal{A}_2+8~\mathcal{C}_2}{1+\ggb}\right)\tau^{-2/3}\right),\nonumber\\
T_{x_{\perp}x_{\perp}} &=& \frac{3\sqrt{2}w^4}{(1+\ggb)^{3/2}\kappa_5^2}\left(\frac{1}{3}\tau^{-4/3}
-\frac{2\ggb}{3}\left(\frac{\mathcal{A}_2-4~\mathcal{C}_2}{1+\ggb}\right)\tau^{-8/3}\right),\label{eq:GB-boundary-stress-results}
\end{eqnarray}
where we identify $\tpp$ with the proper time, $\tau$, at the boundary.\\

Before analyzing $T_{ab}$, we note three immediate observations:
\begin{enumerate}
\item $T_{ab}$ is traceless:
\begin{equation}
\eta^{ab}T_{ab} = 0\label{eq:GB-boundary-Tmn-traceless}
\end{equation}
with $\eta_{ab}$ given by Eq. \eqref{flat-metric-T-Y}.
\item Conservation implies a relationship between $\mathcal{A}_2$ and $\mathcal{C}_2$:
\begin{equation}
\partial_{a}T^{ab} = 0 \implies\mathcal{C}_2 = \frac{\mathcal{A}_2}{2} \,.\label{eq:GB-boundary-stress-constraint-1}
\end{equation}
\item The stress-energy tensor is completely specified by a single time-dependent function, $\varepsilon(\tau)\equiv T_{\tpp\tpp}$:
\begin{align}
 T_{\tpp\tpp} = \varepsilon\,, && T_{yy} = -\tau^2\left(\tau \partial_\tau \varepsilon +\varepsilon \right)\,,&& T_{x_{\perp}x_{\perp}} = \varepsilon+\frac{1}{2}\tau \partial_\tau \varepsilon \,.\label{eq:GB-boundary-Tmn-engdens}
\end{align}
\end{enumerate}
The three properties above are the defining properties of the hydrodynamic description of a relativistic, conformal Bjorken fluid. The only thing that remains to be specified is a single integration constant $\mathcal{A}_2$ (see discussion below Eq. \eqref{eq:ord2-sol-I}).

Now, the energy density of a Bjorken fluid, given by Eq. \eqref{Bjorken3rdFT}, can be written to second order in the hydrodynamic gradient expansion as
\begin{equation}
\varepsilon (\tau) = \frac{C}{\tau^{4/3}}\left(1-2\frac{\bar{\eta}}{\tau^{2/3}}+\frac{\bar{\Sigma}_{(2)}}{\tau^{4/3}} \right),\label{eq:GB-epsilon}
\end{equation}
where $\bar{\Sigma}_{(2)}$ represents the relevant linear combination of second-order transport coefficients:
\begin{equation}
\bar{\Sigma}_{(2)} = \frac{3\bar{\eta}^2}{2}-\frac{2\bar{\eta}\bar{\tau}_{\text{II}}}{3}+\frac{2\bar{\lambda}_1}{3} \,.
\end{equation}
By comparing the energy density of the Gauss-Bonnet fluid derived in the previous section with that of the Bjorken fluid, we identify
\begin{align}
C = \frac{3\sqrt{2}w^4}{(1+\ggb)^{3/2}\kappa_5^2}\,, && \bar{\eta} = \frac{\ggb^2}{3 w}\,, &&\bar{\Sigma}_{(2)} = -\mathcal{A}_2\left(\frac{2\ggb}{1+\ggb}\right)\,.\label{eq:GB-Transport-Coeffs}
\end{align}
At zeroth order in the hydrodynamic expansion, the energy density of our plasma is, as required,
\begin{equation}
\varepsilon_0 = \frac{C}{\tau^{4/3}} = \frac{3\sqrt{2}\pi^4 T^4 }{(1+\ggb)^{3/2}\kappa_5^2} \,, \label{eq:GB-eng-dens}
\end{equation}
where we have used Eq. \eqref{eq:hawking-temp_boosted} to express our answer in terms of $T$. The shear viscosity is then
\begin{equation}
\eta = C \bar{\eta}\left(\frac{\varepsilon}{C}\right)^{3/4} = \frac{\sqrt{2}\pi^3}{\kappa_5^2}\frac{T^3\ggb^2}{(1+\ggb)^{3/2}} \,, \label{eq:GB-shear-viscosity}
\end{equation}
which agrees with Eq. \eqref{gb-visc}. At second order, we find
\begin{equation}
\lambda_1-\eta\tau_{\text{II}} = C\left(\bar{\lambda}_1-\bar{\eta}\bar{\tau}_{\text{II}}\right)\left(\frac{\varepsilon}{C}\right)^{1/2} \,,
\end{equation}
matches the known result (see Eqs. \eqref{gb-visc}--\eqref{l1}),
\begin{equation*}
\lambda_1-\eta \tau_{\text{II}} = \frac{\sqrt{2}\pi^2}{8\kappa_5^2}\frac{T^2}{(1+\ggb)^{3/2}}\left\{
6+\ggb^2\bigg((3\ggb-2)\ggb-11\bigg)+2\ggb^2\ln\left[2+\frac{2}{\ggb}\right]
\right\}
\end{equation*}
provided
\begin{equation}
\mathcal{A}_2 = -\frac{1}{72w^2}\left(\frac{1+\ggb}{\ggb}\right)\left(6+\ggb^2(1+\ggb)(9\ggb-11)+2\ggb^2\ln\left[2+\frac{2}{\ggb}\right]\right).\label{eq:GB-boundary-stress-constraint-2}
\end{equation}
Collecting our results, the energy density, as a function of proper time, takes the final form:
\begin{align}
\varepsilon (\tau) =& ~ \frac{3\sqrt{2}}{(1+\ggb)^{3/2}\kappa_5^2}\left(\frac{w^4}{\tau^{4/3}}\right)\left[
1-\frac{2\ggb^2}{3w}\tau^{-2/3}\right.\nonumber\\
&+\left.\frac{1}{36 w^2}\left(
6+\ggb^2(1+\ggb)(9\ggb-11)+2\ggb^2\ln\left[2+\frac{2}{\ggb}\right]
 \right)\tau^{-4/3}
\right]\,.
\label{eq:GB-energy-density-full}
\end{align}


\section{Breakdown of non-local observables\label{sec:break}}

In this section we study various non-local observables in the boost-invariant backgrounds described
above. As advertised in the Introduction, we will see that requiring a physically sensible behavior
for the observables leads to several constraints on the regime of validity of hydrodynamic gradient
expansions at a given order.

\subsection{Two-point functions}\label{Sec:2ptFn}

According to the holographic dictionary \cite{Gubser:1998bc,Witten:1998qj}, bulk fields
$\phi$ are dual to gauge-invariant operators $\mathcal{O}$ with conformal dimension $\Delta$,
specified by their spin $s$, the mass $m$ and the number of dimensions $d$. For scalar fields, the relation is given by $\Delta(\Delta-d)=m^2$. The equivalence between the two sides of the correspondence can be made more precise by the identification:
\be\label{holography}
\mathcal{Z}_{\text{Bulk}}[\phi_\epsilon]=\left\langle e^{\int d^dx \phi_\epsilon(x)\mathcal{O}(x)}\right\rangle_{\text{CFT}}.
\ee
The left-hand-side of the above equation is  the  bulk  partition  function, where we impose the boundary condition $\phi\to\epsilon^{d-\Delta}\phi_\epsilon$. The right-hand-side is the generating functional of correlation functions of the CFT, where the boundary value $\phi_\epsilon$ acts as a source of
the dual operator $\mathcal{O}$. The equivalence (\ref{holography}) becomes handy by treating the bulk path integral in the saddle point approximation.  In
this regime, the above relation becomes
\be
S_{\text{on-shell}}[\phi_\epsilon]=-\Gamma_{\text{CFT}}[\phi_\epsilon]\,,
\ee
where on the left-hand side we have the bulk action evaluated on-shell and the right-hand side is
the generating functional of connected correlation functions of the CFT.
For instance, two-point functions can be computed by differentiating two times with respect to the source:
\be
\langle\mathcal{O}(x)\mathcal{O}(x')\rangle=-\frac{\delta S_{\text{on-shell}}}{\delta \phi_\epsilon(x)\delta \phi_\epsilon(x')}\bigg|_{\phi_\epsilon=0}.
\ee
For operators with large conformal dimension $\Delta$ (or equivalently, bulk fields with large mass $m$) the above problem simplifies even further. It can be shown that, in this limit, the relevant two-point functions reduce to the computation of geodesics in the given background geometry \cite{Balasubramanian:1999zv,Louko:2000tp}, i.e.
\be
\langle \cO(x) \cO(x') \rangle \sim e^{- \Delta\, \mathcal{S}_{\rm reg}(x,x')}\,,
\ee
where $\mathcal{S}_{\rm reg}$ is the regularized length of a geodesic connecting the boundary points $x$ and $x'$.

\subsubsection{Perturbative expansion: Eddington-Finkelstein vs. Fefferman-Graham}

We can now compute the late-time behavior of scalar two-point functions probing the out-of-equilibrium Bjorken flow. In order to do so we will follow the approach of \cite{Pedraza:2014moa}.\footnote{See \cite{Kundu:2016cgh} for a more detailed explanation.} Consider the functional $\mathcal{L}[\phi(y);\alpha]$ for the geodesic length, i.e. $\mathcal{S}\equiv\int dy\, \mathcal{L}[\phi(y);\alpha]$. Here, $\phi(y)$ denotes collectively all of the embedding functions, $y$ is the affine parameter and $\alpha$ is a small parameter related to the hydrodynamic gradient expansion in which the perturbation is carried out. Its precise definition will be given below. We can expand both $\mathcal{L}$ and $\phi(y)$ as:
\be
\begin{split}
\mathcal{L}[\phi(y);\alpha]&=\mathcal{L}^{(0)}[\phi(y)]+\alpha\mathcal{L}^{(1)}[\phi(y)]+\mathcal{O}(\alpha^2)\,,\\
\phi(y)&=\phi^{(0)}(y)+\alpha \phi^{(1)}(y)+\mathcal{O}(\alpha^2)\,.
\end{split}
\ee
The functions $\phi^{(n)}(y)$ can in principle be found by solving the geodesic equation order-by-order in $\alpha$. However, the embedding equations are in most cases highly non-linear making closed form solutions difficult to find. The key point here is that at first order in $\alpha$,
\be\label{Slambdaonshell}
\begin{split}
&\mathcal{S}_{\text{on-shell}}(x,x')=\int dy\,\mathcal{L}^{(0)}[\phi^{(0)}(y)]+\alpha\int dy\,\mathcal{L}^{(1)}[\phi^{(0)}(y)]\\
&\qquad\qquad\qquad\qquad\,\,\,+\alpha \int dy\, \phi^{(1)}_i(y)\left[\cancel{\frac{d}{dy}\frac{\partial\mathcal{L}^{(0)}}{\partial \phi_i'(y)}-\frac{\partial\mathcal{L}^{(0)}}{\partial \phi_i(y)}}\right]_{\phi^{(0)}}\!\!\!+\cdots\,,
\end{split}
\ee
so we only need $\phi^{(0)}(y)$ to obtain the first correction to the geodesic length.

Let us now discuss the expansion parameter $\alpha$ in more detail. In particular, what we will see is that there is a natural choice for $\alpha$ depending on whether we work in Eddington-Finkelstein or Fefferman-Graham coordinates, so we must proceed with some care before we interpret our results.\footnote{In Appendix \ref{expansions}, we provide details of the metric expansions that we use in these two coordinate systems.} Let us start with the Fefferman-Graham expansion, which was first considered in \cite{Pedraza:2014moa}. In this case, the metric coefficients can be expanded as in Eq. \eqref{FGexp} so each hydrodynamic order is suppressed by a factor of the dimensionless quantity $\tilde{u}=\tau^{-2/3}w^{-1}$, where $w$ is the same dimensionful parameter that appears in the energy density. On the other hand, the near-boundary expansion stipulates that we can alternatively expand all metric coefficients in powers of $\tilde{v}\equiv z\tau^{-1/3}w$. This is the expansion that will be relevant for our perturbative calculation (\ref{Slambdaonshell}). Notice that when $\tilde{v}\to0$, we recover pure AdS,
for which the embedding function $\phi^{(0)}(y)$ is analytically known. The first correction in this expansion enters at order $\cO(\tilde{v}^4)$ so we can identify $\alpha\sim\tilde{v}^4$. Now, according to the UV/IR connection \cite{Peet:1998wn,Hatta:2010dz,Agon:2014rda}, the bulk coordinate $z$ can roughly be mapped to the length scale $z\sim\ell$ in the boundary theory. In our setup, the only length scale of the problem is given by the separation the two points $(x,x')$ so $\ell\sim\Delta x\equiv|x-x'|$.\footnote{More precisely, we will see that $\ell$ can be naturally identified with the maximal depth of the geodesic $z_*$, which at leading order is given by $z_*=\frac{\Delta x}{2}$.} Therefore, in terms of CFT data, our expansion parameter in Fefferman-Graham coordinates is given by
\be
\alpha=\ell^4\tau^{-4/3}w^4\qquad\text{(Fefferman-Graham)}.
\ee
As mentioned already in Appendix \ref{expansions}, the leading correction to the metric in the
near-boundary expansion receives contributions at all orders in hydrodynamics, so one can obtain
non-trivial results by studying contributions to the two-point correlators to only first order. For
instance, as found in Ref. \cite{Pedraza:2014moa}, in order to have a well behaved late-time relaxation of longitudinal two-point functions, first-order hydrodynamics puts a constraint on the regime of validity of $\tilde{u}$. Namely, the approximation breaks down when\footnote{The results of \cite{Pedraza:2014moa} are written in terms of $\epsilon=\frac{3w^4}{4}$.}
\be\label{constraint1storder}
\tilde{u}>1/2\qquad\Longrightarrow\qquad\tau<\tau_{\text{crit}}^{1\text{st}}=2^{3/2}w^{-3/2}\approx2.828w^{-3/2}\,.
\ee
In this work, we are interested in studying both $i)$ higher order hydrodynamic corrections
and $ii)$ (inverse) coupling constant corrections in the $\mathcal{N}=4$ plasma and a hypothetical dual of Gauss-Bonnet theory.

In Eddington-Finkelstein coordinates, the hydrodynamic expansion is performed in terms of $u$, and the near boundary expansion in terms of $v$, both given in Eq. \eqref{vuinEF}. However, notice that these
definitions involve $\tau_+$ instead of $\tau$, which at the leading order becomes Eq. \eqref{tauptotauz}. If we perform a similar analysis in Fefferman-Graham coordinates, we find that in Eddington-Finkelstein coordinates
the expansion parameter is given by $\alpha\sim v^{-4}$, or equivalently,
\be\label{lambdaEF}
\alpha=\ell^4(\tau-\ell)^{-4/3}w^4\qquad\text{(Eddington-Finkelstein)}.
\ee
Notice that in this case, truncating the expansion (\ref{Slambdaonshell}) at the leading order in $\alpha$ is problematic for $\tau<\ell$. Furthermore, if we expand
(\ref{lambdaEF}) for $\ell\ll\tau$, even the first subleading term is not complete since, due to the coordinate mixing, we would require higher order terms in the near-boundary expansion to have a full result at the given order in $\ell/\tau$. Thus, in Eddington-Finkelstein coordinates the results can only be trusted in the limit $\ell/\tau\to0$.\footnote{For longitudinal correlators, this would imply that only the $\Delta y\to0$ limit is valid. Fortunately, this is exactly the limit for which the constraint (\ref{constraint1storder}) was found.} To avoid this issue we will convert first to Fefferman-Graham coordinates and perform our calculations in that chart.\footnote{We explicitly checked that the results in both coordinate systems agree at the leading order in $\ell/\tau$.} Explicit expressions for the metric functions are given in Appendix \ref{FGexpr}.

\subsubsection{Transverse correlator}\label{Sec:2ptTransverse}

In Fefferman-Graham coordinates, a generic bulk metric dual to Bjorken hydrodynamics can be written as follows:
\begin{equation}\label{FGgen}
ds^2 = \frac{1}{z^2}\left(-e^{\tilde{a}}d\tau^2+e^{\tilde{b}}\tau^2dy^2+e^{\tilde{c}}d\vec{x}_{\perp}^2+dz^2\right)\,,
\end{equation}
where $\{\tilde{a},\tilde{b},\tilde{c}\}$ are functions of $(\tau,z)$ that can be
expanded in terms of $\tilde{u}= \tau^{-2/3}w^{-1}$ and $\tilde{v}= z\tau^{-1/3}w\ll1$ as in
(\ref{FGnearB}), i.e., $ \tilde{a}(\tilde{v},\tilde{u}) =
\tilde{\mathfrak{a}}_4(\tilde{u})\tilde{v}^{4}+\ldots\,$, and similarly for $\tilde{b}$ and
$\tilde{c}$. Notice that we have set the AdS radius to unity $L=1$. The AdS radius generally depends
on the cosmological constant $\Lambda$ as well as all higher derivative couplings of the gravity
theory that we consider. Since $L$ is just an overall factor of our metric, it will only appear as an overall factor in the various observables we study, and can be easily restored via dimensional analysis.

Let us begin by considering space-like geodesics connecting two boundary points separated in the transverse plane: $(\tau_0,x)$ and $(\tau_0, x')$, where $x\equiv x_1$ and all other spatial directions are identical. Because the metric (\ref{FGgen}) is invariant under translations in $x$, we can parameterize the geodesic by two functions $\tau(z)$ and $x(z)$, satisfying the following UV boundary conditions:
\be\label{bcstrans}
\tau(0) = \tau_0 \ , \quad x(0) = \pm \frac{\Delta x}{2} \ .
\ee
At the end of the calculation, we can shift our coordinate $x\to x+x_0$, where $x_0=\frac{1}{2}(x+x')$, and express the results in terms of $\Delta x=|x-x'|$, for any $x$ and $x'$.
The length of such a geodesic is given by:
\be\label{actiongeo1}
\mathcal{S}=2\int_{0}^{z_*} \frac{dz}{z} \sqrt{1+e^{\tilde{c}}x'^2-e^{\tilde{a}}\tau'^2}\,.
\ee
We can now use (\ref{FGnearB}) and expand the above as: $\mathcal{S}=\mathcal{S}^{(0)}+\mathcal{S}^{(1)}+\cdots\,$, where
\be
\mathcal{S}^{(0)}=2\int_{0}^{z_*} \frac{dz}{z} \sqrt{1+x'^2-\tau'^2}\,,\qquad \mathcal{S}^{(1)}=w^4\int_{0}^{z_*} dz \frac{z^3(\tilde{\mathfrak{c}}_4x'^2-\tilde{\mathfrak{a}}_4\tau'^2)}{\tau^{4/3}\sqrt{1+x'^2-\tau'^2}}\,.\label{mathcalL01}
\ee
The first term is just the pure AdS contribution, which is UV divergent. To see this, we can use the zeroth order embeddings:
\be\label{embedding0}
\tau(z)=\tau_0\,,\qquad\qquad\,x(z)=\sqrt{z_*^2-z^2},
\ee
with $z_*=\frac{\Delta x}{2}$. Integrating from $\epsilon\to0$ to $z_*$ and subtracting the divergence $\mathcal{S}_{\rm div}=-2 \ln \epsilon$, we obtain:
\be
\mathcal{S}^{(0)}_{\rm reg}=2\ln\Delta x\,,\qquad\qquad \langle\cO(x)\cO(x')\rangle\sim\frac{1}{|x-x'|^{2\Delta}}\,,
\ee
which is the expected result for a two-point correlator in the vacuum of a CFT. At next order, the correlator can be written as follows:
\be
\langle\cO(x)\cO(x')\rangle\sim\frac{1}{|x-x'|^{2\Delta}}e^{-\Delta \mathcal{S}^{(1)}(\tilde{\mathfrak{a}}_4,\tilde{\mathfrak{c}}_4)}\,,
\ee
where $\mathcal{S}^{(1)}$ is given in (\ref{mathcalL01}). The functions $\{\tilde{\mathfrak{a}}_4(\tilde{u}),\tilde{\mathfrak{b}}_4(\tilde{u}),\tilde{\mathfrak{c}}_4(\tilde{u})\}$ are generically theory-dependent (see Appendix \ref{FGexpr} for explicit expressions) and contain information about all orders in hydrodynamics. On general grounds, we expect $\mathcal{S}^{(1)}$ to be positive definite at late times, so the correlator relaxes from above as the plasma cools down. Below, we will use the explicit form of $\{\tilde{\mathfrak{a}}_4(\tilde{u}),\tilde{\mathfrak{b}}_4(\tilde{u}),\tilde{\mathfrak{c}}_4(\tilde{u})\}$ to put constraints on the regime of validity of hydrodynamics, at each order in the derivative expansion.

For the transverse correlator, there is a very drastic simplification: once we evaluate $\mathcal{S}^{(1)}$ using the zeroth order embeddings (\ref{embedding0}), we have:
\be
\mathcal{S}^{(1)}=\frac{w^4\Delta x^4\tilde{\mathfrak{c}}_4(\tilde{u}_0)}{16\tau_0^{4/3}}\int_{0}^{1} dx \frac{x^5}{\sqrt{1-x^2}}=\frac{w^4\Delta x^4\tilde{\mathfrak{c}}_4(\tilde{u}_0)}{30\tau_0^{4/3}}\,,
\ee
where $x=z/z_*$ and $\tilde{u}_0=\tau_0^{-2/3}w^{-1}$. Therefore, the positivity of $\mathcal{S}^{(1)}$ follows directly from the positivity of $\tilde{\mathfrak{c}}_4(\tilde{u})$. Let us specialize to the particular cases of interest: Einstein gravity (which is dual to a Bjorken flow at infinite coupling), and higher derivative gravities with $\alpha'$- and $\lgb$-corrections (two different models of Bjorken flow with finite coupling corrections).

\begin{figure}[t!]
\begin{center}
\includegraphics[angle=0,width=0.48\textwidth]{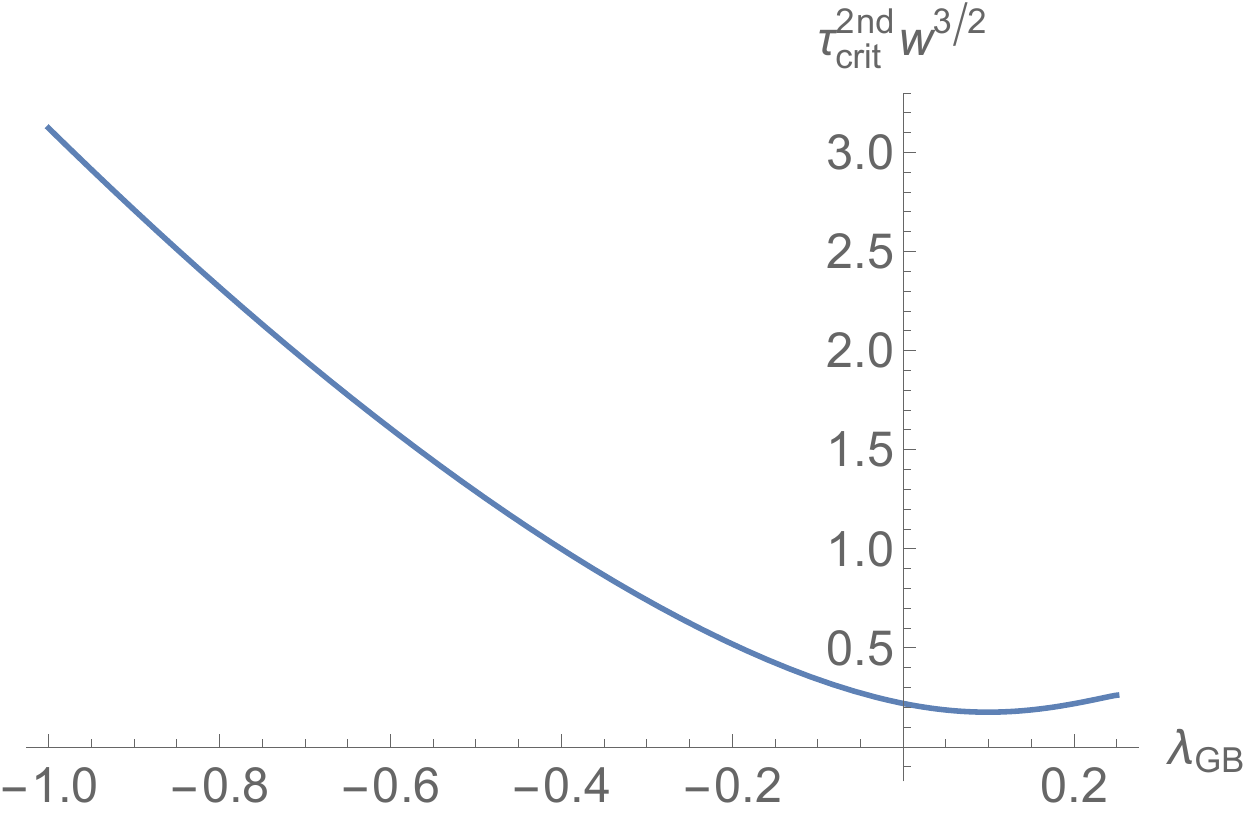}
\caption{Behavior of $\tau_\text{crit}^{2\text{nd}}(\lgb)$, non-perturbative in $\lgb$, coming from the transverse correlator. Negative values of $\lgb$ resemble qualitatively the expected behavior as we flow from strong to weak coupling. \label{tcr2GB}}
\end{center}
\end{figure}

\begin{itemize}
  \item \emph{Einstein gravity.} The function $\tilde{\mathfrak{c}}_4(\tilde{u})$ is known up to third order in hydrodynamics and is given by equation (\ref{3rdEGcoeffs}).
  Up to first order in hydrodynamics $\tilde{\mathfrak{c}}_4(\tilde{u})$  is positive definite but it becomes negative for $\tau<\tau_\text{crit}^{2\text{nd}}$ and $\tau<\tau_\text{crit}^{3\text{rd}}$ in second- and third-order hydrodynamics, respectively, where
  \begin{align}\label{transboundsEG}
  \tau_\text{crit}^{2\text{nd}}=0.219w^{-3/2}\,,&& \tau_\text{crit}^{3\text{rd}}=0.403w^{-3/2}\,.
  \end{align}
It is interesting to note that for this particular obsevable, the above criterion would naively
imply that third-order hydrodynamics is more constraining than second-order hydrodynamics. However,
as we will see below, the most stringent bound on the applicability of hydrodynamics will come from the longitudinal correlator, which decreases at each order in hydrodynamics (up to third order), as expected.
  \item \emph{$\alpha'$-corrections.} The function $\tilde{\mathfrak{c}}_4(\tilde{u})$ is known to linear order in $\gamma = \alpha'^3 \zeta(3) / 8=\lambda^{-3/2}\zeta (3) L^6/8$, and up to second order in hydrodynamics, and is given by equation (\ref{2ndapcoeffs}). The coefficient $\tilde{\mathfrak{c}}_4(\tilde{u})$ is positive definite for first-order hydrodynamics, but becomes negative for $\tau<\tau_\text{crit}^{2\text{nd}}(\gamma)$ in second order hydrodynamics, where
      \be\label{transboundsap}
      \tau_\text{crit}^{2\text{nd}}(\gamma)= \left(0.219+45.711\, \gamma +\CO(\gamma^2)\right)w^{-3/2}\,.
      \ee
      Finite coupling corrections ($\gamma>0$) are shown to increase $\tau_\text{crit}^{2\text{nd}}$, which is in accordance with our expectations that they should reduce the regime of validity of hydrodynamics. As we will see below, the most stringent bound will again come from the longitudinal correlator.
  \item \emph{$\lgb$-corrections.} The function $\tilde{\mathfrak{c}}_4(\tilde{u})$ is known non-perturbatively in $\lgb$ and up to second order in hydrodynamics, and is given by equation (\ref{2ndgbcoeffs}). $\tilde{\mathfrak{c}}_4(\tilde{u})$ is positive definite for first-order hydrodynamics, but becomes negative for $\tau<\tau_\text{crit}^{2\text{nd}}(\lgb)$ in second-order hydrodynamics, where
      \be\label{transboundsgb}
      \tau_\text{crit}^{2\text{nd}}(\lgb)=\left(0.219-0.866\lgb+\CO(\lgb^2)\right)w^{-3/2}\,.
      \ee
      Negative values of $\lgb$ tend to increase $\tau_\text{crit}^{2\text{nd}}$ so they reduce the
      regime of validity of hydrodynamics. This is indeed the expected behavior as we flow from
      strong to weak coupling. It is also interesting to study the full dependence of
      $\tau_\text{crit}^{2\text{nd}}$ on $\lgb\in(-\infty,1/4]$, which we plot in Figure
      \ref{tcr2GB}. For negative $\lgb$, we observe that $\tau_\text{crit}^{2\text{nd}}$ increases
      monotonically. However, for positive $\lgb$, $\tau_\text{crit}^{2\text{nd}}$ is
      non-monotonic. We note that, also for this case, the true bound will come from the longitudinal correlator.
\end{itemize}

Finally, it is worth noting that the results above can be expressed generically in terms of a few theory-specific constants $\{\hat{\Sigma},\,\hat{\Sigma}^{(\gamma)}_\epsilon,\,\hat{\Sigma}^{(\lgb)}_\epsilon,\,\hat{\Lambda}\}$, which can be found in Appendix \ref{crit-analytic}. At second order in the hydrodynamic expansion, the critical time is given by
\begin{equation}
\tau_{\text{crit}}^{2\text{nd}} = \hat{\Sigma}^{3/4}w^{-3/2}\,.\label{crit-t-trans-generic-2nd}
\end{equation}
Expressing our coupling constants $\gamma$ and $\lgb$ collectively as $\beta$, first-order corrections to $\tau_{\text{crit}}^{2\text{nd}}$ then take the form
\begin{equation}
\tau_{\text{crit}}^{2\text{nd}}(\beta)=  \tau_{\text{crit}}^{2\text{nd}}\left(1+\frac{3\beta}{4}\, \hat{\Sigma}^{(\beta)}_{\epsilon}+\mathcal{O}(\beta^2)\right)\,.\label{crit-t-trans-2nd-correction}
\end{equation}
The expressions for $\tau_{\text{crit}}^{3\text{rd}}$ are complicated, but correspond to the smallest real root of the equation
\begin{equation}
1-\hat{\Sigma}\xi^{4/3}-2\hat{\Lambda}\xi^2 = 0 \,, \label{crit-t-trans-3rd}
\end{equation}
where $\xi=\tau_0^{-1}w^{-3/2}$.

\subsubsection{Longitudinal correlator}

We are now interested in a space-like geodesic connecting two boundary points in the longitudinal plane: $(\tau_0,y)$ and $(\tau_0,y')$ for any $y$ and $y'$. We can make use of the invariance under translations in $y$ and parameterize the geodesic by functions $\tau(z)$ and $y(z)$ with boundary conditions
\begin{align}\label{realbcB}
\tau(0) = \tau_0 \, , \qquad \qquad y(0) = \pm \frac{\Delta y}{2} \, .
\end{align}
At the end, if desired, we can simply shift our rapidity coordinate $y\to y+y_0$, where $y_0=\frac{1}{2}(y+y')$, and express our results in terms of
$x_3=\tau_0 \sinh (y_0+\tfrac{\Delta y}{2})$ and $x_3'=\tau_0 \sinh (y_0-\tfrac{\Delta y}{2})$. The length of such a geodesic is given by:
\be\label{actiongeoB}
\mathcal{S}=2\int_{0}^{z_*} \frac{dz}{z} \sqrt{1+e^{\tilde{b}}\tau^2y'^2-e^{\tilde{a}}\tau'^2}\,.
\ee
We can now use (\ref{FGnearB}) and expand the above as: $\mathcal{S}=\mathcal{S}^{(0)}+\mathcal{S}^{(1)}+\cdots\,$, where
\be
\mathcal{S}^{(0)}=2\int_{0}^{z_*} \frac{dz}{z} \sqrt{1+\tau^2y'^2-\tau'^2}\,,\qquad \mathcal{S}^{(1)}=w^4\int_{0}^{z_*} dz \frac{z^3(\tilde{\mathfrak{b}}_4\tau^2y'^2-\tilde{\mathfrak{a}}_4\tau'^2)}{\tau^{4/3}\sqrt{1+\tau^2y'^2-\tau'^2}}\,.\label{mathcalL01B}
\ee
Again, the first term gives the pure AdS contribution. To see this, we can use the zeroth order embeddings, which in this case are given by:
\be\label{embedding0B}
\tau(z)=\sqrt{\tau_0^2+z^2}\,,\qquad \qquad y(z)=\mathrm{arccosh}\left(\frac{\tau_0 \cosh (\frac{\Delta y}{2})}{\tau(z)}\right)\,.
\ee
Integrating from $\epsilon\to0$ up to $z_*=\frac{\Delta x_3}{2}=\tau_0 \sinh (\frac{\Delta y}{2})$ and subtracting the divergent part $\mathcal{S}_{\rm div}=-2 \ln \epsilon$, we obtain:
\be
\mathcal{S}^{(0)}_{\rm reg}=2\ln\Delta x_3\,,\qquad\qquad \langle\cO(x_3)\cO(x_3')\rangle\sim\frac{1}{|x_3-x_3'|^{2\Delta}}\,.
\ee
At zeroth order, the longitudinal correlator depends only on $|x_3-x_3'|$. This is expected because this is the result for a two-point correlator in the vacuum of a CFT, which is translationally invariant. At next order, the correlator can be written as follows:
\be
\langle\cO(x)\cO(x')\rangle\sim\frac{1}{|x-x'|^{2\Delta}}e^{-\Delta \mathcal{S}^{(1)}(\tilde{\mathfrak{a}}_4,\tilde{\mathfrak{c}}_4)}\,,
\ee
where $\mathcal{S}^{(1)}$ is given in (\ref{mathcalL01B}). Again, we expect $\mathcal{S}^{(1)}$ to be positive definite at late times, so the correlator relaxes from above as the plasma cools down. However, we will see below that there are crucial differences with respect to the transverse case, which will ultimately lead to stricter bounds on the regime of validity of the hydrodynamic expansion.

The next step is to evaluate $\mathcal{S}^{(1)}$ using the zeroth-order embeddings (\ref{embedding0B}) and then use the explicit forms of $\{\tilde{\mathfrak{a}}_4(\tilde{u}),\tilde{\mathfrak{b}}_4(\tilde{u}),\tilde{\mathfrak{c}}_4(\tilde{u})\}$ which are theory-dependent. Defining a dimensionless variable $x=z/z_*$, we arrive at the following expression:
\be
\mathcal{S}^{(1)}=\frac{w^4 \Delta x_3^4}{\tau_0^{4/3}}\int_0^1dx\frac{x^5 [\tilde{\mathfrak{b}}_4(\tilde{u}(x))\cosh^2(\frac{\Delta y}{2})-\tilde{\mathfrak{a}}_4(\tilde{u}(x))(1-x^2)
\sinh^2(\frac{\Delta y}{2})]}{(1-x^2)^{1/2}[1+x^2 \sinh^2(\frac{\Delta y}{2})]^{5/3}}\,,\label{S1long}
\ee
where
\be
\tilde{u}(x)=\frac{1}{\tau_0^{2/3}w[1+x^2 \sinh^2(\frac{\Delta y}{2})]^{1/3}} \,.
\ee
Let us now consider expanding the functions $\{\tilde{\mathfrak{a}}_4(\tilde{u}),\tilde{\mathfrak{b}}_4(\tilde{u}),\tilde{\mathfrak{c}}_4(\tilde{u})\}$ at different orders in hydrodynamics. From the expansions in (\ref{abc4exps}), or directly from the explicit expressions (\ref{3rdEGcoeffs})--(\ref{2ndgbcoeffs}), it is clear that:
\begin{align}\label{generic-hydro-func-expansion}
\tilde{\mathfrak{a}}_4(\tilde{u})=\sum_{k=0}^{\infty}\tilde{\mathfrak{a}}_4^{(k)}\tilde{u}^k\,,&& \tilde{\mathfrak{b}}_4(\tilde{u})=\sum_{k=0}^{\infty}\tilde{\mathfrak{b}}_4^{(k)}\tilde{u}^k\,,&&
\tilde{\mathfrak{c}}_4(\tilde{u})=\sum_{k=0}^{\infty}\tilde{\mathfrak{c}}_4^{(k)}\tilde{u}^k\,,
\end{align}
for some numbers $\{\tilde{\mathfrak{a}}_4^{(k)},\tilde{\mathfrak{b}}_4^{(k)},\tilde{\mathfrak{c}}_4^{(k)}\}$. Different values of $k$ correspond to contributions from different orders in hydrodynamics; for example, $k=0$ corresponds to the perfect fluid approximation, $k=1$ corresponds to first-order hydrodynamics, and so on. Therefore, we can rewrite $\mathcal{S}^{(1)}$ as follows:
\be
\mathcal{S}^{(1)}=\frac{w^4 \Delta x_3^4}{\tau_0^{4/3}}\sum_{k=0}^{\infty}\tau_0^{-2k/3}w^{-k}\left[\tilde{\mathfrak{b}}_4^{(k)}\mathcal{I}^{(k)}_{-}\cosh^2(\tfrac{\Delta y}{2})-\tilde{\mathfrak{a}}_4^{(k)}\mathcal{I}^{(k)}_{+}\sinh^2(\tfrac{\Delta y}{2})\right]\,,\label{S1long2}
\ee
where
\be
\mathcal{I}^{(k)}_{\pm}=\int_0^1dx\frac{x^5(1-x^2)^{\pm1/2}}{[1+x^2 \sinh^2(\frac{\Delta y}{2})]^{(5+k)/3}}\,.
\ee
Both integrals can be performed analytically for any value of $k$, although we
refrain from writing them out here, since they are not particularly illuminating. Nevertheless, it is interesting to study the $\Delta y\to0$ limit, from which we can extract $\tau_{\text{crit}}$ at different orders in hydrodynamics \cite{Pedraza:2014moa}.
A simple observation is that both of $\mathcal{I}^{(k)}_{\pm}$ are positive definite and decrease monotonically as $\Delta y$
increases. In the limit $\Delta y\to0$, both integrals are finite and independent of $k$:
\be
\mathcal{I}^{(k)}_{\pm}\to \int_0^1x^5(1-x^2)^{\pm 1/2}=\frac{8}{15(4\pm 3)}.
\ee
However, it is clear that the first term of (\ref{S1long2}) dominates since in this limit $\cosh(\tfrac{\Delta y}{2})\to1$, while $\sinh(\tfrac{\Delta y}{2})\to\cO(\Delta y)$. Putting everything together, we find that for $\Delta y\to0$:
\be\label{s1deltay0}
\mathcal{S}^{(1)}\to\frac{8w^4 \Delta
  x_3^4}{15\tau_0^{4/3}}\sum_{k=0}^{\infty}\tilde{\mathfrak{b}}_4^{(k)}\tilde{u}_0^k=\frac{8w^4
  \Delta x_3^4\tilde{\mathfrak{b}}_4(\tilde{u}_0)}{15\tau_0^{4/3}}\, ,
\ee
where $\tilde{u}_0=\tau_0^{-2/3}w^{-1}$. Therefore, in this limit the positivity of $\mathcal{S}^{(1)}$ follows directly from the positivity of $\tilde{\mathfrak{b}}_4(\tilde{u})$. In the cases we considered, this criterion was enough to guarantee the positivity of $\mathcal{S}^{(1)}$ for any other value of $\Delta y$. However this does not trivially follow from (\ref{S1long2}): at finite $\Delta y$, the value of  $\mathcal{S}^{(1)}$  will generally depend on the interplay between the coefficients $\{\tilde{\mathfrak{a}}_4^{(k)},\tilde{\mathfrak{b}}_4^{(k)}\}$. In the following, we will study in more detail the behavior of $\mathcal{S}^{(1)}$ as a function of $\Delta y$ and $\tau_0 w^{3/2}$, specializing to the particular cases of interest: Einstein gravity and higher derivative gravities with $\alpha'$- and $\lgb$-corrections.

\begin{itemize}
  \item \emph{Einstein gravity.}
  The functions $\tilde{\mathfrak{a}}_4(\tilde{u})$ and $\tilde{\mathfrak{b}}_4(\tilde{u})$ are known up to third order in hydrodynamics and are given in (\ref{3rdEGcoeffs}). With these functions at hand we can extract the numbers $\tilde{\mathfrak{a}}_4^{(k)}$ and $\tilde{\mathfrak{b}}_4^{(k)}$ and then use formula (\ref{S1long2}). Figure \ref{LongEin} (left) shows some representative curves for $\tilde{\mathcal{S}}^{(1)}\equiv\mathcal{S}^{(1)}\tau_0^{4/3}/w^4 \Delta x_3^4$ as a function of $\Delta y$ for various values of $\xi=\tau_0^{-1} w^{-3/2}=\{0,0.15,0.3,0.45,0.6\}$ depicted in blue, orange, green, red and purple, respectively. The solid lines correspond to third-order hydrodynamics; the dashed and dotted lines correspond to second- and first-order hydrodynamics, respectively. For $\xi=0.45$ the dotted curve becomes negative for small $\Delta y$, indicating that first-order hydrodynamics is no longer valid. For  $\xi=0.6$ both the dotted and dashed curves are negative for small $\Delta y$. This indicates that second-order hydrodynamics is also invalid at this time. Finally, for all values of $\xi$ that were plotted, the solid lines are always positive, so third-order hydrodynamics is valid for these values. However, if we keep on increasing $\xi$, the solid lines will become unphysical for small $\Delta y$ at some point. We observe the following behavior for any finite value of $\xi$ (in the range of parameters that we plotted): the value of $\tilde{\mathcal{S}}^{(1)}$ increases up to a maximum $\tilde{\mathcal{S}}^{(1)}_{\text{max}}>0$ and then decreases monotonically to zero as $\Delta y \to \infty$. This implies that the positivity of $\tilde{\mathcal{S}}^{(1)}$ at $\Delta y=0$ is enough to guarantee a good physical behavior for any $\Delta y$.
  In Figure \ref{LongEin} (right) we show the behavior of  $\tilde{\mathcal{S}}^{(1)}(0)$ as a function of $\xi$ for first-, second- and third-order hydrodynamics,
  depicted in blue, orange and green, respectively, and we indicate the times at which it becomes negative. From the $\Delta y\to0$ limit of the correlator (\ref{s1deltay0}) we obtain the critical times:
  \be
  \tau_\text{crit}^{1\text{st}}= 2.828w^{-3/2}\,,\qquad \tau_\text{crit}^{2\text{nd}}=1.987w^{-3/2}\,,\qquad \tau_\text{crit}^{3\text{rd}}=1.503w^{-3/2}\,.
  \ee
  These bounds are stricter than the ones derived from the transverse correlator (\ref{transboundsEG}), and decrease at each order in hydrodynamics, as expected.
\begin{figure}[t!]
\begin{center}
\includegraphics[angle=0,width=0.48\textwidth]{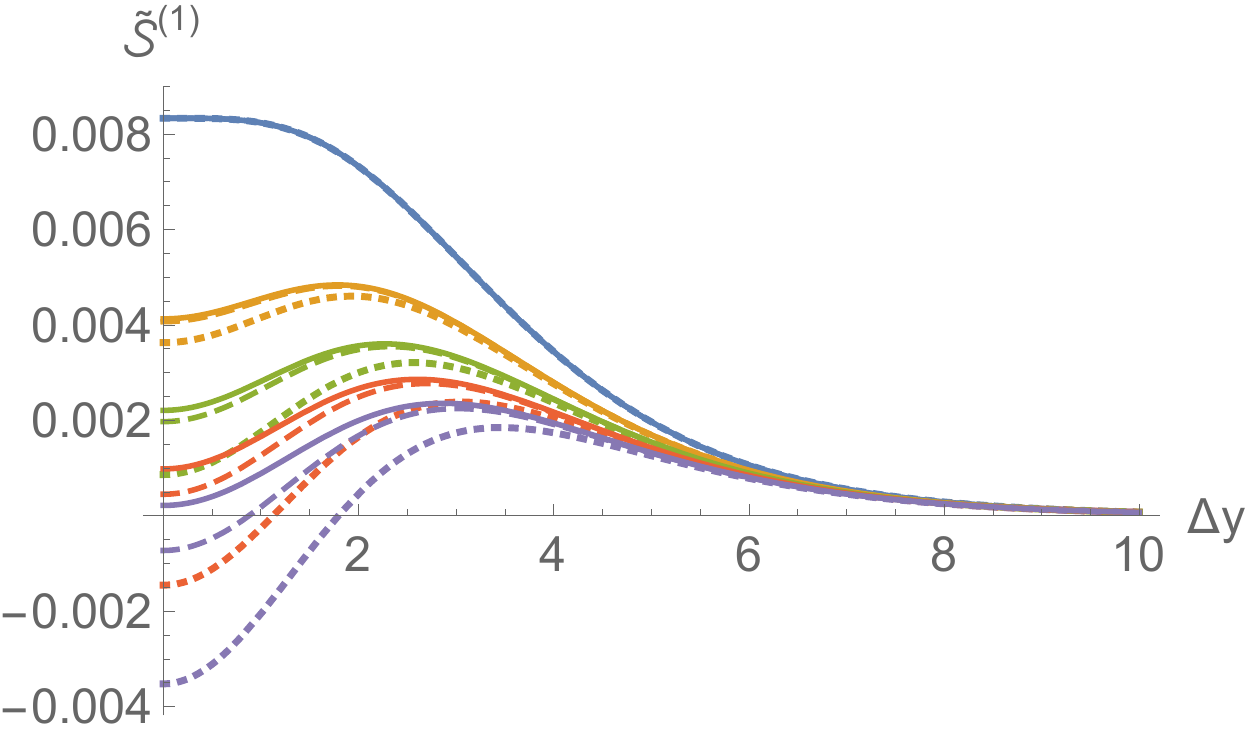}\includegraphics[angle=0,width=0.48\textwidth]{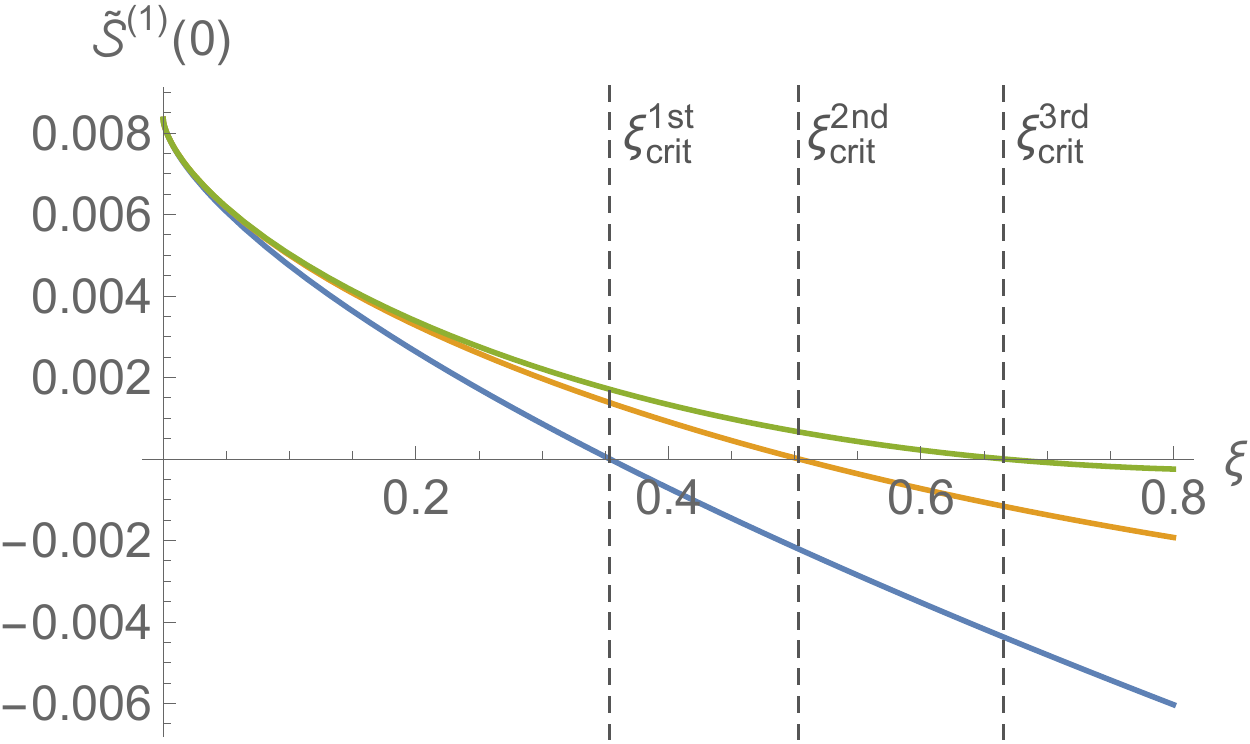}
\caption{ Left: Plots for $\tilde{\mathcal{S}}^{(1)}\equiv\mathcal{S}^{(1)}\tau_0^{4/3}/w^4 \Delta x_3^4$ for various values of $\xi=\tau_0^{-1} w^{-3/2}=\{0,0.15,0.3,0.45,0.6\}$ depicted in blue, orange, green, red and purple, respectively. The solid lines correspond to third-order hydrodynamics; the dashed and dotted lines correspond to second- and first-order hydrodynamics, respectively. Right: Plots for $\tilde{\mathcal{S}}^{(1)}(0)$ for first-, second- and third-order hydrodynamics, depicted in blue, orange and green, respectively. The dashed vertical lines correspond to the critical times at each order in hydrodynamics. \label{LongEin}}
\end{center}
\end{figure}
  \item \emph{$\alpha'$-corrections.}
    The functions $\tilde{\mathfrak{a}}_4(\tilde{u})$ and $\tilde{\mathfrak{b}}_4(\tilde{u})$ are known to linear order in $\gamma = \alpha'^3 \zeta(3) / 8=\lambda^{-3/2}\zeta (3) L^6/8$ and up to second order in hydrodynamics, and are given in (\ref{2ndapcoeffs}). With these functions in hand, we can extract the numbers $\tilde{\mathfrak{a}}_4^{(k)}$ and $\tilde{\mathfrak{b}}_4^{(k)}$ and then use the formula (\ref{S1long2}). Figure \ref{Longap} (left) shows some representative curves for $\tilde{\mathcal{S}}^{(1)}\equiv\mathcal{S}^{(1)}\tau_0^{4/3}/w^4 \Delta x_3^4$ as a function of $\Delta y$ for various values of $\xi=\tau_0^{-1} w^{-3/2}=\{0,0.12,0.25,0.28,0.5\}$ depicted in blue, orange, green, red and purple, respectively. The solid lines correspond to $\gamma=0$ (Einstein gravity) while the dashed lines correspond to $\gamma=10^{-3}$, both for second-order hydrodynamics. For all the $\xi$ that were plotted the solid lines are well behaved because we have chosen $\xi<\xi_\text{crit}^{2\text{nd}}(\gamma=0)=0.503$. For $\xi=0.5$ the dashed curve becomes negative for small $\Delta y$, indicating that second-order hydrodynamics becomes invalid faster at finite coupling. We observe the same behavior as in Einstein gravity, namely that the positivity of $\tilde{\mathcal{S}}^{(1)}$ at $\Delta y=0$ is enough to guarantee a good physical behavior for any $\Delta y$. In Figure \ref{Longap} (right) we show the behavior of  $\tilde{\mathcal{S}}^{(1)}(0)$ both for $\gamma=0$ and $\gamma=10^{-3}$ as a function of $\xi$ for first- and second-order hydrodynamics, depicted in blue and orange, respectively, and we indicates the times at which it becomes negative. From the $\Delta y\to0$ limit of the correlator (\ref{s1deltay0}) we obtain the following critical times:
\begin{align}
\tau_\text{crit}^{1\text{st}}(\gamma)&=\left(2.828+474.115\gamma +\CO(\gamma^2) \right)w^{-3/2}\,,\\
\tau_\text{crit}^{2\text{nd}}(\gamma)&=\left(1.987+275.079\gamma +\CO(\gamma^2)\right )w^{-3/2}\,.
\end{align}
  These bounds increase as we increase the value of $\gamma$ and are stricter than the ones derived from the transverse correlator (\ref{transboundsap}).
  Based on this, we can conclude that finite coupling corrections indeed tend to reduce the regime of validity of hydrodynamics.
  \begin{figure}[t!]
\begin{center}
\includegraphics[angle=0,width=0.48\textwidth]{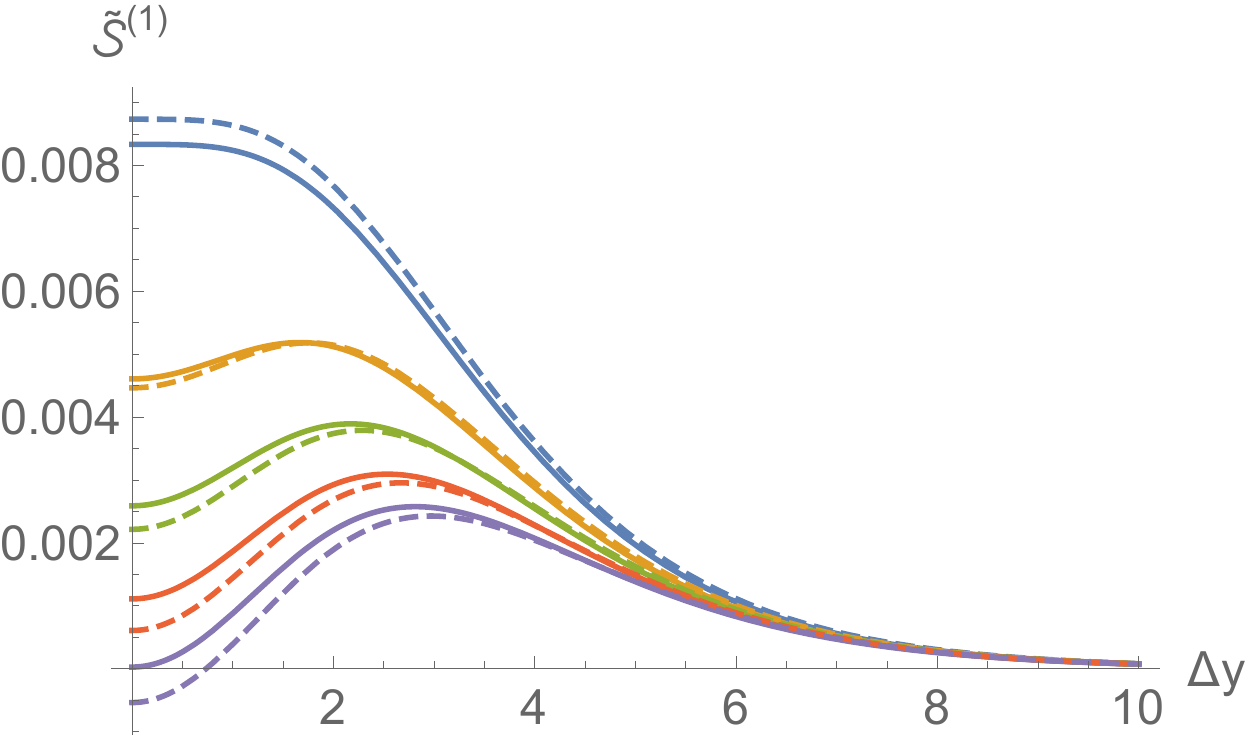}\includegraphics[angle=0,width=0.48\textwidth]{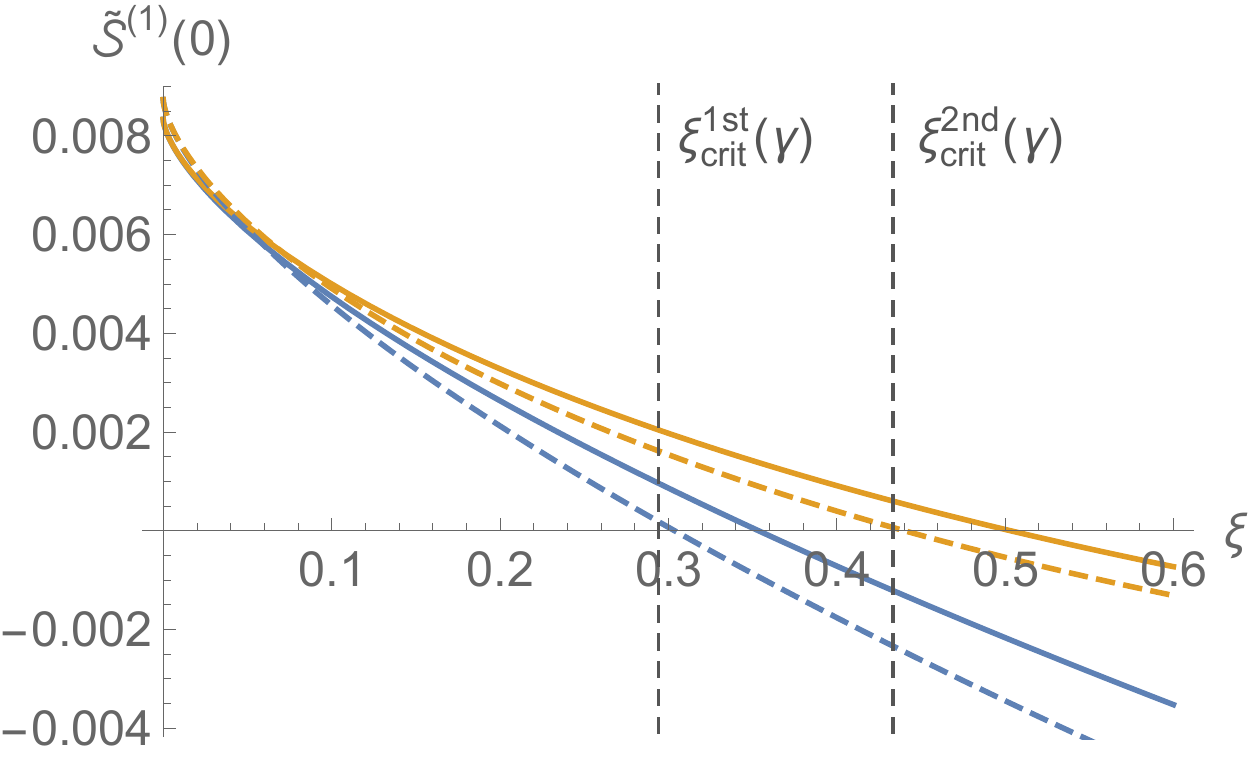}
\caption{ Left: Plots for $\tilde{\mathcal{S}}^{(1)}\equiv\mathcal{S}^{(1)}\tau_0^{4/3}/w^4 \Delta x_3^4$ for various values of $\xi=\tau_0^{-1} w^{-3/2}=\{0,0.12,0.25,0.38,0.5\}$ depicted in blue, orange, green, red and purple, respectively. Solid lines correspond to $\gamma=0$ (Einstein gravity) while the dashed lines correspond to $\gamma=10^{-3}$ ($\alpha'$-corrections), in both cases for second-order hydrodynamics. Right: Plots for $\tilde{\mathcal{S}}^{(1)}(0)$ for first- and second-order hydrodynamics, depicted in blue and orange, respectively. Solid lines correspond to $\gamma=0$ while dashed lines correspond to $\gamma=10^{-3}$.
The dashed vertical lines correspond to the critical times at each order in hydrodynamics, including the leading $\alpha'$-corrections.  \label{Longap}}
\end{center}
\end{figure}
  \item \emph{$\lgb$-corrections.}
  The functions $\tilde{\mathfrak{a}}_4(\tilde{u})$ and $\tilde{\mathfrak{b}}_4(\tilde{u})$ are
  known non-perturvatively in $\lgb$ and up to second order in hydrodynamics, and are given in
  (\ref{2ndgbcoeffs}). With these functions at hand we can extract the numbers
  $\tilde{\mathfrak{a}}_4^{(k)}$ and $\tilde{\mathfrak{b}}_4^{(k)}$ and then use the formula
  (\ref{S1long2}). For small and negative values of $\lgb$ we observe qualitatively the same
  behavior as for the $\gamma-$corrections: the critical time below which first- and second-order hydrodynamics break down increases, which is the expected behavior for a theory that flows from strong to weak coupling. On the other hand, positive values of $\lgb$ behave in the opposite way, and thus appear unphysical for $\lgb$ interpreted as a coupling constant. From the $\Delta y\to0$ limit of the correlator (\ref{s1deltay0}) we obtain the following critical times:
\begin{align}
\tau_\text{crit}^{1\text{st}}(\lgb)&=\left(2.828-16.971\lgb+\cO(\lgb^2)\right)w^{-3/2}\,,\\
\tau_\text{crit}^{2\text{nd}}(\lgb)&=\left(1.987-14.876 \lgb+\cO(\lgb^2)\right)w^{-3/2}\,.
\end{align}
  \begin{figure}[t!]
\begin{center}
\includegraphics[angle=0,width=0.5\textwidth]{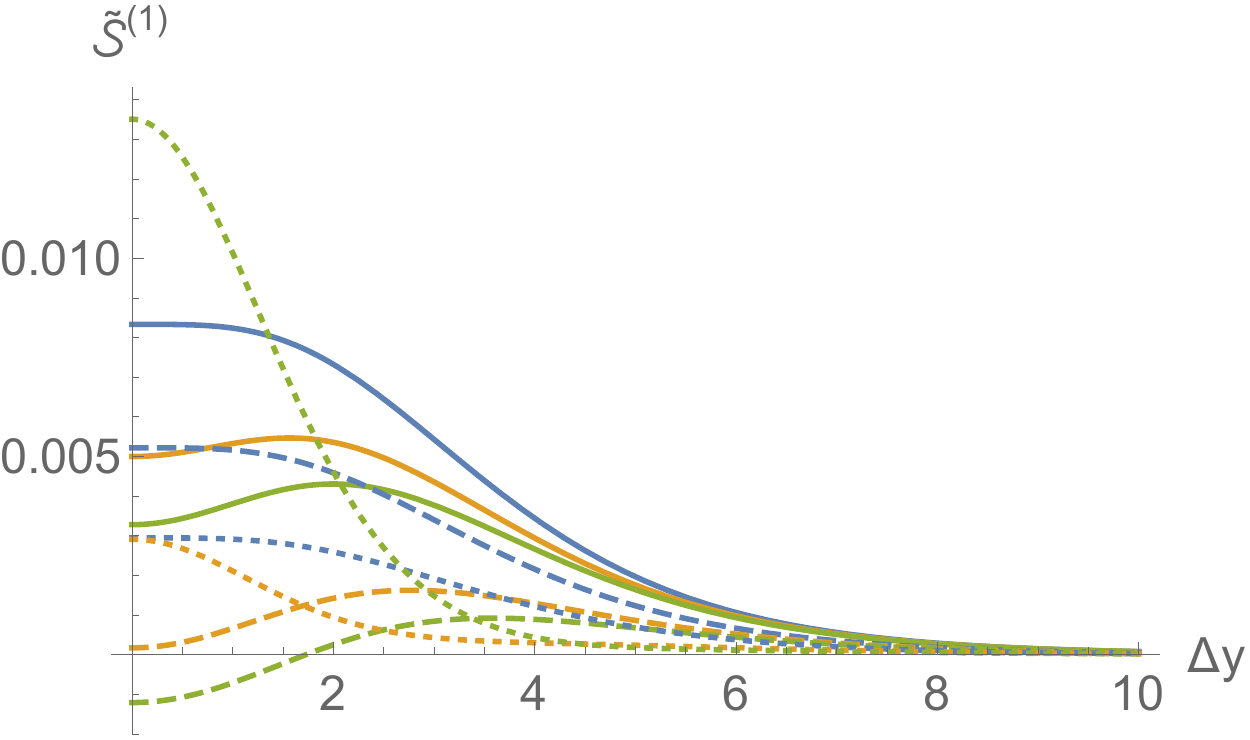}\includegraphics[angle=0,width=0.47\textwidth]{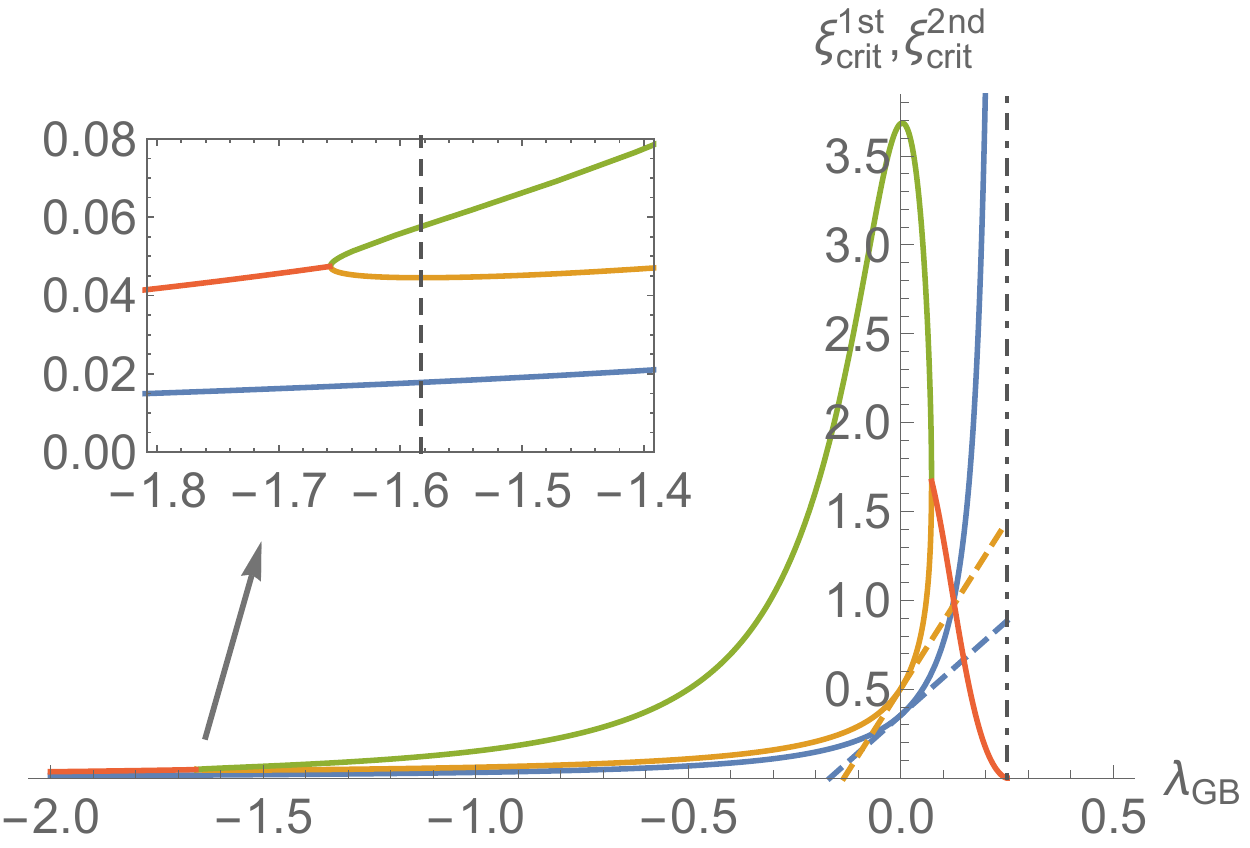}
\caption{ Left: Plots for $\tilde{\mathcal{S}}^{(1)}\equiv\mathcal{S}^{(1)}\tau_0^{4/3}/w^4 \Delta
  x_3^4$ for some representative values of $\xi=\tau_0^{-1} w^{-3/2}=\{0,0.1,0.2\}$ depicted in
  blue, orange and green, respectively. Solid lines correspond to $\lgb=0$ (infinite coupling
  result) while the dashed and dotted lines correspond to $\lgb=-0.5$ and $\lgb=-2$, respectively,
  all cases for second-order hydrodynamics. Right: Plot of $\xi_{\text{crit}}^{1\text{st}}$ (blue) and the
  two branches of $\xi_{\text{crit}}^{2\text{nd}}$ (orange and green) as a function of $\lgb$. In the ranges
  of $\lgb\in(-\infty,-1.657)$ and $\lgb\in(0.073,1/4]$, the correlator is positive but non-monotonic as a function of $\xi$.
  Here, $\xi_{\text{crit}}^{2\text{nd}}$ is found instead by requiring a monotonic decay at late times and
  is depicted in red. The dashed blue and orange lines correspond to the perturbative results to leading order in $\lgb$. The
  vertical line indicates the maximum allowed value for $\lgb=1/4$. The behavior observed for
  negative values of $\lgb$ in the range $\lgb\in(-1.583,0)$ is what is expected for a theory that flows from strong to weak coupling, i.e. $\xi_{\text{crit}}^{2\text{nd}}$ decreases as the coupling decreases. However, $\xi_{\text{crit}}^{2\text{nd}}$ increases in the range $\lgb\in(-1.657,-1.583)$. The small square on top of the figure is a zoomed-in version of the same around this region. The dashed vertical line there signals the value of $\lgb=-1.583$ for which $d\xi_{\text{crit}}^{2\text{nd}}/d\lgb=0$. The discontinuous jump in the derivative of $\xi_{\text{crit}}^{2\text{nd}}$ at $\lgb=-1.657$ is likely to be an artifact of a truncated hydrodynamic gradient expansion or a truncated gravitational derivative expansion.
\label{Longlgb}}
\end{center}
\end{figure}
It is interesting to consider the behavior of the correlator for negative values of $\lgb$ in the non-perturbative regime. Figure \ref{Longlgb} (left) shows $\tilde{\mathcal{S}}^{(1)}\equiv\mathcal{S}^{(1)}\tau_0^{4/3}/w^4 \Delta x_3^4$ plotted as a function of $\Delta y$ for a few representative values of $\xi=\tau_0^{-1} w^{-3/2}=\{0,0.1,0.2\}$, depicted in blue, orange and green, respectively. The solid lines correspond to $\lgb=0$ (infinite coupling limit) while the dashed and dotted lines correspond to $\lgb=-0.5$ and $\lgb=-2$, respectively, all for second-order hydrodynamics. For all the $\xi$ that were plotted the solid lines are well behaved because we have chosen $\xi<\xi_\text{crit}^{2\text{nd}}(\lgb=0)=0.503$. For $\xi=0.2$ the dashed curve becomes negative for small $\Delta y$, indicating that second-order hydrodynamics becomes invalid faster for $\lgb=-0.5$. As mentioned earlier, this is what is indeed expected as the theory flows to weak coupling. However, the dotted curves are always positive in this range of $\xi$, which means that something qualitatively different is happening for sufficiently negative values of $\lgb$. In Figure \ref{Longlgb} (right) we investigate this behavior in more detail. In this plot we show the behavior of $\xi_{\text{crit}}^{1\text{st}}$ and $\xi_{\text{crit}}^{2\text{nd}}$ as a function of $\lgb$. The blue curve corresponds to $\xi_{\text{crit}}^{1\text{st}}$ and has precisely the expected behavior: it decreases monotonically as we decrease the value of $\lgb$. However, we observe something different for $\xi_{\text{crit}}^{2\text{nd}}$: it has two branches for each value of $\lgb$, depicted in orange and green, respectively, which merge at two values of the coupling, $\lgb=-1.657$ and $\lgb=0.073$. For values of the coupling within the ranges $\lgb\in(-\infty,-1.657]$ and $\lgb\in[0.073,1/4]$ the correlator is always positive, however, non-monotonic with respect to $\xi$. In these ranges of $\lgb$ we can find $\xi_{\text{crit}}^{2\text{nd}}$ by requiring monotonicity of the late-time correlator. The result of applying the latter criterion is depicted in red in Figure \ref{Longlgb} (right). Combining these two criteria, we find that $\xi_{\text{crit}}^{2\text{nd}}$ decreases monotonically as $\lgb$ varies from 0 to $-1.583$, but then increases again as $\lgb$ goes from $-1.583$ to $-1.657$. Moreover, the derivative of $\xi_{\text{crit}}^{2\text{nd}}$ is discontinuous at $\lgb=-1.657$. Such behavior does not match the expectations for a theory that flows from infinite to zero coupling. It is likely that the inclusion of higher-than-second-derivative terms in the gravity action (beyond $R^2$ Gauss-Bonnet terms) or a higher-order hydrodynamic expansion would cure these problems. As a result, we conclude that the qualitative resemblance between non-perturbative $\lgb$-corrections and (non-perturbative) finite coupling corrections to the longitudinal two-point correlator, to second order in the hydrodynamic gradient expansion, is restricted to the range of $\lgb\in(-1.583,0]$.
\end{itemize}

The critical times found for the longitudinal correlator can also be expressed generically in terms of a few theory-specific constants $\{\hat{\eta},\,\hat{\eta}^{(\gamma)}_\epsilon,\,\hat{\eta}^{(\lgb)}_\epsilon,\,\hat{\Sigma},\,\hat{\Sigma}^{(\gamma)}_\epsilon,\,\hat{\Sigma}^{(\lgb)}_\epsilon,\,\hat{\Lambda}\}$, defined in Appendix \ref{crit-analytic}, and take the form:
\begin{eqnarray}
\tau^{\text{1st}}_{\text{crit}} &=& \left(\frac{6 \hat{\eta}}{w}\right)^{3/2}\,,\\
\tau^{\text{2nd}}_{\text{crit}} &=& \frac{1}{3}\left(\frac{5\hat{\Sigma}}{w}\right)^{3/2}
\left[
12\hat{\eta}^{3} +\left(\frac{5\hat{\Sigma}}{9}-4\hat{\eta}^2\right)\sqrt{9\hat{\eta}^2-5\hat{\Sigma}}-5\hat{\eta}\hat{\Sigma}
\right]^{-1/2}\,.
\end{eqnarray}
Expressing our coupling constants $\gamma$ and $\lgb$ collectively as $\beta$, first order corrections to $\tau^{1\text{st}}_{\text{crit}}$ and $\tau^{2\text{nd}}_{\text{crit}}$ take the form:
\begin{eqnarray}
\tau^{\text{1st}}_{\text{crit}}(\beta) &=& \tau^{\text{1st}}_{\text{crit}}\left[1+\frac{3\beta}{2}\hat{\eta}^{(\beta)}_{\epsilon}+\mathcal{O}(\beta^2)\right],\\
\tau^{\text{2nd}}_{\text{crit}}(\beta) &=& \tau^{\text{2nd}}_{\text{crit}}\left[1+\frac{3\beta}{4}\left(\hat{\Sigma}^{(\beta)}_{\epsilon}+\frac{3\hat{\eta}}{\sqrt{9\hat{\eta}^{2}-5\hat{\Sigma}}}\left(2\hat{\eta}^{(\beta)}_{\epsilon}-\hat{\Sigma}^{(\beta)}_{\epsilon}\right)\right)+\mathcal{O}(\beta^2)\right].
\end{eqnarray}
The expression for $\tau^{\text{3rd}}_{\text{crit}}$ now corresponds to the smallest real root of the equation
\begin{equation}
1-6\hat{\eta}~\xi^{2/3}+5\hat{\Sigma}~\xi^{4/3}+7\hat{\Lambda}\xi^2 = 0 \,,
\end{equation}
where $\xi=\tau_0^{-1}w^{-3/2}$.
%
%
\subsection{Wilson loops}\label{Sec:Wilson}

Wilson loops are another phenomenologically relevant non-local observable that can be studied within the framework explored in this work. The Wilson loop operator is a path-ordered integral of the gauge field, defined as
\begin{equation}
  W(\mathcal{C})=\frac{1}{N_c}\mathrm{tr}\left(\mathcal{P}e^{i\oint_\mathcal{C} A}\right)\,,
\end{equation}
where the trace runs over the fundamental representation and $\mathcal{C}$ is a closed loop in
spacetime. In AdS/CFT, the recipe for computing the expectation value of a Wilson loop, in the
strong-coupling limit, is given by \cite{Maldacena:1998im}
\begin{equation}
  \langle W(\mathcal{C}) \rangle = e^{-\mathcal{S}_{\text{NG}}(\Sigma)}\,,
\end{equation}
where $\mathcal{S}_{\text{NG}}=(2\pi\alpha')^{-1}\times\mathrm{Area}(\Sigma)$ is the Nambu-Goto
action and $\Sigma$ is an extremal surface with boundary condition $\partial \Sigma=\mathcal{C}$.

Here, we consider two separate cases. The first case consists of a rectangular loop in the
plane transverse to the boost-invariant direction of the Bjorken flow, where $x_1\in[-\frac{\Delta x}{2},\frac{\Delta x}{2}]$, $x_2\in[-\frac{\ell}{2},\frac{\ell}{2}]$ and $\ell\to\infty$. In the
second case, we consider a rectangular loop with two sides extended along the longitudinal (beam) direction, $y\in[-\frac{\Delta y}{2},\frac{\Delta y}{2}]$, $x_1\in[-\frac{\ell}{2},\frac{\ell}{2}]$ and $\ell\to\infty$.

The calculation of the Wilson loop is qualitatively similar to that of the two-point function, so we will omit some of the redundant details below.
\subsubsection{Transverse Wilson loop}
The Nambu-Goto action for the transverse Wilson loop in the Fefferman-Graham chart is
\begin{equation}
  \mathcal{S}_{\text{NG}}=\frac{\ell}{\pi\alpha'}\int_0^{z_*}\frac{dz}{z^2}\sqrt{e^{\tilde{c}}(1+e^{\tilde{c}} x'^2-e^{\tilde{a}}\tau'^2)} \,.
\end{equation}
Using Eq. (\ref{FGnearB}), we can expand this expression as
$\mathcal{S}_{\rm{NG}} = \mathcal{S}_{\rm{NG}}^{(0)} + \mathcal{S}_{\rm{NG}}^{(1)} + \dots$, where
\begin{align}
\mathcal{S}_{\text{NG}}^{(0)}&=\frac{\ell\sqrt{\lambda}}{\pi}\int_{0}^{z_*} \frac{dz}{z^2}
\sqrt{1+x'^2-\tau'^2}\,,\\
\mathcal{S}_{\text{NG}}^{(1)}&=\frac{w^4\ell\sqrt{\lambda}}{2\pi}\int_{0}^{z_*} dz
\frac{z^2(\tilde{\mathfrak{c}}_4(1+2
  x'^2-\tau'^2)-\tilde{\mathfrak{a}}_4\tau'^2)}{\tau^{4/3}\sqrt{1+x'^2-\tau'^2}} \,.
\end{align}
and we used $\alpha' = \lambda^{-1/2}$. The first term is the pure AdS contribution, which we can see by using the zeroth-order embeddings:
\begin{align}\label{embeddingW0}
\tau(z)=\tau_0\,, && \,x(z)=\frac{\sqrt{2}\pi^{3/2} z_*}{\Gamma [1/4]^2}-\frac{z^3 \, _2F_1\left(\frac{1}{2},\frac{3}{4};\frac{7}{4};\frac{z^4}{z_*^4}\right)}{3 z_*^2} \, ,
\end{align}
with
$z_*=\Delta x \, \Gamma [1/4]^2/(2\pi)^{3/2} $. Integrating from $\epsilon\to0$ to $z_*$, and subtracting the divergent part, $\mathcal{S}_{\rm div}=\ell\sqrt{\lambda}/\pi\epsilon$, we obtain
\begin{equation}
  \mathcal{S}_{\text{NG}_{\rm{reg}}}^{(0)}=-\frac{4 \pi^2\ell\sqrt{\lambda} }{\Delta x\, \Gamma [1/4]^4}\,,
\end{equation}
which gives the vacuum expectation value of the Wilson loop,
\begin{align}
\langle W^{(0)}\rangle=\exp \left\{ \frac{4 \pi^2\ell \sqrt{\lambda}  }{\Delta x\, \Gamma [1/4]^4}\right\} \,.
\end{align}
At next order, after using the zeroth-order embeddings and defining a dimensionless variable $x=z/z_*$, we find
\begin{align}
\mathcal{S}_{\text{NG}}^{(1)} &=\frac{w^4\ell \sqrt{\lambda} \Delta x^3 \,\Gamma [1/4]^6 } {32\pi^{11/2}\sqrt{2}}\frac{\tilde{\mathfrak{c}}_4(\tilde{u}_0)}{\tau_0^{4/3}} \int_0^1 dx\,\frac{x^2(1+x^4)}{\sqrt{1-x^4}} \nn
&=\frac{w^4 \ell \sqrt{\lambda}\,\Delta x^3\, \Gamma [1/4]^4\,}{20 \pi^4}\frac{ \tilde{\mathfrak{c}}_4(\tilde{u}_0)}{\tau_0^{4/3}} \,,
\end{align}
where $\tilde{u}_0=\tau_0^{-2/3}w^{-1}$. We observe that
$\mathcal{S}_{\text{NG}}^{(1)}$ depends linearly on $\tilde{\mathfrak{c}}_4(\tilde{u})$, similarly
to $\mathcal{S}^{(1)}$ for the transverse two-point function. Therefore the resulting values of
$\tau_{\text{crit}}^{2\text{nd}}$ and $\tau_{\text{crit}}^{3\text{rd}}$ will be the same as those obtained in that case, for
both Einstein gravity and the higher derivative gravities with $\alpha'$ and $\lambda_{GB}$
corrections. As a result, the transverse Wilson loop provides no new bounds on the validity of the hydrodynamic description.

\subsubsection{Longitudinal Wilson loop}
The Nambu-Goto action for the longitudinal Wilson loop is

\begin{equation}
  \mathcal{S}_{\text{NG}}=\frac{\ell\sqrt{\lambda}}{\pi} \int_0^{z_*}\frac{dz}{z^2}\sqrt{e^{\tilde{c}}(1+e^{\tilde{b}}\tau^2 y'^2-e^{\tilde{a}}\tau'^2)} \,,
\end{equation}
which gives via (\ref{FGnearB})
\begin{eqnarray}
  \mathcal{S}_{\text{NG}}^{(0)}&=&\frac{\ell\sqrt{\lambda}}{\pi}\int_{0}^{z_*} \frac{dz}{z^2}
                                   \sqrt{1+\tau^2 y'^2-\tau'^2}\,,\\
  \mathcal{S}_{\text{NG}}^{(1)}&=&\frac{w^4\ell\sqrt{\lambda}}{2\pi}\int_{0}^{z_*} dz \frac{z^2(\tilde{\mathfrak{c}}_4(1+\tau^2 y'^2-\tau'^2)+\tilde{\mathfrak{b}}_4 \tau^2 y'^2
                                   -\tilde{\mathfrak{a}}_4\tau'^2)}{\tau^{4/3}\sqrt{1+\tau^2 y'^2-\tau'^2}}\,.
\end{eqnarray}
Again, the first expression gives the pure AdS embedding when we use the zeroth-order embeddings:
\begin{align}\label{embeddingW1}
\tau(z)=\sqrt{t_0^2 - x(z)^2}\,, &&  y(z)=\mathrm{arccosh}\left(\frac{t_0}{\tau(z)}\right) \,,
\end{align}
with
\begin{align}\label{xOfz}
x(z)=\frac{\sqrt{2}\pi^{3/2} z_*}{\Gamma [1/4]^2}-\frac{z^3 \, _2F_1\left(\frac{1}{2},\frac{3}{4};\frac{7}{4};\frac{z^4}{z_*^4}\right)}{3 z_*^2}\,,&& z_*=\frac{\Gamma [1/4]^2}{(2\pi)^{3/2}}\Delta x \,.
\end{align}
Integrating from $\epsilon\to0$ to $z_*$, and subtracting the divergent part
$\mathcal{S}_{\rm div}=\ell\sqrt{\lambda}/\pi\epsilon$, we find
\begin{equation}
  \mathcal{S}_{\text{NG}_{\rm{reg}}}^{(0)}=-\frac{4 \pi^2\ell \sqrt{\lambda} }{\Delta x\, \Gamma [1/4]^4}\,,
\end{equation}
i.e. the same result as in the transverse case.

The next step is to evaluate $\mathcal{S}_{\text{NG}}^{(1)}$ using the zeroth-order embeddings
(\ref{embeddingW1})--(\ref{xOfz}) along with the explicit forms of
$\{\tilde{\mathfrak{a}}_4(\tilde{u}),\tilde{\mathfrak{b}}_4(\tilde{u}),\tilde{\mathfrak{c}}_4(\tilde{u})\}$. Defining the dimensionless variable $x=z/z_*$ and expanding $\{\tilde{\mathfrak{a}}_4(\tilde{u}),\tilde{\mathfrak{b}}_4(\tilde{u}),\tilde{\mathfrak{c}}_4(\tilde{u})\}$
as in (\ref{generic-hydro-func-expansion}), we find
\begin{align}
  \label{Long1stActionOrders}
  \mathcal{S}_{\text{NG}}^{(1)}=& \,\frac{w^4 \ell\sqrt{\lambda}   \Delta x_3^3  \,\Gamma [1/4]^6 }{32 \sqrt{2} \pi^{11/2} \tau_0^{4/3}}\sum_{k=0}^{\infty}\tau_0^{-2k/3}w^{-k}\nonumber\\
                               & \times \bigg[\left(\tilde{\mathfrak{c}}_4^{(k)}\mathcal{I}^{(k)}_1+\tilde{\mathfrak{b}}_4^{(k)}\mathcal{I}^{(k)}_2\right)\cosh ^2\left(\tfrac{\Delta y}{2}\right)- \left(\tilde{\mathfrak{c}}_4^{(k)}\mathcal{I}^{(k)}_{1\mathfrak{h}}+\tilde{\mathfrak{a}}_4^{(k)}\mathcal{I}^{(k)}_{2\mathfrak{h}}\right)\sinh^2\left(\tfrac{\Delta y}{2}\right)\bigg] \,,
\end{align}
where
\begin{eqnarray}
  \mathcal{I}^{(k)}_1&=&\int_0^1 dx~\mathcal{F}^{(k)}_1=\int_0^1 dx\frac{x^2}{\sqrt{1-x^4}\left(\cosh ^2\left(\frac{\Delta y}{2}\right)- \mathfrak{h}(x){}^2\sinh ^2\left(\frac{\Delta y}{2}\right)\right)^{(5+k)/3}}
  \,,\\
  \mathcal{I}^{(k)}_2&=&\int_0^1 dx~\mathcal{F}^{(k)}_2=\int_0^1 dx \frac{x^6}{\sqrt{1-x^4}\left(\cosh ^2\left(\frac{\Delta y}{2}\right)- \mathfrak{h}(x){}^2\sinh ^2\left(\frac{\Delta y}{2}\right)\right)^{(5+k)/3}}\,,
\end{eqnarray}
\begin{equation}
\mathcal{I}^{(k)}_{j\mathfrak{h}} = \int_0^1\mathfrak{h}(x)^2\mathcal{F}^{(k)}_{j} \,,
\end{equation}
and
\begin{equation}
  \mathfrak{h}(x) \equiv \frac{x^3 \,\Gamma [1/4]^2}{3 \sqrt{2}\pi^{3/2}} \, _2F_1\left(\tfrac{1}{2},\tfrac{3}{4};\tfrac{7}{4};x^4\right)-1\,.
\end{equation}

We can extract $\tau_{\text{crit}}$ at different orders in the hydrodynamic expansion by studying the $\Delta y\to 0$ behavior of $\mathcal{S}_{\text{NG}}^{(1)}$. In this limit, the $\sinh^2(\Delta y/2)$ term of (\ref{Long1stActionOrders}) vanishes and the relevant $\mathcal{I}^{(k)}$ integrals are finite and independent of $k$:
\begin{align}\label{LongWLSmallyI}
  \mathcal{I}^{(k)}_1 &\to \int_0^1 dx\frac{x^2}{\sqrt{1-x^4}}=\frac{\sqrt{2}\pi^{3/2}}{\Gamma [1/4]^2}\,, \\
  \mathcal{I}^{(k)}_2 &\to \int_0^1 dx\frac{x^6}{\sqrt{1-x^4}}=\frac{3\sqrt{2}\pi^{3/2}}{5\Gamma [1/4]^2}\,,
\end{align}
Collecting our results, we find that for $\Delta y\to 0$:
\begin{align}\label{s1deltay0WL}
\mathcal{S}_{\text{NG}}^{(1)} &\to \frac{w^4 \ell\sqrt{\lambda}  \Delta x_3^3\, \Gamma [1/4]^4}{32 \pi^4 \tau_0^{4/3}}\sum_{k=0}^{\infty}\left(\tilde{\mathfrak{c}}_4^{(k)}+\frac{3}{5}\tilde{\mathfrak{b}}_4^{(k)}\right)\tilde{u}_0^k \nn
&=\frac{w^4 \ell\sqrt{\lambda} \Delta x_3^3 \,\Gamma [1/4]^4 }{ 32 \pi^4 \tau_0^{4/3}} \left(\tilde{\mathfrak{c}}_4(\tilde{u}_0)+\frac{3}{5}\tilde{\mathfrak{b}}_4(\tilde{u}_0)\right)\,.
\end{align}
where $\tilde{u}_0 = \tau_0^{-2/3}w^{-1}$.

Unlike the longitudinal correlator, positivity of $\mathcal{S}_{\text{NG}}^{(1)}(0)$ itself does not provide a useful criterion for establishing the regime of validity of the hydrodynamic description at all orders in the hydrodynamic expansion, so we have to also impose monotonicity. The positivity criterion is enough only at first order, however $\mathcal{S}_{\text{NG}}^{(1)}(0)$ is strictly positive at second and third order in the backgrounds we consider. In these cases, we find that $\mathcal{S}_{\text{NG}}^{(1)}(0)$ decreases with decreasing $\tau$ until it reaches some minimum value, $\mathcal{S}_{\text{NG,min}}^{(1)}(0,\tau=\tau_{\text{min}})$, and then turns around and grows without bound (this behavior is demonstrated in Figure \ref{LongWLEin} for Einstein gravity). Therefore, for $\tau<\tau_{\text{min}}$, the longitudinal Wilson loops are unphysical. This will be our criterion for establishing $\tau_{\text{crit}}$ for the higher order hydrodynamic descriptions.

In the following, we will study the full behavior of $\mathcal{S}_{\text{NG}}^{(1)}$ as a function of $\Delta y$ and $\xi=\tau_0^{-1}w^{-3/2}$ for our three cases of interest. In each case, the bounds on the validity of the hydrodynamic description are less constraining than those coming from the longitudinal correlator.
\begin{itemize}
\item \emph{Einstein gravity.}  Using the expansions of $\tilde{\mathfrak{a}}_4(\tilde{u})$,
  $\tilde{\mathfrak{b}}_4(\tilde{u})$ and $\tilde{\mathfrak{c}}_4(\tilde{u})$ up to third order in
  hydrodynamics in (\ref{3rdEGcoeffs}), we evaluate $\mathcal{S}_{\text{NG}}^{(1)}$ via
  (\ref{Long1stActionOrders}), and plot the results for some representative values of $\xi$ in
  Figure \ref{LongWLEin}. From the $\Delta y\to0$ limit of $\mathcal{S}_{\text{NG}}^{(1)}$, we find:
\begin{align}
\tau_\text{crit}^{1\text{st}}= 0.650 w^{-3/2}\,, && \tau_\text{crit}^{2\text{nd}}= 0.294 w^{-3/2}\,, && \tau_\text{crit}^{3\text{rd}}= 0.669 w^{-3/2}\,.
\end{align}

\begin{figure}[t!]
\begin{center}
\includegraphics[angle=0,width=0.48\textwidth]{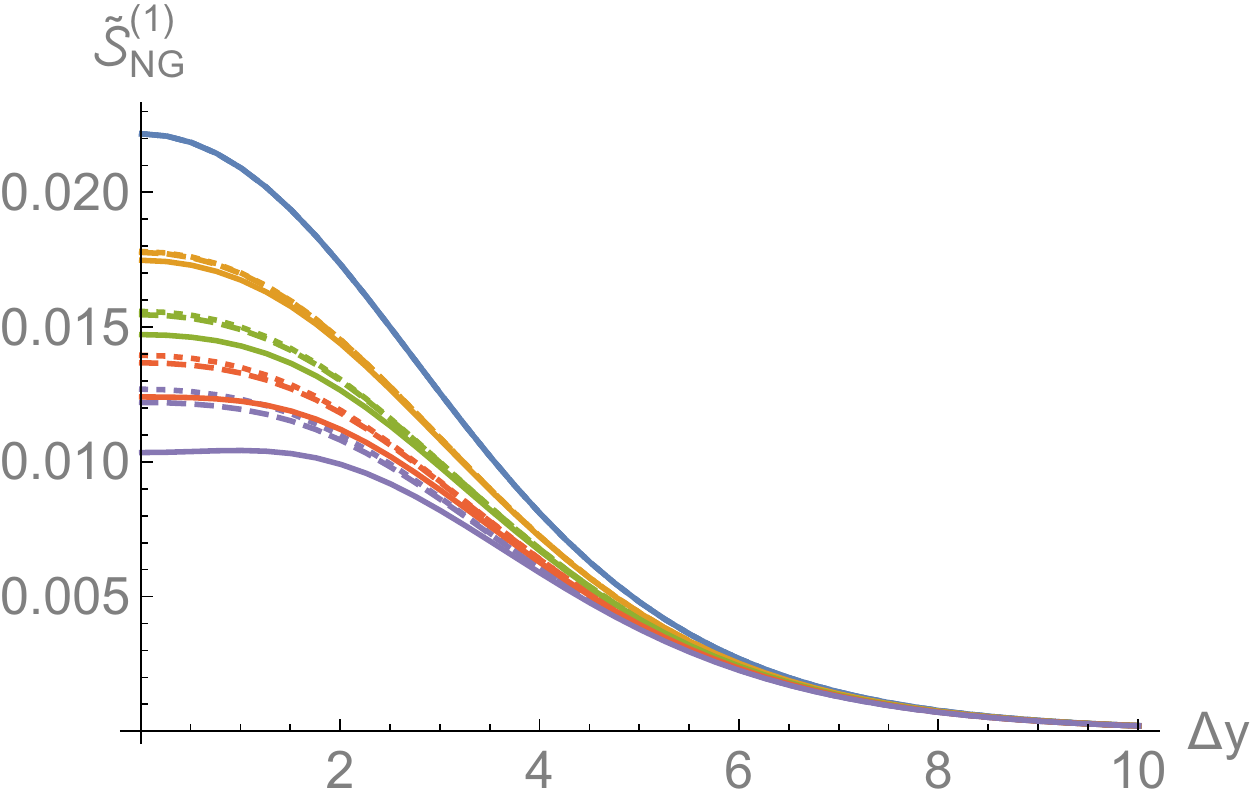}\includegraphics[angle=0,width=0.48\textwidth]{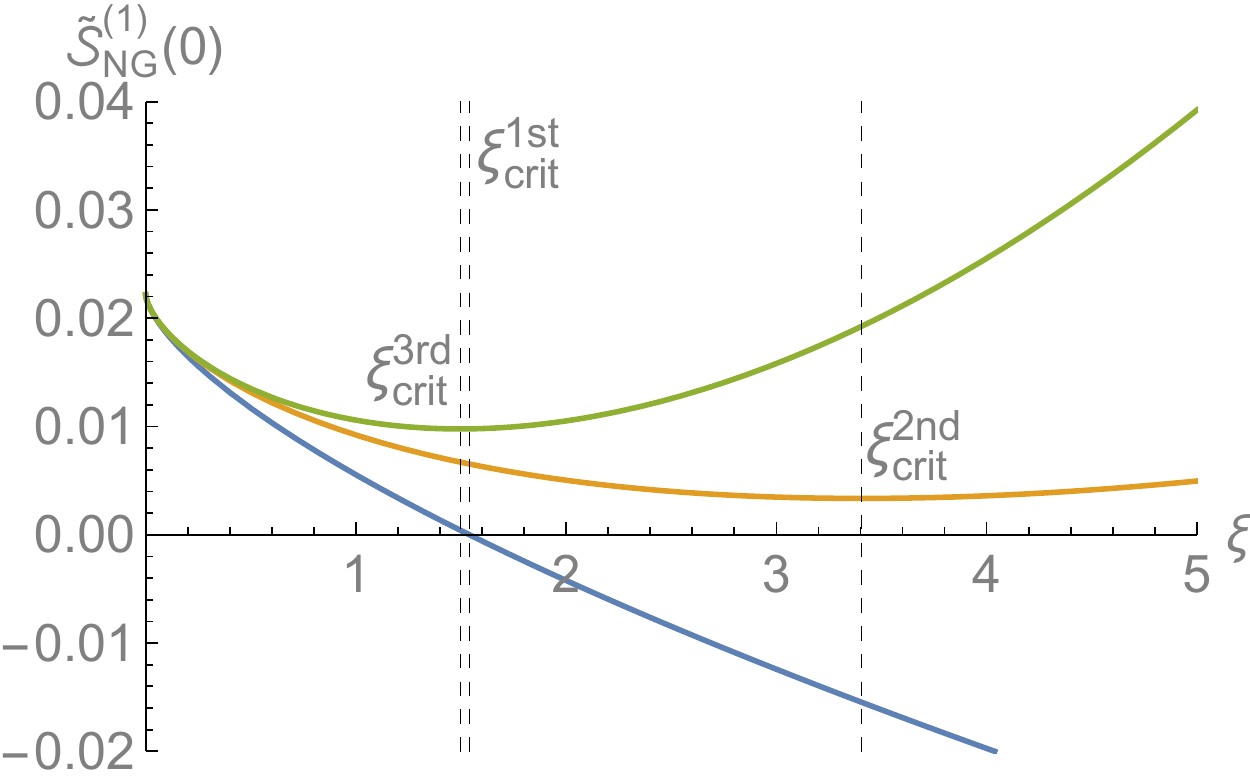}
      \caption{Left: Plots for
        $\tilde{\mathcal{S}}_{\text{NG}}^{(1)}\equiv\mathcal{S}_{\text{NG}}^{(1)}\tau_0^{4/3}/w^4
        \ell \sqrt{\lambda} \Delta x_3^3$ for various values of
        $\xi=\tau_0^{-1} w^{-3/2}=\{0,0.15,0.3,0.45,0.6\}$ depicted in blue, orange, green, red and
        purple, respectively.  The solid lines correspond to $3^{\text{rd}}$ order hydrodynamics; the
        dashed and dotted lines correspond to $2^{\text{nd}}$ and $1^{\text{st}}$ order hydrodynamics,
        respectively. Right: Plots for $\tilde{\mathcal{S}}_{\text{NG}}^{(1)}(0)$ for $1^{\text{st}}$, $2^{\text{nd}}$ and $3^{\text{rd}}$ order hydrodynamics, depicted in green, orange and blue, respectively. The dashed vertical lines correspond to the critical times at each order in hydrodynamics.
         \label{LongWLEin}}
\end{center}
\end{figure}

\item \emph{$\alpha'$-corrections.} Using the expansions of $\tilde{\mathfrak{a}}_4(\tilde{u})$,
  $\tilde{\mathfrak{b}}_4(\tilde{u})$ and $\tilde{\mathfrak{c}}_4(\tilde{u})$ up to second order in
  hydrodynamics in (\ref{2ndapcoeffs}) we evaluate $\mathcal{S}_{\text{NG}}^{(1)}$ via
  (\ref{Long1stActionOrders}), the results of which are shown in Figure \ref{LongWLString}. The
  solid lines correspond to $\gamma=0$ (Einstein gravity) while the dashed lines correspond to
  $\gamma=10^{-3}$. From the $\Delta y\to0$ limit of $\mathcal{S}_{\text{NG}}^{(1)}$, we find:
\begin{eqnarray}
\tau_\text{crit}^{1\text{st}}(\gamma)&=& \left(0.650+108.876\gamma+\mathcal{O}(\gamma^2)\right)w^{-3/2} \,,\\
\tau_\text{crit}^{2\text{nd}}(\gamma)&=& \left(0.294+72.997\gamma+\mathcal{O}(\gamma^2)\right)w^{-3/2}\,.
\end{eqnarray}
\begin{figure}[t!]
    \begin{center}
      \includegraphics[angle=0,width=0.48\textwidth]{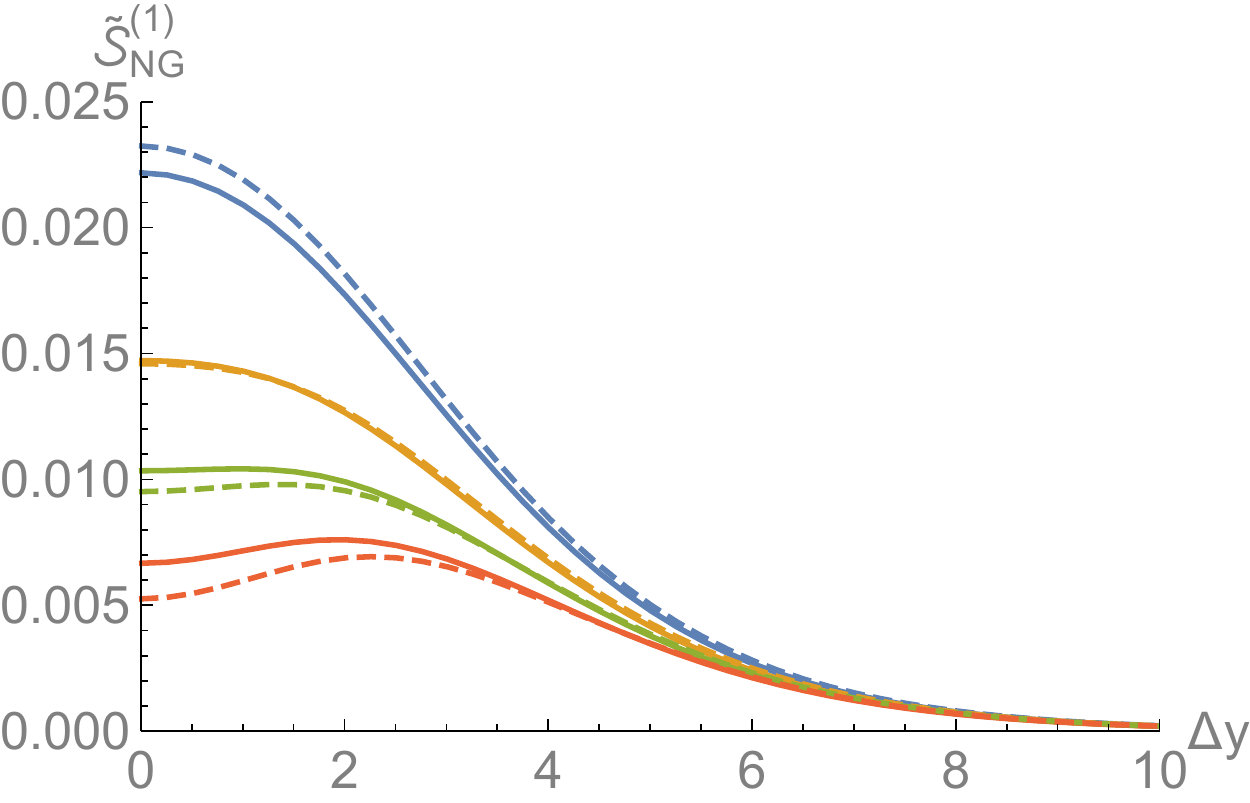}
       \includegraphics[angle=0,width=0.48\textwidth]{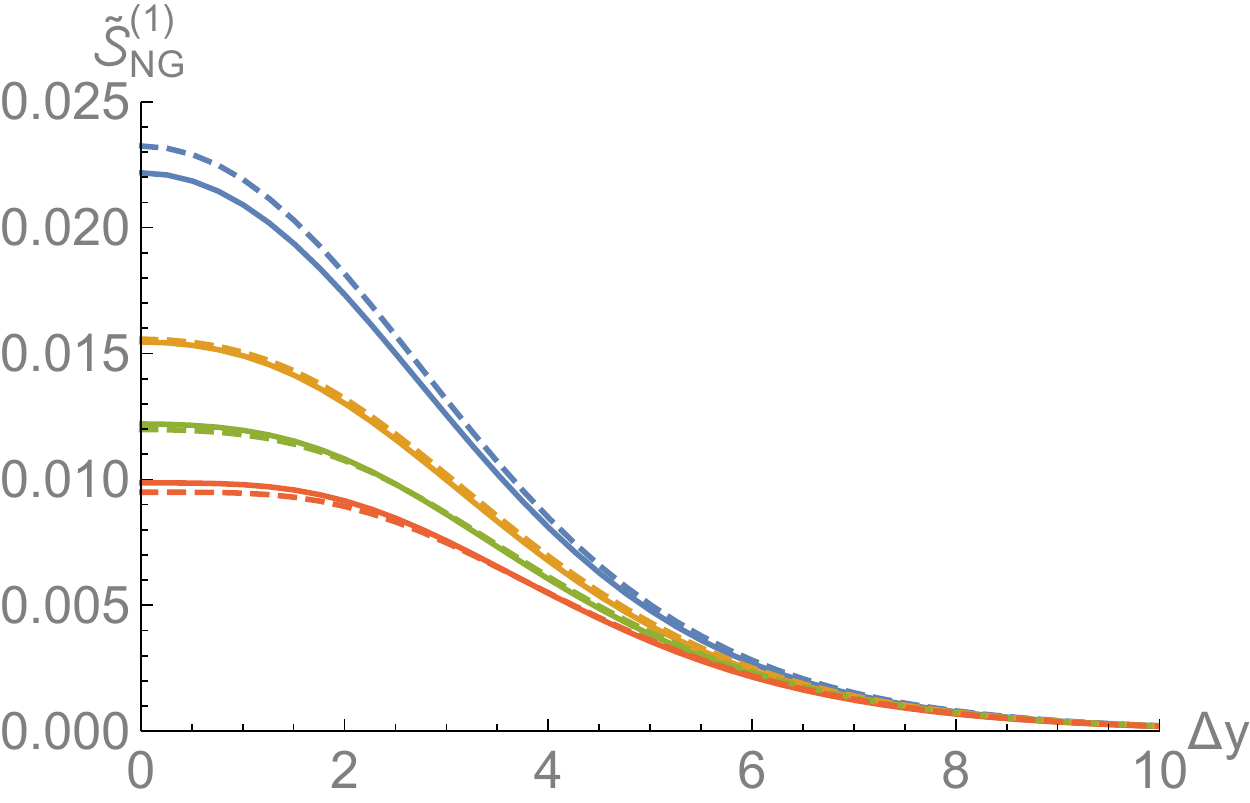}
      \caption{Plots for
        $\tilde{\mathcal{S}}_{\text{NG}}^{(1)}\equiv\mathcal{S}_{\text{NG}}^{(1)}\tau_0^{4/3}/w^4
        \ell \sqrt{\lambda} \Delta x_3^3$ for various values of
        $\xi=\tau_0^{-1} w^{-3/2}=\{0,0.3,0.6,0.9\}$ depicted in blue, orange, green and red,
        respectively. Solid lines correspond to $\gamma=0$ (Einstein gravity) while the
        dashed lines correspond to $\gamma=10^{-3}$ ($\alpha'$-corrections). The plot on the left corresponds to
        first-order hydrodynamics, while the plot on the right corresponds to second-order
        hydrodynamics. \label{LongWLString}}
    \end{center}
\end{figure}
\item \emph{$\lgb$-corrections.} Using the expansions of $\tilde{\mathfrak{a}}_4(\tilde{u})$, $\tilde{\mathfrak{b}}_4(\tilde{u})$ and $\tilde{\mathfrak{c}}_4(\tilde{u})$ up to second-order hydrodynamics in (\ref{2ndgbcoeffs}) we evaluate $\mathcal{S}_{\text{NG}}^{(1)}$ via (\ref{Long1stActionOrders}), and plot the results in Figure \ref{LongWLGB}. The solid lines correspond to $\lgb=0$ (Einstein gravity) while the dashed lines correspond to $\lgb=-0.2$. Following the same line of reasoning as in the previous two cases, we find:
\begin{align}
   \tau_\text{crit}^{1\text{st}}(\lgb)&= \left(0.650-3.897\lgb +\CO(\lgb^2) \right)w^{-3/2} \,,\\
   \tau_\text{crit}^{2\text{nd}}(\lgb)&= \left(0.294-0.554\lgb + \CO(\lgb^2) \right)w^{-3/2}\,.
\end{align}
\begin{figure}[t!]
    \begin{center}
      \includegraphics[angle=0,width=0.52\textwidth]{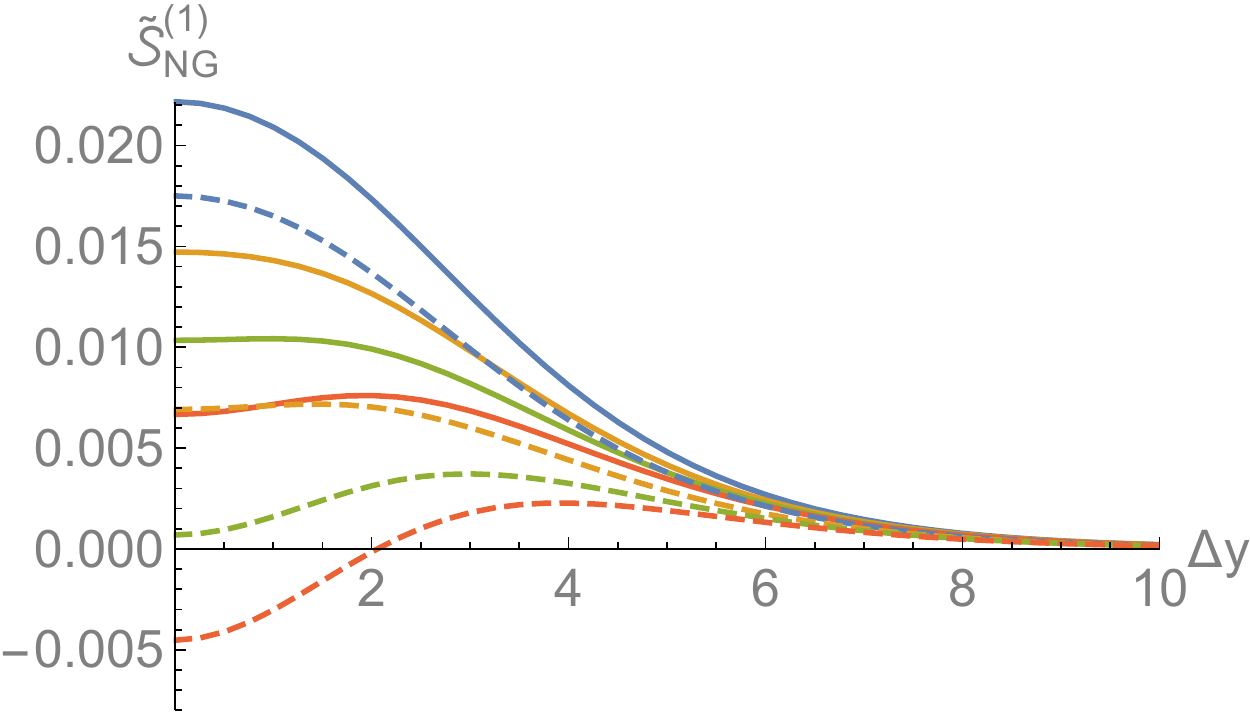}
       \includegraphics[angle=0,width=0.47\textwidth]{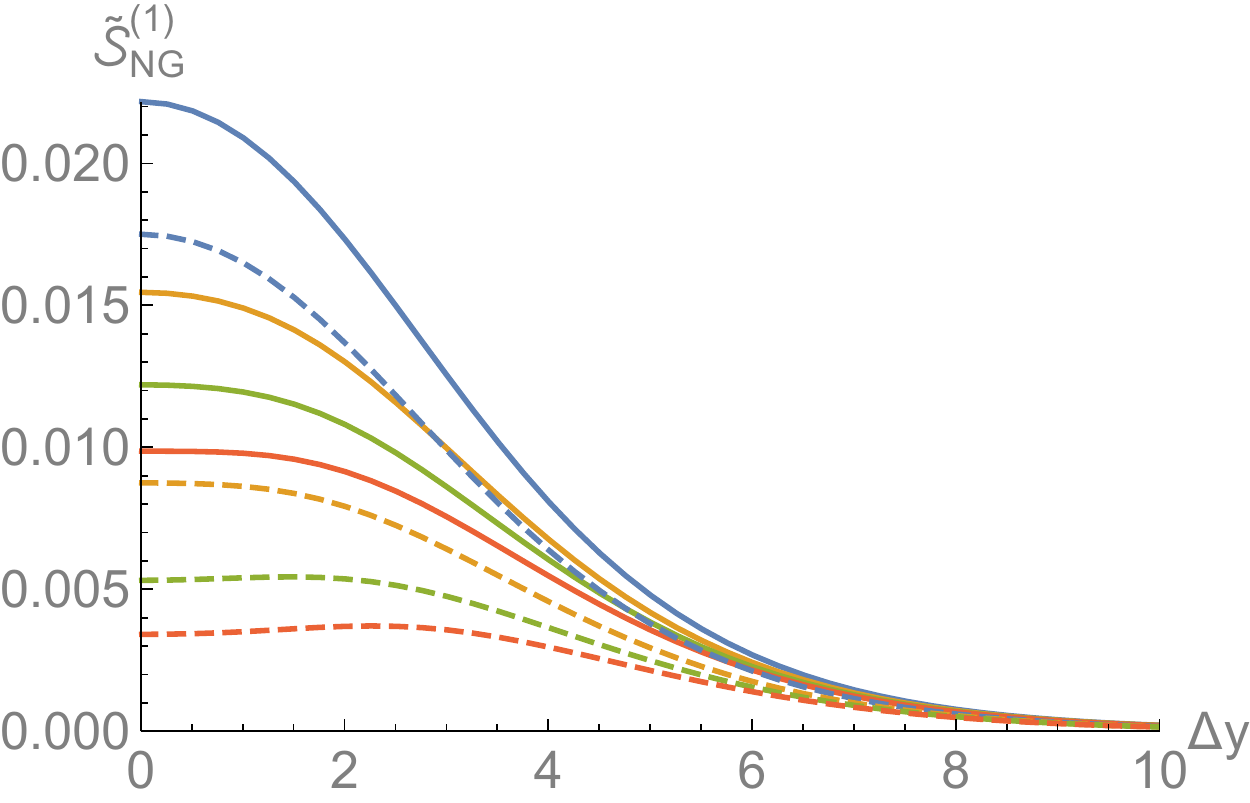}
      \caption{Plots for
        $\tilde{\mathcal{S}}_{\text{NG}}^{(1)}\equiv\mathcal{S}_{\text{NG}}^{(1)}\tau_0^{4/3}/w^4
        \ell \sqrt{\lambda} \Delta x_3^3$ for various values of
        $\xi=\tau_0^{-1} w^{-3/2}=\{0,0.3,0.6,0.9\}$ depicted in blue, orange, green and red,
        respectively. Solid lines correspond to $\lgb=0$ (Einstein gravity) while the
        dashed lines correspond to $\lgb=-0.2$. The plot on the
        left corresponds to first-order hydrodynamics, while the plot on the right corresponds to
        second-order hydrodynamics. \label{LongWLGB}}
    \end{center}
\end{figure}
\end{itemize}

Finally, the critical times found above can be expressed generically in terms of the theory-specific constants defined in Appendix \ref{crit-analytic}, and take the form:
\begin{eqnarray}
\tau^{\text{1st}}_{\text{crit}} &=& \left(\frac{9\hat{\eta}}{4 w}\right)^{3/2}, \qquad\qquad \tau^{\text{2nd}}_{\text{crit}} = \left(\frac{10}{9 w}\cdot\frac{\hat{\Sigma}}{\hat{\eta}}\right)^{3/2}\,.
\end{eqnarray}
Expressing our coupling constants $\gamma$ and $\lgb$ collectively as $\beta$, first order corrections to $\tau^{1\text{st}}_{\text{crit}}$ and $\tau^{2\text{nd}}_{\text{crit}}$ take the form:
\begin{eqnarray}
\tau^{\text{1st}}_{\text{crit}}(\beta) &=& \tau^{\text{1st}}_{\text{crit}}\left[1+\frac{3\beta}{2}\hat{\eta}_{\epsilon}^{(\beta)}+\mathcal{O}(\beta^2)\right],\\
\tau^{\text{2nd}}_{\text{crit}}(\beta) &=& \tau^{\text{2nd}}_{\text{crit}}\left[1+\frac{3\beta}{2}\left(\hat{\Sigma}_{\epsilon}^{(\beta)}-\hat{\eta}_{\epsilon}^{(\beta)} \right)+\mathcal{O}(\beta^2)\right].
\end{eqnarray}
The expressions for $\tau^{\text{3rd}}_{\text{crit}}$ correspond to the smallest real root of the equation
\begin{equation}
\hat{\eta}-\frac{10}{9}\hat{\Sigma}~\xi^{2/3}-\frac{11}{6}\hat{\Lambda}~\xi^{4/3} = 0\,,
\end{equation}
where $\xi = \tau_0^{-1}w^{-3/2}$.
%
%
\section{Discussion}\label{Sec:Discussion}

This work provides a new tile in the mosaic of recent developments on coupling-dependent thermal
physics from the point of view of holography. With a view towards a better understanding of
heavy ion collisions, the goal of this program has been to uncover qualitative and quantitative
features of physical phenomena across a wide range of coupling constants---an understanding of which
will likely require an interpolation between weakly-coupled perturbative field theory and
strongly-coupled holographic techniques.

Non-linear shock wave collisions were recently analyzed in perturbative Gauss-Bonnet theory
to, for the first time, numerically model coupling-dependent heavy ion collisions
\cite{Grozdanov:2016zjj} and, for example, compute the corrected hydrodynamization time. The
extension of those results to either non-perturbative Gauss-Bonnet gravity or to type IIB
supergravity is technically demanding. Therefore, it is useful to also study other, simpler models and probes
of phenomena related to hydrodynamization. In this paper, we studied the gravity backgrounds dual to a boost-invariant Bjorken flow, which are good models for the late time dynamics of heavy ion collisions, at least in the regime of mid-rapidities. We considered non-perturbative Gauss-Bonnet gravity, studied in the present context for the first time, and type IIB supergravity (to leading order in $\alpha'$), both to second order in hydrodynamics. Following up on \cite{Pedraza:2014moa}, we provided an example of an analytically-tractable computation of a critical time defined through relaxation properties of non-local observables (equal-time correlators and Wilson loops), after which hydrodynamics becomes a good description.

Numerical estimates of the critical times obtained for second-order hydrodynamics---computed to
leading order in inverse 't Hooft coupling corrections in $\CN = 4$ theory and non-perturbatively in $\lgb$ in Gauss-Bonnet theory---are summarized in Tables \ref{table:CT1} and \ref{table:CT2}, where we show the increase of the critical time at decreased field theory coupling corresponding to a $10\%$ and an $80\%$ increase of $\eta/s$ compared to its infinitely strongly coupled value of $\eta / s = 1 / 4\pi$. In both theories, the most stringent critical time is set by the longitudinal two-point correlator, $\la \phi\phi \ra_{\parallel}$.

\begin{table}[tbh]
\begin{center}
    \begin{tabular}{| c | c | c | c |}
    \hline
     & $\CN = 4$ to $\CO(\lambda^{-3/2})$ & GB to $\CO(\lgb)$ & non-perturbative GB \\ \hline
     $\la \phi\phi \ra_{\perp}$, $ \langle W(\mathcal{C}) \rangle_{\perp}$ & 17.3\% & 9.9\% & 10.9\% \\ \hline
     $\la \phi\phi \ra_{\parallel}$ & 11.5\%& 18.7\% & 18.5\%  \\ \hline
     $\langle W(\mathcal{C}) \rangle_{\parallel}$  & 20.7\%  & 4.7\% & 6.7\%\\ \hline
    \end{tabular}
\end{center}
\caption{Increase of the critical time in $\CN =4$ SYM theory at $\gamma \approx 8.33 \times 10^{-4}$ ($\lambda \approx 31.9$) and in a dual of Gauss-Bonnet theory at $\lgb = -0.025$. Both choices of the coupling correspond to a $10\%$ increase of $\eta/s$. We use $\perp$ and $\parallel$ subscripts to denote transverse and longitudinal operators, respectively.}
\label{table:CT1}
\end{table}

\begin{table}[tbh]
\begin{center}
    \begin{tabular}{| c | c | c | c |}
    \hline
     & $\CN = 4$ to $\CO(\lambda^{-3/2})$ & GB to $\CO(\lgb)$ & non-perturbative GB \\ \hline
     $\la \phi\phi \ra_{\perp}$, $ \langle W(\mathcal{C}) \rangle_{\perp}$ & 138.8\% &  78.9\%& 136.4\% \\ \hline
     $\la \phi\phi \ra_{\parallel}$ & 92.3\% & 149.7\%& 145.1\%  \\ \hline
     $\langle W(\mathcal{C}) \rangle_{\parallel}$  &  165.7\% & 37.7 \% & 131.4\%\\ \hline
    \end{tabular}
\end{center}
\caption{Increase of the critical time in $\CN =4$ SYM theory at $\gamma \approx 6.67 \times 10^{-3}$ ($\lambda \approx 7.98$) and in a dual of Gauss-Bonnet theory at $\lgb = -0.2$. In this case, the choices of the coupling correspond to an $80\%$ increase of $\eta/s$.}
\label{table:CT2}
\end{table}

Several interesting features can be extracted from our analysis. One is the possibility of direct
comparison between the size of effects of the 't Hooft coupling in $\CN = 4$ SYM and $\lgb$ in the
hypothetical dual of Gauss-Bonnet gravity. Such results should come in handy when using Gauss-Bonnet
theory for phenomenologically relevant studies. The second is the comparison between the sizes of
perturbative and non-perturbative corrections in Gauss-Bonnet theory. As noted before, in both
$\CN = 4$ SYM and Gauss-Bonnet gravity, the strictest bound on the regime of validity of
hydrodynamics comes from the longitudinal two-point correlator. Since all other bounds are weaker,
their non-convergent behavior in terms of the gradient expansion (third-order hydrodynamics giving a
stricter bound than second-order hydrodynamics for $\la \phi\phi \ra_{\perp}$,
$ \langle W(\mathcal{C}) \rangle_{\perp}$ and $ \langle W(\mathcal{C}) \rangle_{\parallel}$) and in
the perturbative $\lgb$ expansion should not be taken seriously: at their respective critical
times, the hydrodynamic description assumed in the derivation is no longer valid. What is important,
however, is that for the critical time derived from the longitudinal $\la \phi\phi \ra_{\parallel}$,
the perturbative $\lgb$ corrections converge remarkably quickly to the non-perturbative results,
even for the increase of $\eta /s $ by $80\%$. While perhaps surprising at first, this observation
is compatible with the results of \cite{Grozdanov:2016zjj}.

Another interesting consequence of our analysis is the emergent restriction on the range of the
(non-perturbative) Gauss-Bonnet coupling for the second-order hydrodynamic approximation to a boost-invariant
flow. While the Gauss-Bonnet theory with negative $\lgb$ very well reproduces the expected behavior
of a thermal CFT with finite coupling
\cite{Grozdanov:2015asa,Grozdanov:2016vgg,Grozdanov:2016fkt,Andrade:2016rln,Grozdanov:2016zjj}, it
is also known that the theory suffers from instabilities and UV problems for large (or finite)
values of $\lgb$. For the non-linear setup studied in this work, our computations suggest that the
range of the non-perturbative coupling needs to be restricted to the interval
$\lgb\in(-1.583,0]$. If we continue to decrease the Gauss-Bonnet coupling, then the bound on
hydrodynamics becomes weaker, which is incompatible with the expectations for the behavior of a
theory that flows from infinite to zero coupling. As is usual in holographic higher-derivative
theories, we expect that in order to (reliably) flow from an infinitely coupled theory dual to Einstein gravity
to a free thermal CFT, one would need to include an infinite tower of higher-order curvature
corrections, beyond the $R^2$ terms considered in the Gauss-Bonnet theory, or the $R^4$ terms
derived from type IIB string theory. We leave the investigation of these, and issues pertaining to
finding phenomenologically relevant applications of non-local observables and the validity of
hydrodynamics investigated in this work for the future.

\acknowledgments It is a pleasure to thank Elena C\'aceres, Michal Heller and Wilke van der Schee for useful
discussions and comments on the manuscript.  The work of BSD is supported by the National
Science Foundation (NSF) under Grant No PHY-1620610. BSD also acknowledges support from the
$\Delta$-ITP visiting program and would like to thank Umut G\"{u}rsoy and the Institute for
Theoretical Physics at Utrecht University for the warm hospitality during his extended visit. SG and
JFP are partially supported by the $\Delta$-ITP consortium, the Foundation for Fundamental Research
on Matter (FOM) and by the Netherlands Organization for Scientific Research (NWO) under the VENI
scheme, which are funded by the Dutch Ministry of Education, Culture and Science (OCW). JFP would
also like to thank the Centro de Ciencias de Benasque Pedro Pascual as well as the organizers of the
workshop ``Gravity - New perspectives from strings and higher dimensions'' for the welcoming
atmosphere and hospitality during the final stages of this project. SY would like to thank
Noisebridge Hackerspace in San Francisco, Litchee Labs in Shenzhen, and NYC
Resistor in Brooklyn for providing stimulating work environments while this work was being
completed.

%
%
\appendix

\section{Second order solutions in perturbative Gauss-Bonnet gravity}
\label{appendix-second-order}
As discussed in Section \ref{section:GBG}, the Gauss-Bonnet equations of motion can be solved at
second order in the late-time expansion to first order in $\lgb$ by writing the metric functions
$a_2$, $b_2$ and $c_2$ as $a_2 = a_2^0+\lgb\hat{a}_2$ (and similarly for the other two functions) and expanding the equations of motion to first order in $\lgb$. The resulting system of equations is solved by
\begin{eqnarray}
a_2^{0} &=& \frac{w^2}{3 v^4} \ln \left[\frac{w^2}{v^2}+1\right]+\left(\frac{v^4+w^4}{3v^5 w}\right) \arctan\left[\frac{v}{w}\right]\nonumber\\
&-&\frac{1}{18w v^6 }\left(3 \pi  v \left(v^4+w^4\right)-6 v^4 w+v^2 w^3 (3+2\ln 2)+12 v w^4+6 w^5\right),\nonumber
\end{eqnarray}
\begin{eqnarray}
b_2^{0} &=&\frac{1}{3w^2}\ln\left[\frac{(v+w)^{1/2}(v^2+w^2)^{3/4}}{v^2}\right]-\left(\frac{v-3w}{6 v w^2}\right) \arctan\left[\frac{v}{w}\right]\nonumber\\
&-&\frac{4+3 \pi }{12 v w}+\frac{\pi }{12 w^2}+\frac{1}{3 v^2}\,,\nonumber\\
\partial_v c_2^{0} &=&\frac{1}{9 v w \left(v^4-w^4\right)}\left(
v^3\ln\left[\frac{v^4}{(v+w)^2(v^2+w^2)} \right]+w^3\ln\left[\frac{4(v+w)^2}{v^2+w^2} \right]
\right)\nonumber\\
&-&\frac{v^4+2 v w^3-3 w^4}{9 v^2 w \left(v^4-w^4\right)}\arctan\left[\frac{v}{w}\right]+\frac{1}{18 v(v+w)}\left(\frac{\pi}{v}-\frac{1}{v+w}\right)\nonumber\\
&+&\frac{1}{18(v^2+w^2)}\left(\frac{5v-\pi(v-w)}{v^2}-\frac{6v+2w}{v^2+w^2} \right)+\frac{\pi v+4 w}{18 v^3 w}\,,
\label{perturbative-hydro-ord2-A}
\end{eqnarray}
and
\begin{eqnarray}
\hat{a}_2 &=& \frac{w^2 \left(3 v^4+2 w^4\right)}{9 v^8}\ln \left[\frac{2 v^6}{\left(v^2+w^2\right)^3}\right]-\left(\frac{8 v^8+5 v^4 w^4+9 w^8}{9 v^9 w}\right)\arctan\left[\frac{v}{w}\right]\nonumber\\
&+&\frac{7 w^8}{3 v^{10}}+\frac{(20+3 \pi ) w^7}{6 v^9}+\frac{19 w^6}{54 v^8}+\frac{2 w^4}{3 v^6}+\frac{(48+5 \pi ) w^3}{18 v^5}+\frac{17 w^2}{27 v^4}-\frac{8}{9 v^2}+\frac{4 \pi }{9 v w} \,, \nonumber\\
\hat{b}_2 &=& \frac{2}{3w^2}\ln\left[\frac{v^4}{(v+w)(v^2+w^2)^{3/2}}\right]+\left(\frac{2 (v-2 w)}{3 v w^2}\right)\arctan\left[\frac{v}{w}\right]\nonumber\\
&-&\frac{2 w^2}{9 v^4}-\frac{2}{3 v^2}+\frac{2 (2+\pi )}{3 v w}-\frac{\pi }{3 w^2} \,,\nonumber\\
\partial_v \hat{c}_2 &=& \frac{w^2}{3v^5}\ln\left[\frac{v^2+w^2}{4(v+w)^2}\right]+\frac{1}{9wv^2}\ln\left[\frac{(v+w)^2(v^2+w^2)}{v^4}\right]+\frac{2(v^3+w^3)}{3wv^5}\arctan\left[\frac{v}{w}\right]\nonumber\\
&+&\frac{1}{9 v(v+w)}\left(\frac{1}{2v}-\frac{1}{v+w}\right)+\frac{1}{18(v^2+w^2)}\left(\frac{4(v-w)}{v^2+w^2}+\frac{(17v+w)}{v^2}\right)\nonumber\\
&-&\frac{10 w^3}{3 v^6}+\frac{(19-9 \pi ) w^2}{27 v^5}-\frac{2}{3 v^3}-\frac{\pi }{3 v^2 w} \,,
\label{perturbative-hydro-ord2-B}
\end{eqnarray}
where we have presented the solutions for $c_2^0$ and $\hat{c}_2$ as first order derivatives due to the complexity of their integrated forms. Upon integration, the resulting integration constants are set by imposing AdS boundary conditions (see Eq. (\ref{eq:AdS-boundary-conditions})).

\section{Metric expansions}\label{expansions}

In Eddington-Finkelstein coordinates, our background is given by:\footnote{We have set the AdS radius to one, but it can be restored via dimensional analysis whenever needed.}
\begin{equation}
ds^2 = -r^2 a d\tau_{+}^2+2d\tau_{+} dr+\left(1+r\tau_{+}\right)^2e^{2(b-c)}dy^2+r^2 e^{c}d\vec{x}_{\perp}^2\,
\end{equation}
where the coefficients $a$, $b$ and $c$ are expanded as:
\bea
&&a(v,u) = a_0(v)+a_1(v)u+a_2(v)u^2+\ldots\,,\nonumber\\
&&b(v,u) = b_0(v)+b_1(v)u+b_2(v)u^2+\ldots\,,\label{EFexp}\\
&&c(v,u) = c_0(v)+c_1(v)u+c_2(v)u^2+\ldots\,,\nonumber
\eea
where
\be\label{vuinEF}
v\equiv r\tau_+^{1/3}w^{-1}\,,\qquad u\equiv \tau_+^{-2/3}w^{-1}\,.
\ee
Notice that in the above definitions we have included the dimensionful constant $w$ so that both $v$ and $u$ are dimensionless.\footnote{Recall that the energy density scales at late times like $\varepsilon(\tau)\sim\tau^{-4/3}w^4$.}
The expansion here is such that each set of coefficients $\{a_i,b_i,c_i\}$ encodes information of hydrodynamics at the given order. On the other hand, we can also express the coefficients $a$, $b$ and $c$ in a near-boundary expansion. For an asymptotically AdS metric, the coefficients take the form
\bea
&&a(v,u) = 1+\mathfrak{a}_4(u)v^{-4}+\ldots\,,\nonumber\\
&&b(v,u) = \mathfrak{b}_4(u)v^{-4}+\ldots\,,\label{nearbexps}\\
&&c(v,u) = \mathfrak{c}_4(u)v^{-4}+\ldots\,,\nonumber
\eea
so that in the limit $v\to\infty$ ($r\to\infty$) we recover AdS. The terms $\{\mathfrak{a}_4,\mathfrak{b}_4,\mathfrak{c}_4\}$ correspond to the normalizable mode of the metric so they encode information dual to the expectation value of the boundary stress-energy tensor. As such, they receive contributions at all orders in hydrodynamics, which can be seen from their definitions in terms of the $\{a_i,b_i,c_i\}$:
\bea
&&\mathfrak{a}_4(u)=\lim_{v\to\infty}v^4\left(\sum_{k=0}^{\infty}a_k(v)u^k-1\right)\,,\nonumber\\
&&\mathfrak{b}_4(u)=\lim_{v\to\infty}v^4\sum_{k=0}^{\infty}b_k(v)u^k\,,\\
&&\mathfrak{c}_4(u)=\lim_{v\to\infty}v^4\sum_{k=0}^{\infty}c_k(v)u^k\,.\nonumber
\eea
Finally, it can be checked that for empty AdS ($a=1$, $b=c=0$) the coordinate transformation
\begin{equation}\label{tauptotauz}
\tau_{+}\rightarrow \tau-z\,,\qquad r\rightarrow \frac{1}{z}\,,
\end{equation}
brings the metric to the standard form in Poincare coordinates.

Another useful form of the metric is in Fefferman-Graham coordinates:
\begin{equation}\label{FGgeneric}
ds^2 = \frac{1}{z^2}\left(-e^{\tilde{a}}d\tau^2+e^{\tilde{b}}\tau^2dy^2+e^{\tilde{c}}d\vec{x}_{\perp}^2+dz^2\right)\,
\end{equation}
where the coefficients $\tilde{a}$, $\tilde{b}$ and $\tilde{c}$ are of the form:
\bea
&&\tilde{a}(\tilde{v},\tilde{u}) = \tilde{a}_0(\tilde{v})+\tilde{a}_1(\tilde{v})\tilde{u}+\tilde{a}_2(\tilde{v})\tilde{u}^2+\ldots\,,\nonumber\\
&&\tilde{b}(\tilde{v},\tilde{u}) = \tilde{b}_0(\tilde{v})+\tilde{b}_1(\tilde{v})\tilde{u}+\tilde{b}_2(\tilde{v})\tilde{u}^2+\ldots\,,\label{FGexp}\\
&&\tilde{c}(\tilde{v},\tilde{u}) = \tilde{c}_0(\tilde{v})+\tilde{c}_1(\tilde{v})\tilde{u}+\tilde{c}_2(\tilde{v})\tilde{u}^2+\ldots\,,\nonumber
\eea
with
\be
\tilde{v}\equiv z\tau^{-1/3}w\,,\qquad\,\tilde{u}\equiv \tau^{-2/3}w^{-1}\,.
\ee
Near the boundary of AdS $\tilde{v}\to0$ ($z\to0$) we can have an alternative expansion:
\bea
&&\tilde{a}(\tilde{v},\tilde{u}) = \tilde{\mathfrak{a}}_4(\tilde{u})\tilde{v}^{4}+\ldots\,,\nonumber\\
&&\tilde{b}(\tilde{v},\tilde{u}) = \tilde{\mathfrak{b}}_4(\tilde{u})\tilde{v}^{4}+\ldots\,,\label{FGnearB}\\
&&\tilde{c}(\tilde{v},\tilde{u}) = \tilde{\mathfrak{c}}_4(\tilde{u})\tilde{v}^{4}+\ldots\,,\nonumber
\eea
Again, the leading order coefficients $\{\tilde{\mathfrak{a}}_4,\tilde{\mathfrak{b}}_4,\tilde{\mathfrak{c}}_4\}$ can be obtained from:
\bea
&&\tilde{\mathfrak{a}}_4(u)=\lim_{\tilde{v}\to0}\tilde{v}^{-4}\sum_{k=0}^{\infty}\tilde{a}_k(\tilde{v})\tilde{u}^k\,,\nonumber\\
&&\tilde{\mathfrak{b}}_4(u)=\lim_{\tilde{v}\to0}\tilde{v}^{-4}\sum_{k=0}^{\infty}\tilde{b}_k(\tilde{v})\tilde{u}^k\,,\label{abc4exps}\\
&&\tilde{\mathfrak{c}}_4(u)=\lim_{\tilde{v}\to0}\tilde{v}^{-4}\sum_{k=0}^{\infty}\tilde{c}_k(\tilde{v})\tilde{u}^k\,.\nonumber
\eea

An important difference between the Eddington-Finkelstein and Fefferman-Graham expansions is that the latter is directly in terms of the physical $\tau$, while the former is in terms of $\tau_+$, a coordinate that mixes $\tau$ and the radial coordinate $z$. This point will play an important role in the calculation of non-local observables perturbatively.

\subsection{Explicit expansions in Fefferman-Graham coordinates\label{FGexpr}}

We will consider three gravity solutions dual to Bjorken flow: Einstein gravity including 3$^{rd}$ order hydrodynamics, perturbative $\alpha'$-corrections up to second order in hydrodynamics and non-perturbative $\lgb$-corrections up to second order in hydrodynamics:
\begin{itemize}
  \item \emph{Einstein gravity.} The full gravity solution is known analytically only up to second order in hydrodynamics. However, the near-boundary metrics can be easily obtained for 3$^{rd}$ order hydrodynamics from the expected stress-energy tensor and the corresponding transport coefficients \cite{Grozdanov:2015kqa}. In particular, we find that at this order:
\end{itemize}
\bea
&&\tilde{\mathfrak{a}}_4(\tilde{u})=-\frac{3}{4}+\frac{1}{2}\tilde{u}-\frac{1}{24} \left(1+2\ln 2\right) \tilde{u}^2+\frac{1}{648} \left(3-2 \pi ^2-24 \ln 2+24 \ln^2 2\right) \tilde{u}^3\,,\nonumber\label{a4EG}\\
&&\tilde{\mathfrak{b}}_4(\tilde{u})=\frac{1}{4}-\frac{1}{2}\tilde{u}+\frac{5}{72} \left(1+2\ln 2\right) \tilde{u}^2-\frac{7}{1944} \left(3-2 \pi ^2-24 \ln 2+24 \ln^2 2\right) \tilde{u}^3\,,\nonumber\\
&&\tilde{\mathfrak{c}}_4(\tilde{u})=\frac{1}{4}-\frac{1}{72} \left(1+2\ln 2\right) \tilde{u}^2+\frac{1}{972} \left(3-2 \pi ^2-24 \ln 2+24 \ln^2 2\right)  \tilde{u}^3\,.\label{c4EG}\label{3rdEGcoeffs}
\eea
\begin{itemize}
  \item \emph{$\alpha'$-corrections.} The full gravity solution including the leading $\alpha'$-corrections and second-order hydrodynamics was obtained in \cite{Benincasa:2007tp}. Here we just write down the near-boundary coefficients explicitly:
\end{itemize}
\bea
&&\tilde{\mathfrak{a}}_4(\tilde{u})=-\frac{3}{4}+\frac{1}{2}\tilde{u}-\frac{1}{24} \left(1+2\ln 2\right) \tilde{u}^2-\left[36-\frac{639}{8}\tilde{u}+\frac{1}{48} (1133+606 \ln 2) \tilde{u}^2\right]\gamma\,,\nonumber\\
&&\tilde{\mathfrak{b}}_4(\tilde{u})=\frac{1}{4}-\frac{1}{2}\tilde{u}+\frac{5}{72} \left(1+2\ln 2\right) \tilde{u}^2+\left[12 -\frac{639}{8} \tilde{u}+\frac{5}{144} (1133+606 \ln 2) \tilde{u}^2\right]\gamma\,,\nonumber\\
&&\tilde{\mathfrak{c}}_4(\tilde{u})=\frac{1}{4}-\frac{1}{72} \left(1+2\ln 2\right) \tilde{u}^2+\left[12-\frac{1}{144} (1133+606 \ln 2) \tilde{u}^2\right]\gamma\,,\label{2ndapcoeffs}
\eea
where $\gamma = \alpha'^3 \zeta(3) / 8=\lambda^{-3/2}\zeta (3) L^6/8$. As we can see, in the limit of infinite 't Hooft coupling $\lambda\to\infty$ (or $\gamma\to0$) we recover
the coefficients for second-order hydrodynamics in Einstein gravity (\ref{3rdEGcoeffs}).

\begin{itemize}
\item \emph{$\lgb$-corrections.} The full gravity solution including non-perturbative $\lgb$-corrections and first- order hydrodynamics was obtained for the first time in the present paper. Since the transport coefficients are known non-perturbatively up to second order in hydrodynamics \cite{Grozdanov:2016fkt}, we can reconstruct the near-boundary coefficients explicitly. We find that:
\end{itemize}
\bea
\!\!\!\!\!&&\tilde{\mathfrak{a}}_4(\tilde{u})=-\frac{3}{\sqrt{2} (1+\ggb)^{3/2}}\left[1-\frac{2\ggb^2}{3} \tilde{u}+\frac{6-11 \ggb ^2-2 \ggb ^3+9 \ggb ^4+2 \ggb^2 \ln(2+2\ggb^{-1})}{36}\tilde{u}^2\right],\nonumber\\
\!\!\!\!\!&&\tilde{\mathfrak{b}}_4(\tilde{u})=\frac{1}{\sqrt{2} (1+\ggb)^{3/2}}\left[1-2\ggb^2 \tilde{u}+\frac{5(6-11 \ggb ^2-2 \ggb ^3+9 \ggb ^4+2 \ggb^2 \ln(2+2\ggb^{-1}))}{36}\tilde{u}^2\right],\nonumber\\
\!\!\!\!\!&&\tilde{\mathfrak{c}}_4(\tilde{u})=\frac{1}{\sqrt{2} (1+\ggb)^{3/2}}\left[1-\frac{6-11 \ggb ^2-2 \ggb ^3+9 \ggb ^4+2 \ggb^2 \ln(2+2\ggb^{-1})}{36}\tilde{u}^2\right],\label{2ndgbcoeffs}
\eea
where $\ggb=\sqrt{1-4\lgb}$. For $\lgb\to0$ (or $\ggb\to1$) we recover the coefficients for second-order hydrodynamics in Einstein gravity (\ref{3rdEGcoeffs}).
\section{Useful definitions}\label{crit-analytic}
\label{crit-analytic}
We can express the critical times found in the previous sections generically in terms of a few theory-specific constants $\{\hat{\eta}, \hat{\Sigma}, \hat{\Lambda}\}$, which correspond to contributions from first-, second- and third-order hydrodynamics, respectively.
\begin{itemize}
\item \textit{Einstein gravity}
\begin{align}
\hat{\eta} = \frac{1}{3}\,,&&\hat{\Sigma} = \frac{1}{18}\left(1+2\ln 2\right),&&
\hat{\Lambda} = \frac{1}{486}\left[2\pi^2-3\left(1-8(1-\ln 2)\ln 2\right)\right] .
\end{align}
\item \textit{$\alpha$'-corrections}
\begin{eqnarray}
\hat{\eta}^{(\gamma)} &=&\hat{\eta}\left(1+\gamma~\hat{\eta}^{(\gamma)}_{\epsilon}\right)=\hat{\eta}\left(1+\frac{447}{4}\gamma\right),\\
\hat{\Sigma}^{(\gamma)} &=&\hat{\Sigma}\left(1+\gamma\hat{\Sigma}^{(\gamma)}_{\epsilon}\right)=\hat{\Sigma}\left[1+\frac{\gamma}{2}\left(\frac{1037+414\ln 2}{1+2\ln 2}\right)\right],
\end{eqnarray}
where
\begin{align}
\hat{\eta}_{\epsilon}^{(\gamma)}=\frac{447}{4}\,, &&\hat{\Sigma}_{\epsilon}^{(\gamma)}=\frac{1}{2}\left(\frac{1037+414\ln 2}{1+2\ln 2}\right).
\end{align}
\item \textit{$\lgb$-corrections}\\
Here, we will write the $\lgb$ constants in terms of $\ggb = \sqrt{1-4\lgb}$,
\begin{eqnarray}
\hat{\eta}^{(\lgb)} &=&\frac{\ggb^2}{3} \,, \\
\hat{\Sigma}^{(\lgb)} &=& \frac{1}{36}\left[6+\ggb^2(1+\ggb)(9\ggb-11)+2\ggb^2\ln( 2+2\ggb^{-1})\right]\,.
\end{eqnarray}
In the limit $\lgb\rightarrow 0$, we have
\begin{eqnarray}
\hat{\eta}^{(\lgb)}&\underset{\lgb\rightarrow 0}{\approx}&\hat{\eta}\left(1+\lgb\hat{\eta}^{(\lgb)}_{\epsilon}\right)=\hat{\eta}\left(1-4\lgb\right),\\
\hat{\Sigma}^{(\lgb)}&\underset{\lgb\rightarrow 0}{\approx}&\hat{\Sigma}\left(1+\lgb\hat{\Sigma}^{(\lgb)}_{\epsilon}\right)=\hat{\Sigma}\left[1-4\lgb\left(1+\frac{3}{4(1+2\ln 2)}\right)\right],
\end{eqnarray}
where
\begin{align}
\hat{\eta}_{\epsilon}^{(\lgb)} = -4\,, && \hat{\Sigma}_{\epsilon}^{(\lgb)} = -4\left(1+\frac{3}{4(1+2\ln 2)}\right).
\end{align}
\end{itemize}

\bibliographystyle{JHEP}
\bibliography{refs}

\end{document}